\documentclass[aps,prc,reprint,showpacs,superscriptaddress,floatfix]{revtex4-1}
\usepackage{graphicx}
\usepackage{dcolumn}
\usepackage{amssymb}
\usepackage{longtable}
\usepackage{hyperref}

\begin{document}


\title{Sensitivity study of explosive nucleosynthesis in Type Ia supernovae: \\
I. Modification of individual thermonuclear reaction rates}

\author{Eduardo Bravo}
\email[]{eduardo.bravo@upc.edu}
\affiliation{Dept. F\'\i sica i Enginyeria Nuclear, Univ. Polit\`ecnica de
              Catalunya, Carrer Pere Serra 1-15, 08173 Sant Cugat del Vall\`es, Spain}

\author{Gabriel Mart\'\i nez-Pinedo}
\email[]{g.martinez@gsi.de}
\affiliation{Technische Universit{\"a}t Darmstadt, Institut f{\"u}r
  Kernphysik, Schlossgartenstr. 2, 64289 Darmstadt, Germany}
\affiliation{GSI Helmholtzzentrum f\"ur Schwerioneneforschung,
  Planckstr. 1, 64291 Darmstadt, Germany}

\date{\today}

\begin{abstract}
  \begin{description}
  \item[Background] Type Ia supernovae contribute significantly to the
    nucleosynthesis of many Fe-group and intermediate-mass
    elements. However, the robustness of nucleosynthesis obtained via
    models of this class of explosions has not been studied in depth
    until now.
  \item[Purpose] We explore the sensitivity of the nucleosynthesis
    resulting from thermonuclear explosions of massive white dwarfs
    with respect to uncertainties in nuclear reaction rates. We lay
    particular emphasis on indentifying the individual reactions rates
    that most strongly affect the isotopic products of these supernovae.
  \item[Method] We have adopted a standard one-dimensional delayed
    detonation model of the explosion of a Chandrasekhar-mass white
    dwarf, and have post-processed the thermodynamic trajectories of
    every mass-shell with a nucleosynthetic code in order to obtain
    the chemical composition of the ejected matter. We have considered
    increases (decreases) by a factor of ten on the rates of 1196
    nuclear reactions (simultaneously with their inverse reactions)
    repeating the nucleosynthesis calculations after modification of
    each reaction rate pair. We have computed as well hydrodynamic
    models for different rates of the fusion reactions of $^{12}$C and
    of $^{16}$O. From the calculations we have selected the reactions
    that have the largest impact on the supernova yields, and we have
    computed again the nucleosynthesis using two or three alternative
    prescriptions for their rates, taken from the JINA REACLIB
    database. For the three reactions with the largest sensitivity we
    have analyzed as well the temperature ranges where a modifications
    of their rates has the strongest effect on nucleosynthesis.
  \item[Results] The nucleosynthesis resulting from the Type Ia
    supernova models is quite robust with respect to variations of
    nuclear reaction rates, with the exception of the reaction of
    fusion of two $^{12}$C nuclei. The energy of the explosion changes
    by less than $\sim4\%$ when the rates of the reactions
    $^{12}\text{C}+{}^{12}\textrm{C}$ or $^{16}\text{O}+{}^{16}\text{O}$ are
    multiplied by a factor of $\times10$ or $\times0.1$. The changes in
    the nucleosynthesis due to the modification of the rates of these
    fusion reactions are as well quite modest, for instance no species
    with a mass fraction larger than 0.02 experiences a variation of
    its yield larger than a factor of two. 
We provide the sensitivity of the yields of the most abundant species with respect to the rates of
the most intense reactions with protons, neutrons, and alphas.
In general, the yields of Fe-group nuclei are more
    robust than the yields of intermediate-mass elements. 
    Among the species with yields larger than $10^{-8}$~M$_\odot$, $^{35}$S has
    the largest sensitivity to the nuclear reaction rates. 
    It is remarkable that the reactions involving elements with $Z>22$
    have a tiny influence on the supernova nucleosynthesis. Among the
    charged particle reactions, the most influential on supernova
    nucleosynthesis are
    $^{30}\text{Si}+\text{p}\rightleftarrows{}^{31}\text{P}+\gamma$,
    $^{20}\text{Ne}+\alpha\rightleftarrows{}^{24}\text{Mg}+\gamma$,
    and
    $^{24}\text{Mg}+\alpha\rightleftarrows{}^{27}\text{Al}+\text{p}$. The
    temperatures at which a modification of their rate has a larger
    impact are in the range $2\lesssim T\lesssim 4$~GK.
  \item[Conclusions] The explosion model (i.e., the assumed conditions and propagation of the
flame) chiefly determines the
    element production of Type Ia supernovae, and derived quantities
    like their luminosity, while the nuclear reaction rates used in
    the simulations have a small influence on the kinetic energy and
    final chemical composition of the ejecta. Our results show that
    the uncertainty in individual thermonuclear reaction rates cannot
    account for discrepancies of a factor of two between isotopic ratios
    in Type Ia supernovae and those in the solar system, especially
    within the Fe-group
  \end{description}
\end{abstract}



\maketitle

\section{Introduction}

Thanks to their high luminosity, Type Ia supernovae (SNIa) are used routinely
as standard candles to measure cosmological distances. They are
instrumental to our current understanding of the Universe, providing
evidence for its accelerated expansion~\cite{rie98, per98, sch98,
  per99, rie01, ton05}. 
Type Ia supernovae play also an important role
in the chemical evolution of galaxies, being responsible for most of the
Fe-group elements  and smaller amounts of
Silicon, Sulfur, Argon, and Calcium (see e.g.~\cite{tim95,thi03, mat09}). The
elemental composition is evident in optical and infrared spectra
recorded from days to months after the explosion (see e.g.~\cite{ste05, maz07}) and in X-ray
spectra of their remnants visible for
hundreds of years (for a review see~\cite{bad10}). Finally,
SNIa are one of the key targets for $\gamma$-ray astronomy, as a
source of a variety of radioactive isotopes (see e.g.~\cite{ise06}).

This is the first paper of a series in which we will study the sensitivity
of the nucleosynthesis produced in SNIa with respect to uncertainties
in nuclear data. In this paper, we study the sensitivity to variations
in rates of thermonuclear reactions (fusion reactions, radiative
captures, and transfer reactions). In forthcoming publications, we
will study the sensitivity to uncertainties in nuclear masses and in
weak interaction rates. Studies of the effect of nuclear data
uncertainties in different astrophysical scenarios have been published
from time to time during the past few decades, e.g. \cite{bah82,%
kra90, smi93, il02, izz07, the98, hix03, arc10} to cite only a few,
although none of them has dealt with SNIa. These works followed
different methodologies to test the impact of nuclear reaction
rates. For instance, ref.~\cite{the98} varied the rate of individual
nuclear reactions relevant for $^{44}$Ti nucleosynthesis in order to
determine which reactions were a prime target for the experimental
measurement of their cross sections. On the other hand ref.~\cite{hix03}
designed a numerical experiment to measure the uncertainty of the
nucleosynthesis of nova explosions. To this end, they followed a Monte
Carlo approach in which they varied simultaneously by random factors
all the reaction rates in their network.  The focus of this second
approach was on the final nova nucleosynthesis rather than in
determining the individual reactions that are most influential.
Finally, ref.~\cite{arc10} used theoretical nuclear reaction rates based on
four different nuclear mass models to determine their impact on the
r-process abundances. In this case, the emphasis was on testing
different nuclear models. In the present work, we wish to determine
the individual nuclear reactions most influential on the
nucleosynthesis of SNIa, hence we will follow the same strategy as
ref.~\cite{the98}.

At present, the favored model of SNIa is the thermonuclear explosion of
a carbon-oxygen white dwarf (WD) near the Chandrasekhar mass that
accretes matter from a companion star in a close binary
system~\cite{hil00}. Other models, such as the sub-Chandrasekhar
models or the double degenerate scenario, although not completely
ruled out, either have difficulties in explaining the gross
features of the spectrum and light curve of normal SNIa, or face
severe theoretical objections, see e.g. \cite{nug97, woo10, bra11,
  nap02, sai98, seg97}. Super-Chandrasekhar models have been
proposed to explain a few overluminous SNIa \cite{jef06, how06,
  sca10, tan10, sil11} but, given the scarcity of 
observations despite of their high intrinsic luminosity  
they are thought to represent at most a few percent of all
SNIa explosions.  Moreover, the properties of the progenitors of
super-Chandrasekhar SNIa and the explosions themselves are not well
understood. Thus, we will concentrate our efforts on the study of a
reference SNIa Chandrasekhar-mass model \cite{bad05}. 

Even though the hydrostatic evolution of SNIa progenitors lasts for
several Gyrs while the thermonuclear explosion lasts for a few seconds
at most, the outcome is nearly independent of the history of the white
dwarf prior to its explosive ignition. This fact is commonly denoted
as 'stellar amnesia' \cite{hoe03}. The only link between the white
dwarf at ignition time and its previous evolution comes through its
chemical composition ($^{12}$C, $^{16}$O, $^{22}$Ne, and other trace
species) and the distribution of hot spots that are the seeds of the
emerging thermonuclear flame. The influence of uncertain reaction
rates, specifically that of the reaction
$^{12}\text{C}(\alpha,\gamma){}^{16}\text{O}$, on the chemical
composition of massive white dwarfs has been studied by
ref.~\cite{str03}, who found that the central C/O ratio might vary by
a factor of $\sim13$ at ignition time. On the other hand, the effect of
different C/O ratios on supernova luminosity and nucleosynthesis was
studied in ref.~\cite{hoe98}, therefore accounting implicitly for a
variation on the rate of $\alpha$ capture on $^{12}$C. They found that
the C/O ratio can have a sizeable impact on the ejecta composition. On
the contrary, ref.~\cite{rop06b} reached the opposite conclusion after
analyzing the same problem with their three-dimensional deflagration
models of SNIa.

Prior to the SNIa explosion there is a phase of carbon simmering that
lasts $\sim1000$~yrs and involves temperatures below $10^9$~K. During
this phase the neutron excess of matter
can be raised due to electron captures on $^{13}$N and $^{23}$Na. The
leading thermonuclear reactions during carbon simmering are, aside
from $^{12}$C+$^{12}$C, reactions that participate in the
transmutation of $^{12}$C into $^{16}$O:
$^{12}\text{C}(p,\gamma){}^{13}\text{N}$,
$^{12}\text{C}(n,\gamma){}^{13}\text{C}$, and
$^{13}\text{C}(\alpha,n){}^{16}\text{O}$. However, the timescale of
neutronization is controlled by the $^{12}$C fusion reaction and the
rate of electron captures \cite{pir08b, ch08}.  Thus, we do not expect
that a modification of the rates of radiative captures and transfer
reactions can affect appreciably the neutronization of the white dwarf
and, hence, the final supernova composition. In this work, we will
consider only modifications of the thermonuclear reaction rates during
the explosive phase of the supernova.

The temperature range relevant for explosive nucleosynthesis in SNIa
is approximately $10^9$~K to $10^{10}$~K. However, at densities
and temperatures in excess of $\sim10^8$~g~cm$^{-3}$ and
$\sim5.5\times 10^9$~K nuclei attain a nuclear statistical equilibrium
state (NSE) in which the chemical composition, 
for given temperature, density and electron mole fraction, is
determined by nuclear bulk properties (masses and partition
functions), i.e. it does not depend on the reaction rates. In these
conditions, NSE erases any imprint of the previous thermodynamic
evolution of matter, and reaction rates do not play any role until
matter leaves NSE (freeze-out process). The minimum temperature
relevant for nucleosynthesis in SNIa depends on the type of combustion
front.
For a detonation, a shock heats the fuel
to temperatures $\gtrsim2\times 10^9$~K, the precise value depending mainly
on density, {\sl before} nuclear reactions start modifying the chemical
composition. On the other hand, the process of combustion {\sl within} a subsonic flame
presents two different phases. Below a critical temperature,
$T_{\text{crit}}\sim 2$--$5\times 10^9$~K, the matter temperature is set by heat
diffusion from the hot ashes, while above $T_{\text{crit}}$ the
nuclear energy released by combustion 
 dominates over heat diffusion. Thus, we do not
expect modifying the thermonuclear reaction rates below
$\sim 1$--$2\times 10^9$~K to have an impact on the final chemical composition.

The plan of the paper is as follows. In the next Section, we detail
the methodology used to achieve our goals. We describe the
post-processing code used to integrate the nuclear evolutionary
equations, the characteristics of our reference SNIa model, the
selection of the nuclear reactions to test for variations in their
rates, and the ways in which we have modified these rates. In
Section~\ref{sensfus}, we present the results of the sensitivity study
with respect to the fusion reactions of $^{12}$C, $^{16}$O, and the
$3\alpha$ reaction, which are the reactions that rule the initial steps of
thermonuclear combustion in SNIa. We test modifications of the first
two reaction rates for effects on the propagation of the flame during a
SNIa explosion. In Section~\ref{senspar}, we present the results of
the sensitivity study with respect to thermonuclear reaction rates
involving protons, neutrons, and $\alpha$ particles. We have followed
different strategies in modifying these reaction rates, using either a
fixed enhancement factor or a temperature dependent one. We have
tested as well the use of different prescriptions for the most
influential reaction rates, taken from recent literature. For a few
reactions we have explored the temperature range where a modification
of their rates have a stronger impact on the supernova
yields. Finally, in Section~\ref{conc}, we summarize and give our
conclusions.

\section{Methodology}

\subsection{Integration of the nuclear evolutionary equations}

\begin{table*}[!htb]
 \caption{Nuclear network
 \label{tab1}}
 \begin{ruledtabular}
 \begin{tabular}{lrr@{\hspace{0.5truecm}}|lrr@{\hspace{0.5truecm}}|lrr@{\hspace{0.5truecm}}|lrr}
$Z$ & $A_{\text{min}}$ & $A_{\text{max}}$ & $Z$ & $A_{\text{min}}$ & $A_{\text{max}}$ & $Z$ &
$A_{\text{min}}$ & $A_{\text{max}}$ & $Z$ & $A_{\text{min}}$ & $A_{\text{max}}$ \\
 \hline
 n &  1 &  1 & Al & 22 & 36 & Fe & 49 & 63 &  Y & 79 & 101 \\
 H &  1 &  4 & Si & 24 & 38 & Co & 51 & 65 & Zr & 81 & 101 \\
He &  3 &  9 &  P & 26 & 40 & Ni & 53 & 69 & Nb & 85 & 101 \\
Li &  4 & 11 &  S & 28 & 42 & Cu & 55 & 71 & Mo & 87 & 101 \\
Be &  6 & 14 & Cl & 30 & 44 & Zn & 57 & 78 & Tc & 89 & 101 \\
 B &  7 & 17 & Ar & 32 & 46 & Ga & 61 & 81 & Ru & 91 & 101 \\
 C &  8 & 20 &  K & 34 & 49 & Ge & 63 & 83 & Rh & 93 & 101 \\
 N & 10 & 21 & Ca & 36 & 51 & As & 65 & 85 & Pd & 95 & 101 \\
 O & 12 & 23 & Sc & 38 & 52 & Se & 67 & 87 & Ag & 97 & 101 \\
 F & 14 & 25 & Ti & 40 & 54 & Br & 69 & 90 & Cd & 99 & 101 \\
Ne & 16 & 27 &  V & 42 & 56 & Kr & 71 & 93 & In & 101 & 101 \\
Na & 18 & 34 & Cr & 44 & 58 & Rb & 73 & 99 &    &    &    \\
Mg & 20 & 35 & Mn & 46 & 60 & Sr & 77 & 100 &    &    &    \\
 \end{tabular}
\end{ruledtabular}
\end{table*}

We have computed the chemical composition of a reference SNIa model
with the nucleosynthetic code CRANK (Code for the Resolution of an
Adaptive Nuclear networK). CRANK is a post-processing code that
integrates the temporal evolution of a nuclear network for given
thermal and structural (density) time profile, and initial
composition. We have selected the nuclear reactions that
contribute most to the synthesis of abundant species. Then, we have
recomputed the nucleosynthesis modifying the rate of each one of the
selected reactions.

The inputs to CRANK are the nuclear data and the thermodynamic
trajectories, as a function of time, of each mass shell of the
supernova model. The evolutionary equations for the nuclear composition follow the time
evolution of the molar fraction, $Y_i$, or abundance of each species
until the temperature falls below $10^8$~K, after which time the
chemical composition is no longer substantially modified. 
The nuclear network is integrated with an implicit, iterative method with adaptive time steps.
The iterative procedure ends when the molar abundances of all species
with $Y_i>10^{-14}$~mol~g$^{-1}$ have converged to better than a relative variation of  $10^{-6}$.

The nuclear species present in the network are dynamically determined during
the calculation. Initially, the network is defined by those species with
an appreciable abundance ($>10^{-24}$~mol~g$^{-1}$) plus
n, p, and alphas and the nuclei that can be reached from any of the
abundant species by any one of the reactions included in the network.
A reaction rate is included in the network only if the predicted
change of a molar abundance in the next time step, $\Delta t$, is
larger than a threshold:
\begin{equation}
  N_{\text{A}}\rho\langle\sigma v\rangle Y_i Y_j \Delta t > 10^{-20}~\text{mol}~\text{g}^{-1}\,.
\label{eqthreshold}
\end{equation}
\noindent A similar method of integration of the nuclear evolutionary equations using an adaptive
network has been described in ref.~\cite{rau02}.

Our nuclear network consists of a maximum of 722 nuclei, from free
nucleons up to $^{101}$In, linked by three fusion reactions:
$3\alpha$, $^{12}$C+$^{12}$C, and $^{16}$O+$^{16}$O, electron and
positron captures, $\beta^-$ and $\beta^+$ decays, and 12 reactions
per each nucleus with $Z\ge6$: 
$\left(\text{n},\gamma\right)$, $\left(\text{n},\text{p}\right)$,
$\left(\text{n},\alpha\right)$, $\left(\text{p},\gamma\right)$,
$\left(\text{p},\text{n}\right)$, $\left(\text{p},\alpha\right)$,
$\left(\alpha,\gamma\right)$, $\left(\alpha,\text{n}\right)$,
$\left(\alpha,\text{p}\right)$, $\left(\gamma,\text{n}\right)$,
$\left(\gamma,\text{p}\right)$, and $\left(\gamma,\alpha\right)$.  We
show the nuclear network in Table~\ref{tab1}. From the whole set of
reactions that might be included in the calculations, only 3138 enter
effectively into the reaction network equations during the integration
of the thermodynamic trajectories in our SNIa model.

The thermonuclear reaction rates, nuclear masses and partition
functions are taken from the REACLIB compilation~\cite{cyb10}. Both
theoretical and experimental thermonuclear reaction rates are fitted
in the JINA~\footnote{http://groups.nscl.msu.edu/jina/reaclib/db/} REACLIB library by an analytic
function with seven
parameters. The fits are usually better than 5\% although deviations
up to 30\% are possible. The authors estimate an additional
uncertainty typically of order 30\% in the original reaction
rates. Electron screening to thermonuclear reactions in the strong,
intermediate and weak regimes was taken into account \cite{ito79,
  sal69}. In general, in the conditions achieved during thermonuclear
supernova explosions the electron screening factors are small
\cite{kho97}. Weak interaction rates were taken from \cite{ffn82,
  mar00}.

\subsection{Type Ia supernova model}\label{secsnia}

\begin{figure*}[!htb]
\includegraphics[width=8truecm]{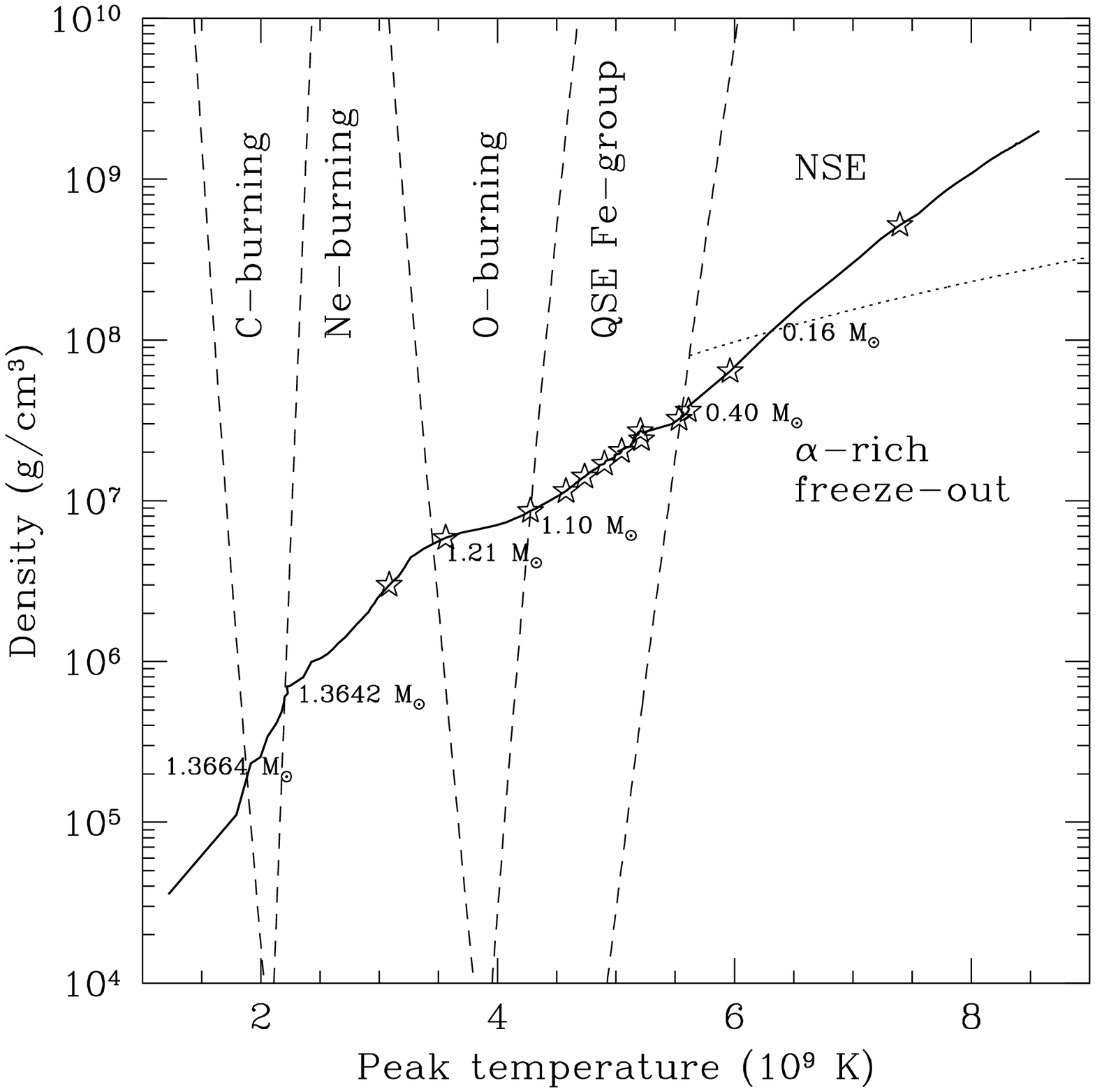}
\includegraphics[width=8truecm]{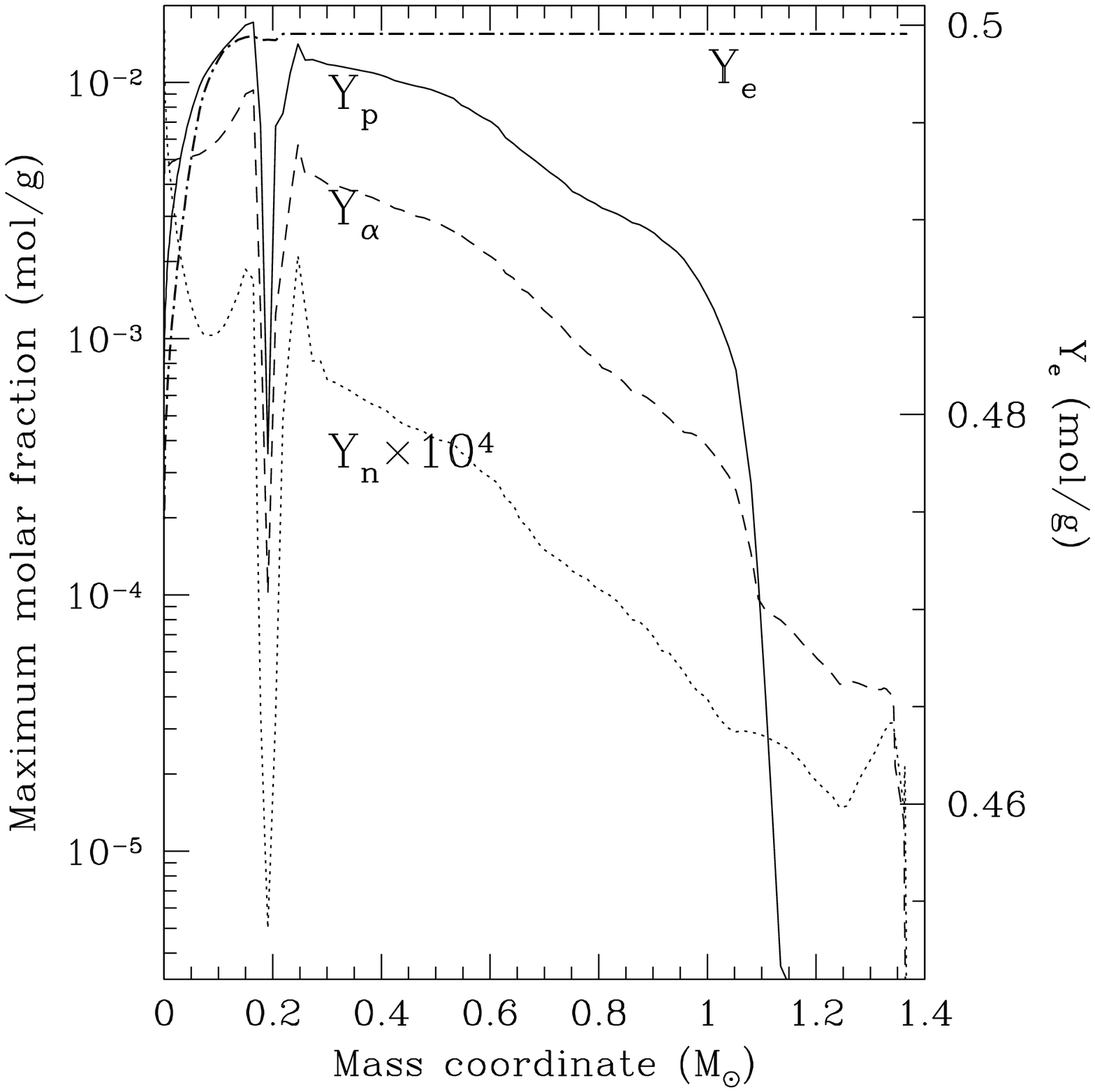}
\caption{
Profiles of physico-chemical properties accross the reference model, as a function of the
Lagrangian mass coordinate (zero at the center).
\textbf{Left}: Peak temperature and density achieved at each mass shell during the supernova
explosion (thick solid line). Star marks have been located every 0.1~M$_\odot$, with the center of
the white dwarf at the top right end of the solid line, and the surface at its bottom left end.
The $\rho-T$ plane has been divided according to approximate locations of different explosive
nucleosynthetic processes (dashed and dotted lines). We indicate as well the Lagrangian mass
coordinate at which the solid line crosses the dashed and dotted lines.
\textbf{Right}: Maximum molar fractions of neutrons, protons, and alphas achieved at a given
mass coordinate at any time during the explosion, and final electron mole number (dot-dashed line).
Note that the neutron molar fraction has been
scaled up by a factor of $10^4$ for presentation purposes.
\label{fig1}}
\end{figure*}

Our reference SNIa model is the one-dimensional delayed-detonation
model DDTc in \cite{bad05c}, characterized by its deflagration-to-detonation (DDT) transition
density, $\rho_{\text{DDT}}= 2.2\times 10^7$~g~cm$^{-3}$. The supernova progenitor is a
Chandrasekhar mass white dwarf of central density $1.8\times
10^9$~g~cm$^{-3}$ and uniform composition: 49.5\% $^{12}$C, 49.5\%
$^{16}$O, and 1\% $^{22}$Ne by mass. In this model the flame begins
as a subsonic deflagration flame near the center of the star. As the
flame propagates through the star, the pressure rises and the star
expands. When the flame reaches a zone with a low enough density,
$\rho_{\text{DDT}}$, there is a
transition to a supersonic detonation that burns most of the remaining
fuel. Finally, the nuclear energy released is enough to unbind the
whole star and eject its matter into the interstellar medium. 
In Fig.~\ref{fig1} we show the profiles of the most relevant physico-chemical quantities affecting
the nucleosynthesis. 

\begin{figure}[!htb]
\includegraphics[width=8truecm]{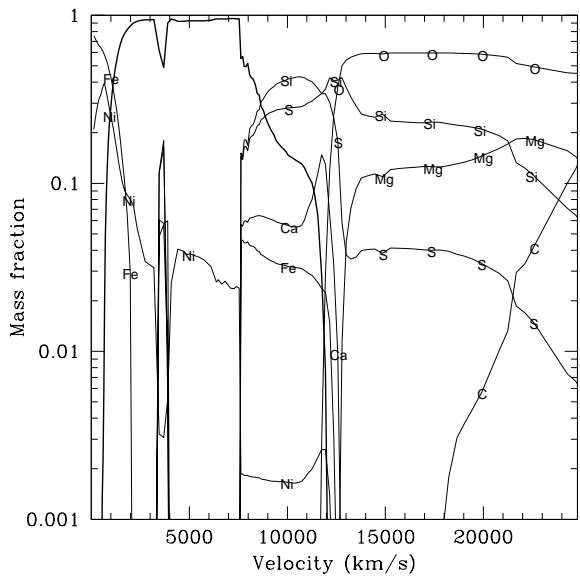}
\caption{
Chemical composition of the reference model as a function of the final velocity. The
curves labelled as Fe and Ni include only stable isotopes. The thick curve is the mass fraction of
$^{56}$Ni.
\label{fig2}}
\end{figure}

This kind of SNIa model generates a layered structure (see Fig.~\ref{fig2}) in which the inner
several tenths of a solar mass achieve maximum temperatures high enough 
($T_\textrm{max}\gtrsim5.5\times10^9$~K)
to process
matter into NSE, undergoing copious
electron captures. When matter expands the composition is relaxed out
of NSE and consists mainly of iron group elements with isotopic
fractions determined by the electron mole number resulting from the
electron captures phase. 
As can be seen in Fig.~\ref{fig1}, in our reference model the electron captures modify the
progenitor electron mole number only in the central $\sim0.1$~M$_\odot$.
The zone were the transition from deflagration to detonation takes place, at a mass coordinate of
$\sim0.2~\mathrm{M}_\odot$, can be identified by the
trough in the $Y_\mathrm{p}$, $Y_\mathrm{\alpha}$, and $Y_\mathrm{n}$ profiles.
The central $0.4~\mathrm{M}_\odot$ reach NSE, from which roughly $0.24~\mathrm{M}_\odot$
experience a moderately $\alpha$-rich freeze-out. 
Shortly after the detonation forms, it propagates fast through the white dwarf, which
has no time to relax its structure before the combustion front burns most of the remaining
fuel (this condition can be identified in Fig.~\ref{fig1} by the crowding of the star
symbols between $T_9\sim4$ and $\sim5.5$). 
Between Lagrangian mass coordinates of $\sim0.4$ and $\sim1.1~\mathrm{M}_\odot$ the peak
temperatures and densities are high enough to experience Si-burning and achieve quasi-statistical
equilibrium (QSE) of the Fe-group, although this group does not achieve equilibrium with the
Si-group. Farther out from the center, a tinier amount of mass is subject to
explosive oxygen and neon burning, and only a few thousandths of a solar mass
experience only explosive carbon burning. The mass of unburned carbon ejected by the supernova
explosion is on the same order, in agreement with the upper limits deduced by \cite{fol11}. We
note that all the nucleosynthetic processes deemed relevant in SNIa feature in our reference model.

For reference, we give in Table~\ref{tab2} the nucleosynthesis
obtained for this supernova model. The composition given in this and
forthcoming tables corresponds to a time of one day after beginning of the
explosion, hence there appear radioactive as well as stable
nuclides. We have included in this table all nuclides whose ejected
mass is $m_{i}>10^{-5}$~M$_\odot$, with the exception of $^{26}$Al,
that has been included because it is an interesting radionuclide.
The ejected mass of $^{56}$Ni is 0.675~M$_\odot$, and the kinetic energy of the ejecta is
$1.16\times10^{51}$~erg, both values deemed typical for normal bright SNIa. The resulting chemical
composition (Fig.~\ref{fig2}) compares well with the abundance stratification induced from
observations of normal SNIa as, for instance, SN2003du (e.g., Fig.~8 in \cite{tan11}, who estimated
that the ejected mass of $^{56}$Ni was 0.65~M$_\odot$). Model DDTc also provides an
excellent match to the X-ray spectrum of the remnant of SN1572 (Tycho), a prototype of SNIa
(see Fig.~7 in \cite{bad06}).

\begin{table*}
 \caption{Nucleosynthesis of the reference Type Ia supernova model
 \label{tab2}}
 \begin{ruledtabular}
 \begin{tabular}{lcc@{\hspace{0.5truecm}}|lcc}
 Nucleus & Ejected mass & $T$ range\footnotemark[1] &
 Nucleus & Ejected mass & $T$
 range\footnotemark[1] \\
 & (M$_\odot$) & (GK)  &   &  (M$_\odot$) & (GK) \\
 \hline
 $^{12}$C  &  $2.71\times 10^{-3}$ & destroyed & $^{39}$K  &
 $3.98\times 10^{-5}$ & 2.6--4.0 \\
 $^{16}$O  &  $1.12 \times 10^{-1}$ & destroyed & $^{40}$Ca &  $3.62 \times 10^{-2}$ & 4.0--5.2 \\
 $^{20}$Ne &  $2.16 \times 10^{-3}$ & 2.0--2.8 & $^{44}$Ti &  $3.25 \times 10^{-5}$ & 3.8--5.6 \\
 $^{23}$Na &  $1.60 \times 10^{-5}$ & 2.0--3.2 & $^{48}$V  &  $3.31 \times 10^{-4}$ & 4.2--5.2 \\
 $^{24}$Mg &  $1.80 \times 10^{-2}$ & 2.4--3.4 & $^{49}$V  &  $1.55 \times 10^{-5}$ & 4.2--5.2 \\
 $^{25}$Mg &  $1.55 \times 10^{-5}$ & 2.0--3.4 & $^{50}$Cr &  $1.15 \times 10^{-4}$ & 4.0--5.2 and
$>6.0$ \\
 $^{26}$Mg &  $2.72 \times 10^{-5}$ & 2.0--3.2 & $^{51}$Cr &  $3.82 \times 10^{-5}$ & 3.8--5.6 \\
 $^{26}$Al &  $1.21 \times 10^{-7}$ & 2.0--3.0 & $^{52}$Mn &  $6.15 \times 10^{-3}$ & 4.2--5.2 \\
 $^{27}$Al &  $4.58 \times 10^{-4}$ & 2.2--3.4 & $^{53}$Mn &  $6.13 \times 10^{-4}$ & 4.4--5.2 and
$>6.0$ \\
 $^{28}$Si &  $2.29 \times 10^{-1}$ & 2.8--5.0 & $^{54}$Fe &  $3.91 \times 10^{-2}$ & 4.2--5.2 and
$>6.0$ \\
 $^{29}$Si &  $4.51 \times 10^{-4}$ & 2.2--3.6 & $^{55}$Fe &  $5.27 \times 10^{-3}$ & 4.2--5.2 and
$>6.0$ \\
 $^{30}$Si &  $8.22 \times 10^{-4}$ & 2.4--3.6 & $^{56}$Fe &  $2.69 \times 10^{-3}$ & $>4.8$ \\
 $^{31}$P  &  $2.42 \times 10^{-4}$ & 2.4--3.8 & $^{57}$Co &  $1.37 \times 10^{-2}$ & $>5.0$ \\
 $^{32}$S  &  $1.43 \times 10^{-1}$ & 3.2--5.0 & $^{56}$Ni &  $6.75 \times 10^{-1}$ & $>4.8$ \\
 $^{33}$S  &  $1.78 \times 10^{-4}$ & 2.6--4.0 & $^{58}$Ni &  $3.08 \times 10^{-2}$ & $>5.2$ \\
 $^{34}$S  &  $1.05 \times 10^{-3}$ & 2.6--3.8 & $^{59}$Ni &  $4.34 \times 10^{-4}$ & $>5.0$ \\
 $^{35}$Cl &  $5.51 \times 10^{-5}$ & 2.4--4.0 & $^{60}$Ni &  $6.37 \times 10^{-3}$ & $>5.2$ \\
 $^{36}$Ar &  $3.42 \times 10^{-2}$ & 3.6--5.0 & $^{61}$Ni &  $1.58 \times 10^{-4}$ & $>5.2$ \\
 $^{37}$Ar &  $1.27 \times 10^{-5}$ & 2.6--4.2 & $^{62}$Ni &  $7.00 \times 10^{-4}$ & $>5.2$ \\
 $^{38}$Ar &  $3.47 \times 10^{-4}$ & 3.2--4.0 & & & \\
 \end{tabular}
\end{ruledtabular}
\footnotetext[1]{Range of maximum temperatures achieved in the shells in which 90\%
of each nuclide is produced.}
\end{table*}

The maximum abundances of free protons, neutrons, and
$\alpha$-particles attained during the explosion are shown in the
right panel of Fig.~\ref{fig1}, as a function of the Lagrangian mass
coordinate within the exploding white dwarf. These profiles can be
used to gain insight into the expected sensitivities of the
nucleosynthesis with respect to different types of nuclear reactions,
to be discussed in the next chapters. Neutrons are always the less
abundant nucleons by $\sim5-6$ orders of magnitude, thus we expect
that the nucleosynthesis will not be too sensitive to reactions with
neutrons except, perhaps, in the outer $\sim0.1$~M$_\odot$. Note that
neutrons are relatively abundant in the very center of the white
dwarf, because of the lower $Y_\mathrm{e}$ that results from efficient
electron captures in NSE matter at high density, but nucleosynthesis
in these layers is not expected to be sensitive to the rate of any
particular reaction with neutrons because the chemical composition
there is controlled by the Saha equation until matter cools to low
temperatures.  Protons and $\alpha$-particles have similar abundances
within the inner $\sim1.1$~M$_\odot$, although their maximum molar
fractions decrease steadily outwards within the detonated matter
($M\gtrsim0.25$~M$_\odot$). Beyond $\sim1.1$~M$_\odot$, the maximum
abundance achieved by protons is much lower than that of
$\alpha$-particles. The maximum temperatures attained in these layers
stay below $\sim4.4\times10^9$~K, implying that the thermonuclear
combustion hardly goes beyond O-burning. Thus, we expect that the
products of O-burning will be mostly sensitive to reactions with
$\alpha$-particles.

\begin{figure*}[!htb]
\includegraphics[width=8truecm]{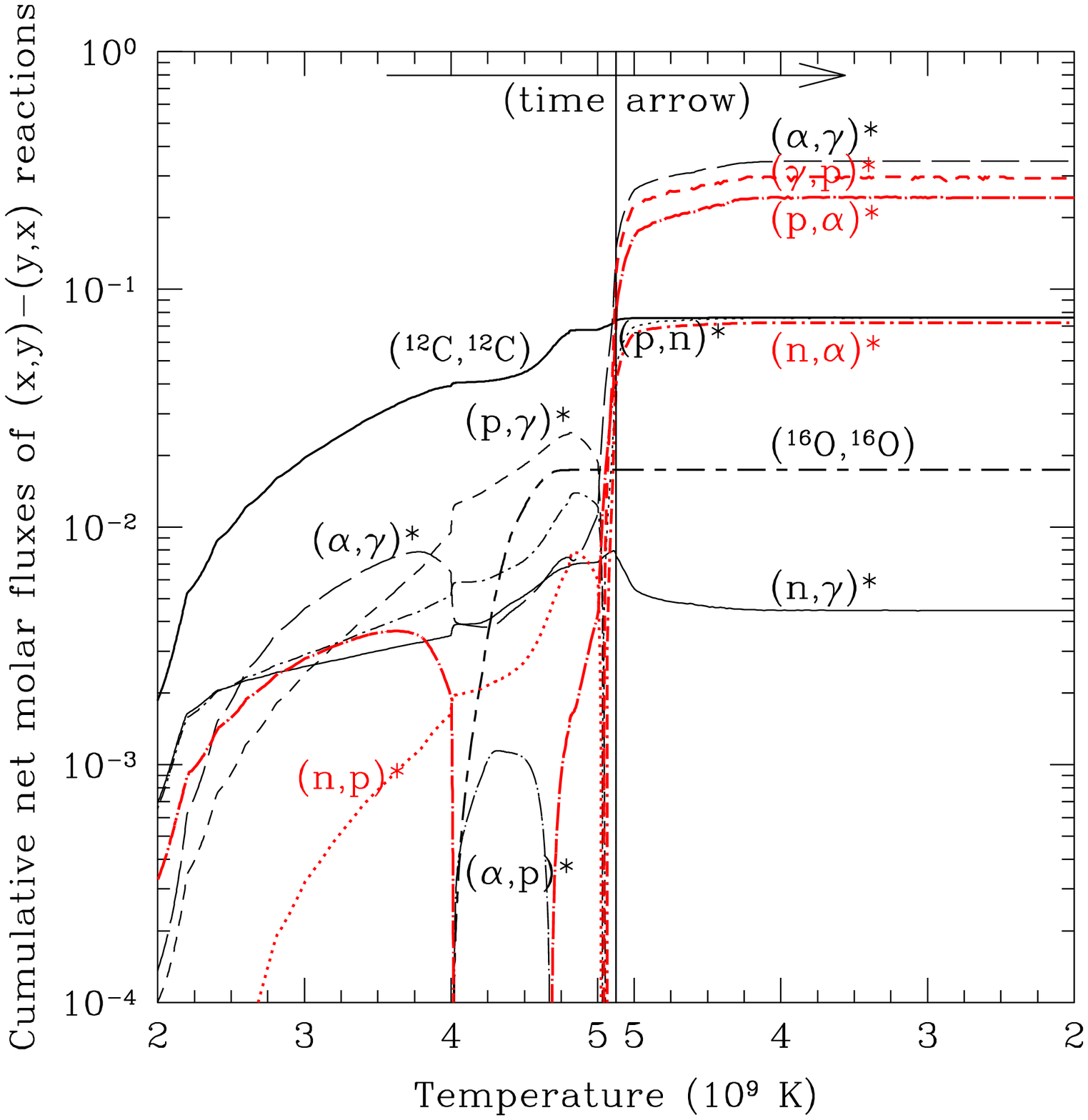}
\includegraphics[width=8truecm]{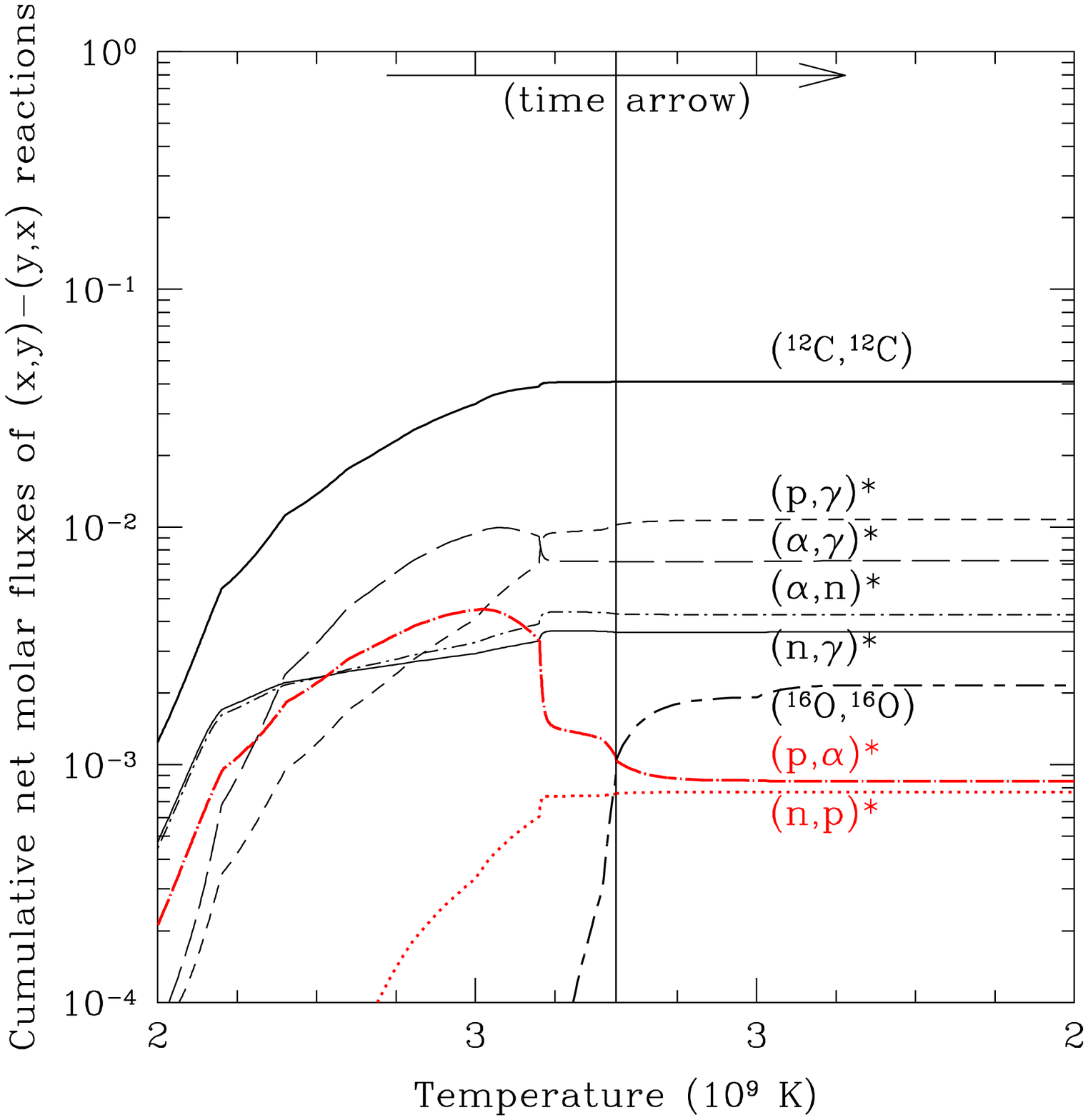}
\caption{(Color online) Cumulative net molar fluxes of direct reactions minus inverse reactions,
grouped by
reaction type, compared to the $^{12}$C-fusion and $^{16}$O-fusion reaction fluxes, as
functions of temperature. Each reaction type is identified by a different line type and a label, in
each label an $\ast$ is drawn to recall that the fluxes take into account direct and inverse
reactions.
Thin (black) curves represent those cases in which the (accumulated) molar flux of the direct
reaction is
larger than that of the inverse reaction, while thick (red) curves belong to the opposite case.
Vertical
lines are drawn to separate the heating phase from the cooling phase. 
\textbf{Left:} Evolution at Lagrangian mass coordinate of $0.5$~M$_\odot$. 
\textbf{Right:} Evolution at Lagrangian mass coordinate of $1.2$~M$_\odot$. 
\label{fig3}}
\end{figure*}

The above analysis can be complemented with an examination of the molar fluxes due to different
reaction types, e.g. $\left(\textrm{p},\textrm{n}\right)$, etc.
In Fig.~\ref{fig3} we show the evolution of the net molar fluxes in two representative mass shells
of our SNIa model, grouped by reaction type. The net molar fluxes of a given reaction type in a
mass shell are accumulated in time according to:
\begin{equation}
 \sum \left[ \int\rho N_{\text{A}}
\langle\sigma v\rangle_{jk} Y_j Y_k 
\text{d}t\right]\,,
\end{equation}
where $\rho$ is density, $N_{\text{A}}$ is Avogadro's number, $Y_j$ is
the molar fraction of species $j$, and the time integral extends from thermal runaway until the
temperature goes below $10^8$~K. The summation extends to all reactions of the given type, from
which their inverse reactions are subtracted, e.g. in the computation of the net molar fluxes of
the $\left(\textrm{p},\textrm{n}\right)$ type reactions all the
$\left(\textrm{n},\textrm{p}\right)$ reactions are considered inverse reactions and their
contributions are deducted from those of the direct, $\left(\textrm{p},\textrm{n}\right)$,
reactions. 

In the left panel of Fig.~\ref{fig3} we show the net molar fluxes in a
mass shell located at a Lagrangian mass coordinate of
$0.5$~M$_\odot$. This layer was hit by the detonation wave $\sim2.5$~s
after central thermal runaway, when its density was
$1.17\times10^7$~g~cm$^{-3}$, and heated to $\gtrsim2\times10^9$~K by
the shock front associated with the detonation. Above this
temperature, it is the energy release by nuclear reactions which
controls the evolution of temperature. The temperature rises very fast
at the beginning due to rapid burning of carbon and oxygen, mainly to
produce silicon and sulfur. About 1~ms after being shocked, a maximum
temperature of $5.12\times10^9$~K is achieved. Later, matter expands
and cools with a longer timescale (it takes 0.1~s to cool by
$2\times10^9$~K) while most of the nuclear reactions are nearly in
equilibrium with their inverse reactions. During the heating phase, it
is the $^{12}\textrm{C}+{}^{12}$C reaction which dominates the nuclear
fluxes, followed by radiative captures of protons and
$\alpha$-particles once the temperature exceeds
$\sim2.5\times10^9$~K. Compared to the plethora of reactions with
light particles unleashed by the carbon fusion reaction, the
contribution of the $^{16}\textrm{O}+{}^{16}$O reaction is quite
modest until the temperature exceeds $\gtrsim4\times10^9$~K. Above
$\sim5\times10^9$~K there is a sharp increase in the cumulative molar
fluxes belonging to $\left(\alpha,\gamma\right)$,
$\left(\gamma,\textrm{p}\right)$, and $\left(\textrm{p},\alpha\right)$
reactions, which reach similar levels. On the other hand, during the
cooling phase there is little additional contribution to the net molar
fluxes, and $\left(\textrm{p},\textrm{n}\right)$ and
$\left(\textrm{n},\alpha\right)$ reactions attain a level similar to
that of $^{16}\textrm{O}+{}^{16}$O, while
$\left(\textrm{n},\gamma\right)$ reactions are the ones that process
the smallest mass. Note that, in this mass shell, the final cumulative
molar flux due to the $^{16}$O fusion reaction is about a factor four
smaller than that due to the $^{12}$C fusion reaction. Taking this
mass shell as representative of layers that experience incomplete
Si-burning, we expect that the products of this nucleosynthetic
process will be most sensitive to $\left(\alpha,\gamma\right)$,
$\left(\gamma,\textrm{p}\right)$, and $\left(\textrm{p},\alpha\right)$
reactions, and their inverses. Note that in shells that achieve a
temperature high enough to reach NSE all the molar fluxes established
prior to NSE are irrelevant, because NSE erases all memory
of previous nuclear processes with the exception of weak interactions.

The right panel of Fig.~\ref{fig3} shows the net molar fluxes in a
mass shell located at a Lagrangian mass coordinate of $1.2$~M$_\odot$. In
this case the maximum temperature achieved was $3.45\times10^9$~K,
because the density at the time of detonation impact (at
$t\sim2.64$~s) was only $2.4\times10^6$~g~cm$^{-3}$. Due to the small
value of the maximum temperature, the $^{12}\textrm{C}+{}^{12}$C
reaction dominates the molar fluxes at all times.  Oxygen burning is
incomplete, the final molar flux due to the $^{16}$O fusion reaction
being about 20 times smaller than that due to the $^{12}$C fusion
reaction. Even $\left(\textrm{n},\gamma\right)$ reactions process more
matter than the $^{16}$O fusion reaction. Taking this mass shell as
representative of layers that do not go beyond carbon burning, we
expect that the products of this nucleosynthetic process will be most
sensitive to the rate of the $^{12}\textrm{C}+{}^{12}$C reaction and,
to a lesser extent, to $\left(\textrm{p},\gamma\right)$ and
$\left(\alpha,\gamma\right)$ reactions.

\subsection{Selection of the nuclear reactions}\label{secselec}

As explained before, only 3138 nuclear reactions exceed the threshold
of Eq.~\ref{eqthreshold} and are actually included in the
nucleosynthesis calculation. However, most of these reactions
contribute negligibly to the determination of the final chemical
composition of the supernova ejecta. In order to determine the most
relevant reactions, we define the total mass processed by a nuclear
reaction, between particle $k$ and nucleus $j$, in all the mass shells
of the supernova model $\mathfrak{M}_{jk}$: 

\begin{equation}
 \mathfrak{M}_{jk} = \sum_{\alpha} \left[M_{\alpha}\int\rho \langle\sigma v\rangle_{jk}
N_{\text{A}} Y_j Y_k
\left(A_j + A_k\right)
\text{d}t\right]\,,
\label{eqmassflow}
\end{equation}
\noindent where $M_{\alpha}$ is the mass of shell $\alpha$ of the supernova model, and $A_j$ is the
baryon
number of species $j$.
In the computation of the integral we have not taken into account
reactions above $5\times 10^9$~K, because at such temperatures the
direct and inverse reactions are in equilibrium, causing the nuclear abundances to be determined by
properties of the nuclei involved (mass,
partition function) instead of the reaction rates.
For mass shells that went through NSE, the computation of the integral in
Eq.~\ref{eqmassflow} starts when the temperature drops below $5\times 10^9$~K, since their
chemical composition is insensitive to the nuclear history prior to the NSE state (with the
exception of weak interactions, whose effect is not addressed in the present work).

The reactions we have selected for careful study are the three fusion reactions plus
those for which $\mathfrak{M}_{jk} \ge 10^{-8}$~M$_\odot$. This
warrants that we test all the reactions able to contribute
significantly to the synthesis of every species whose yield is larger
than the chosen $10^{-8}$~M$_\odot$. Each time we integrate the
nuclear evolutionary equations we modify by the same factor the direct and inverse reactions. 
Following this procedure, we find that the nucleosynthesis at this chosen level could be
sensitive to 1096 (pairs of) reactions in addition to the above mentioned three
fusion reactions.

Table~\ref{tab3} gives the masses processed by the three fusion
reactions and the top ten radiative captures and transfer
reactions, where the masses processed by the inverse reactions have been subtracted from those of
the direct reactions. 
The quoted values
of $\mathfrak{M}_{jk}$ give a quite generous upper limit of the
impact these reactions might have on the resulting nucleosynthesis of
the supernova, as the subsequent nuclear reactions destroy the products of earlier reactions.
As we will see in the following, the top ten reactions
listed in Table~\ref{tab3} are not in fact the most influential
reactions.

\begin{table*}
\caption{Masses processed by the fusion reactions and the top ten radiative captures and transfer
reactions 
\label{tab3}}
\begin{ruledtabular}
\begin{tabular}{lc@{\hspace{0.5truecm}}|lc}
Reaction & $\mathfrak{M}_{jk}$ $\left(\text{M}_\odot\right)$ &
Reaction & $\mathfrak{M}_{jk}$ $\left(\text{M}_\odot\right)$ \\
\hline
$^{12}\textrm{C}+{}^{12}$C & 0.524 &
$^{29}\text{Si}+\alpha\rightleftarrows{}^{32}\text{S}+\text{n}$ & 0.70 \\
$^{16}\textrm{O}+{}^{16}$O & 0.198 &
$^{30}\text{P}+\alpha\rightleftarrows{}^{33}\text{S}+\text{p}$ & 0.68 \\
$3\alpha$ & $6.7\times10^{-4}$ & 
$^{29}\text{Si}+\text{p}\rightleftarrows{}^{30}\text{P}+\gamma$ & 0.67 \\
$^{28}\text{Si}+\alpha\rightleftarrows{}^{32}\text{S}+\gamma$ & 0.93 &
$^{32}\text{S}+\alpha\rightleftarrows{}^{35}\text{Cl}+\text{p}$ & 0.65 \\
$^{28}\text{Si}+\text{p}\rightleftarrows{}^{29}\text{P}+\gamma$ & 0.84 &
$^{33}\text{S}+\alpha\rightleftarrows{}^{36}\text{Ar}+\text{n}$ & 0.64 \\
$^{29}\text{P}+\alpha\rightleftarrows{}^{32}\text{S}+\text{p}$ & 0.83 &
$^{27}\text{Al}+\alpha\rightleftarrows{}^{30}\text{Si}+\text{p}$ & 0.63 \\
$^{28}\text{Si}+\alpha\rightleftarrows{}^{31}\text{P}+\text{p}$ & 0.77 \\
\end{tabular}
\end{ruledtabular}
\end{table*}

\subsection{Modification of the reaction rates}

As a first approach to study the sensitivity to the different reaction
rates, we modify them, one by one, by a fixed factor,
either equal to $f_0=10$ or $f_0=0.1$, repeating the
nucleosynthesis calculation for each variation. As mentioned
previously, each time we modify the rate of a reaction we modify as
well by the same factor the rate of the inverse reaction, in order to
maintain detailed balance.

The Gamow energies in the reactions that play a significant role in
the nucleosynthesis of Type Ia supernovae go from a few tenths of a
MeV (for instance, $E_0=0.39$~MeV for the
$^{12}\text{C}+\text{p}\rightleftarrows{}^{13}\text{N}+\gamma$
reaction at $T=10^9$~K) to nearly ten MeV (e.g., $E_0=8.50$~MeV for the
$^{62}\text{Zn}+\alpha\rightleftarrows{}^{65}\text{Ga}+\text{p}$
reaction at $T=5\times10^9$~K). 
It is expected, both from theoretical and experimental arguments, that the uncertainties in the
rates at low temperatures are larger than at high temperatures 
\cite{hof99}.
Most of the theoretical reaction rates we have used are based on an
statistical model of nuclei, which assumes formation of a compound nucleus with a high level
density, a condition generally satisfied at high temperatures. Furthermore, experimental
measurements of nuclear cross sections involving high-Z nuclei are generally difficult to perform
at energies below the Coulomb barrier.
Consequently, we use a second approach in which the
reaction rates are modified by applying a factor that is a monotonic decreasing
(exponential) function of the temperature. We have applied the
following temperature dependent factor to each reaction rate:

\begin{equation}
 f(T) = 1+\left(f_0-1\right)\exp\left(-\frac{T}{3\times 10^9\ \text{K}}\right)\,,
\label{eq2}
\end{equation}
where $f_0=10$ or $0.1$, is the fixed factor applied in the first
approach. Of course, we are not trying to convey that Eq.~\ref{eq2} is
representative of the uncertainty of all the reactions studied here
(see Sections~\ref{prescriptions} and \ref{windows}), but it provides
a convenient way to invetigate the effects of a temperature dependent 
rate error.

\section{Sensitivity to the rate of fusion of carbon and of oxygen, and the triple--alpha
reaction.\label{sensfus}}

We have checked the effect of varying each fusion reaction rate by the
factors given above, either taking them fixed or as function of the
temperature. Because the fusion reactions are relevant for the nuclear
energy generation in the supernova explosion, we have recomputed the
hydrodynamics with the modified reaction rates and give the results in
Section \ref{senshydro}. When we kept unchanged the thermodynamic
trajectories of the reference model, but the reaction rates were
modified in the nucleosynthetic code, we obtained the results shown in
Section \ref{sensrea}.

\subsection{Rate modified in the hydrodynamic explosion model}\label{senshydro}

\begin{figure}[!htb]
\includegraphics[width=8truecm]{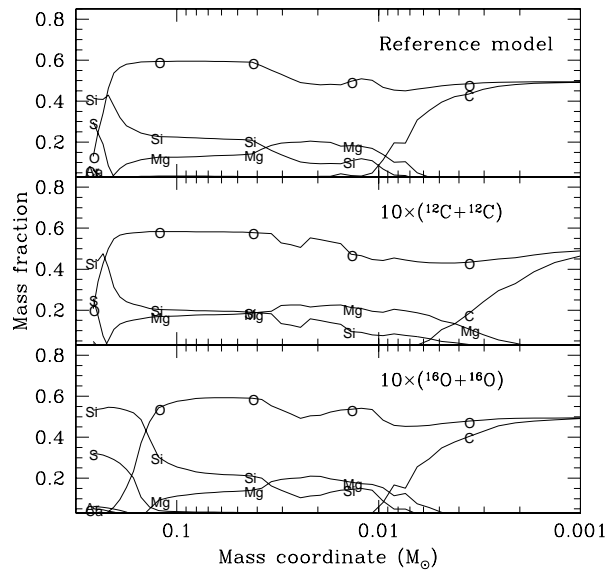}
\caption{Final chemical profile within the outer layers of the SNIa ejecta for three of the
computed models: our reference model (top), the model with the $^{12}\textrm{C}+{}^{12}\textrm{C}$
reaction increased by a constant factor of ten (middle), and the model with the
$^{16}\textrm{O}+{}^{16}\textrm{O}$ reaction increased by a constant factor of ten (bottom).
In this plot, the mass coordinate is zero at the white dwarf surface and increases inwards.
\label{fig4}}
\end{figure}

The nuclear energy release of the supernova is more sensitive to the
rate of the $^{16}$O fusion reaction than to that of $^{12}$C. The
final kinetic energy of the ejecta varies by less than 1\% when the
$^{12}\textrm{C}+{}^{12}$C reaction rate is varied by a factor of $10$ or
$0.1$, either fixed or as a function of the temperature given by
Eq.~(\ref{eq2}). In contrast, the same relative variation in the rate
of the $^{16}\textrm{O}+{}^{16}$O reaction produces a change of
kinetic energy of up to $\pm4\%$. 
We ascribe this lack of sensitivity to the relatively small amount of mass that does not experience
complete carbon or oxygen burning. Figure~\ref{fig4} shows the final chemical profiles
in the outermost $0.2$~M$_\odot$ of ejecta, where the changes in the $^{12}\textrm{C}+{}^{12}$C
and the $^{16}\textrm{O}+{}^{16}$O reaction rates are most influential. The three panels show the
profiles belonging to our reference model and the models in which either the carbon or the oxygen
fusion rates are increased by a factor of ten, both in the hydrodynamic as well as in the
nucleosynthetic codes. As can be seen, increasing the $^{12}\textrm{C}+{}^{12}$C rate by a factor
of ten barely affects the limits of the region undergoing carbon burning, which move outwards 
$\sim0.004$~M$_\odot$. On the other hand, when the $^{16}\textrm{O}+{}^{16}$O reaction rate is
enhanced by the same factor the limits of the oxygen burning region move outwards
$\sim0.082$~M$_\odot$.
We conclude that the impact of the rates uncertainties on the energy of the supernova is
negligible.

\begin{figure}[!htb]
\includegraphics[width=8truecm]{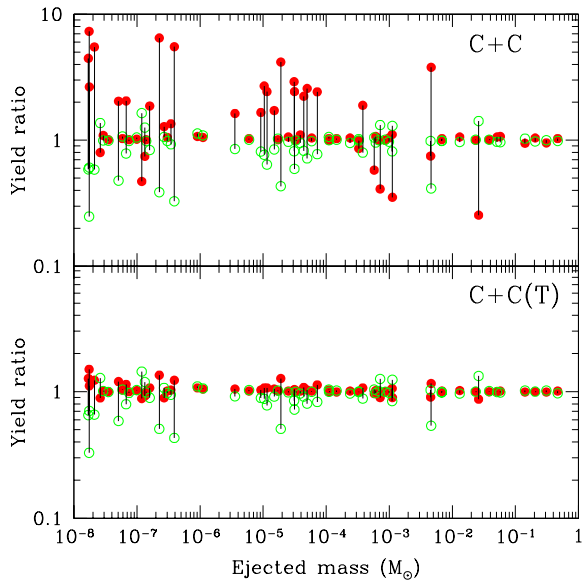}
\caption{(Color online) Ratio of mass ejected for each nuclide with a modified
  $^{12}$C+$^{12}$C reaction rate 
with respect to the mass ejected in the reference model, as a function of the mass fraction in the
reference model (most abundant species are located to the right of each figure). 
Note that the species included in Table \ref{tab4a} are those with an ejected mass larger than
$10^{-5}$~M$_\odot$.
Vertical lines
link the results obtained for the same nuclide when the rate is either increased or decreased. The
reaction rate
was modified both in the hydrodynamics calculation as well as in the nucleosynthetic code. 
\textbf{Top}: Rate multiplied by a fixed factor, either $\times10$ (green empty circles) or
$\times0.1$ (red filled circles). 
\textbf{Bottom}: Rate multiplied by a factor function of temperature given by
Eq.~(\ref{eq2}),
with either$f_0=10$ (green empty circles) or $f_0=0.1$ (red filled circles).
\label{fig5}}
\end{figure}

\begin{figure}[!htb]
\includegraphics[width=8truecm]{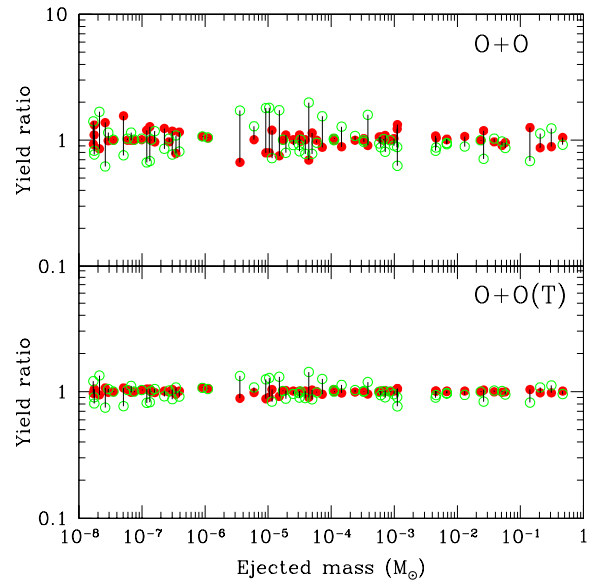}
\caption{(Color online) Ratio of mass ejected for each nuclide with a modified $^{16}$O+$^{16}$O
reaction rate
with respect to the mass ejected in the reference model, as a function of the mass fraction in the
reference model (most abundant species are located to the right of each figure). 
Note that the species included in Table \ref{tab4a} are those with an ejected mass larger than
$10^{-5}$~M$_\odot$.
Vertical lines
link the results obtained for the same nuclide when the rate is either increased or decreased. We
modified the
reaction rate both in the hydrodynamics calculation as well as in the nucleosynthetic code.
\textbf{Top}: Rate
multiplied by a fixed factor, either $\times10$ (green empty circles) or $\times0.1$ (red filled
circles).
\textbf{Bottom}: Rate multiplied by a factor function of temperature given by Eq.~(\ref{eq2}),
with
either
$f_0=10$ (green empty circles) or $f_0=0.1$ (red filled circles).
\label{fig6}}
\end{figure}

\begin{figure}[!htb]
 \includegraphics[width=8truecm]{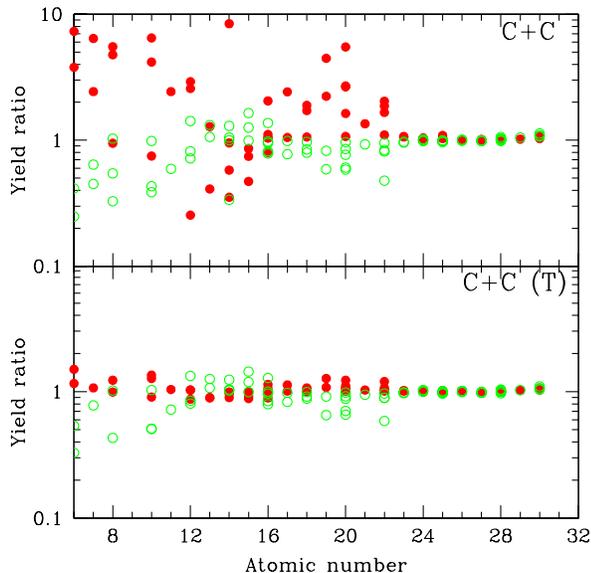}
\caption{(Color online) Same as Fig.~\ref{fig5} but plotted as a function of the
  atomic number of the product nucleus. Note that not all the isotops shown here appear in
Table~\ref{tab4a}.
 \textbf{Top}: Rate multiplied
  by a fixed factor, either $\times10$ (green empty circles) or
  $\times0.1$ (red filled circles). \textbf{Bottom}: Rate multiplied
  by a factor function of temperature given by Eq.~(\ref{eq2}), with
  either $f_0=10$ (green empty circles) or $f_0=0.1$ (red filled
  circles). \label{fig7}}
\end{figure}

\begin{figure}[!htb]
 \includegraphics[width=8truecm]{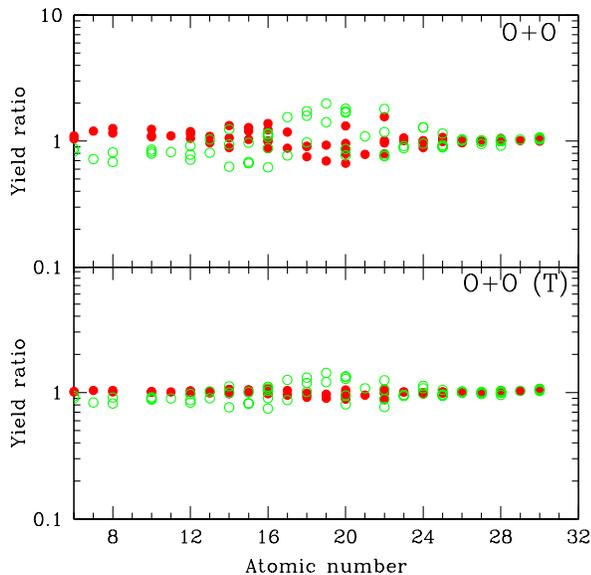}
\caption{(Color online) Same as Fig.~\ref{fig6} but plotted as a function of the atomic number of
the product
nucleus. Note that not all the isotops shown here appear in
Table~\ref{tab4a}. \textbf{Top}: Rate multiplied by a fixed factor, either $\times10$ (green empty
circles)
or $\times0.1$ (red filled circles). \textbf{Bottom}: Rate multiplied by a factor function of
temperature given by Eq.~(\ref{eq2}), with either $f_0=10$ (green empty circles) or $f_0=0.1$
(red
filled circles).
\label{fig8}}
\end{figure}

Figures~\ref{fig5} to \ref{fig8} and Table~\ref{tab4a} show the impact of the changes in the
fusion rates of $^{12}$C and of $^{16}$O on the nucleosynthesis of the
Type Ia supernova, when we modified the rates in the 
full supernova simulation.  Figures~\ref{fig5}
and~\ref{fig6} show the results sorted by final mass fraction of the 
product species. The mass fractions of the most abundant species are
insensitive to the rate of fusion of $^{12}$C. As one goes to smaller
abundances, the scatter of the yield ratio is larger.  Among the
species with mass fraction greater than $0.01$ there is only one
nuclide that is significantly affected by the modification of the rate
of $^{12}\textrm{C}+{}^{12}$C: not surprisingly it is $^{24}$Mg. When the
factor that modifies the $^{12}$C fusion rate is a function of
temperature, Eq.~(\ref{eq2}), 
the effect on the yields of all species is dramatically reduced
(bottom frame in Fig.~\ref{fig5}): no species experiences an increase
larger than a factor of two in its abundance, and only a few species
with quite small mass fractions
($X_{i}<10^{-6}$) experience a reduction of more than a factor of two in
their yields when the $^{12}$C fusion reaction rate is multiplied by a
factor of ten.

When the $^{16}\textrm{O}+{}^{16}$O rate is modified
(Fig.~\ref{fig6}) the impact is in general smaller than when the
$^{12}$C fusion rate was modified. However, many of the most abundant species are
more sensitive to the $^{16}$O fusion rate than to the $^{12}$C rate
because the products of C-burning (mainy $^{16}$O, $^{20}$Ne, and $^{24}$Mg) are in general less
abundant than the products of O-burning (mainly $^{28}$Si, $^{32}$S, $^{36}$Ar, and $^{40}$Ca).

Figure~\ref{fig7} presents the same results as Fig.~\ref{fig5} from
another perspective: the impact of the modification of the
$^{12}\textrm{C}+{}^{12}$C reaction rate is shown against the element
atomic number. The trend that can be observed in this figure is that
increasing the $^{12}$C fusion rate (green empty circles) decreases
the abundances both of CNO nuclei and of IMEs
between Phosphorus and Titanium, and increases the abundances of
Magnesium, Aluminum, and Silicon, while elements beyond Vanadium are
scarcely affected at all. If the rate of $^{12}$C fusion is decreased
(red filled circles) the trend is inverted, but the yields are in
general more sensitive to a decrease in this rate than to an increase
by the same factor.

As can be seen in Fig.~\ref{fig8}, an increase in the
$^{16}\textrm{O}+{}^{16}$O reaction rate results in a small decrease in the
production of elements up to Magnesium and an increase in elements
from Chlorine to Chromium. The effect on the mass fractions is
much smaller than that due to variations in the $^{12}$C fusion rate.

We give in Table~\ref{tab4a} the sensitivity of the yield of each one
of the species included in Table~\ref{tab2} to the rate of fusion
reactions. There, $D_i$ is the logarithmic derivative of the mass
ejected of species $i$ with respect to the enhancement factor of each
fusion reaction, $f_0$ (note that when using Eq.~\ref{eq2}, $f_0$
represents the maximum enhancement factor, attained at low
temperatures),
\begin{equation}
 D_{i} = \frac{\text{d}\log m_{i}}{\text{d}\log f_0} \approx 0.5 \log\left(\frac{m_{i,10}}
{m_{i,0.1}}\right)\,,
\label{eqdi}
\end{equation}
where $m_{i,10}$ is the mass ejected of species $i$ for $f_0=10.$, and
$m_{i,0.1}$ is the corresponding mass when $f_0=0.1$. According to
this definition, a value of $D_i\approx0.3$ means that the abundance of species $i$
approximately doubles for a constant enhancement factor of $f_0=10$ in
the corresponding fusion reaction rate. Similarly, a relative change in the
abundance of a species by $12\%$ would correspond to $D_i\approx0.05$,
and a change by $2\%$ would derive from $D_i\approx0.01$. 

Most notable is the robustness of the production of most Fe-group isotopes,
notably of $^{56}$Ni. When the
enhancement factor is computed from Eq.~\ref{eq2}, there is no species
with $|D_i|>0.1$, neither with respect to the rate of
\mbox{$^{12}\textrm{C}+{}^{12}$C} nor with respect to the \mbox{$^{16}\textrm{O}+{}^{16}$O} rate,
with the exceptions of $^{12}$C and $^{39}$K, respectively. 

\begin{table*}
\caption{Sensitivity of the nucleosynthesis to the rate of fusion reactions: Rate modified in the
hydrodynamic and nucleosynthetic codes (see also Figs.~\ref{fig5} to \ref{fig8})\footnote{Values of
$D_{i}$ less than 1.0E-3 have been put to 0.}.
\label{tab4a}}
\begin{ruledtabular}
\begin{tabular}{lrrrr}
Nucleus & $D_{i}\left(^{12}\text{C}+{}^{12}\text{C}\right)$ &
$D_{i}\left(^{12}\text{C}+{}^{12}\text{C}\right)$\footnotemark[2] &
$D_{i}\left(^{16}\text{O}+{}^{16}\text{O}\right)$ & 
$D_{i}\left(^{16}\text{O}+{}^{16}\text{O}\right)$\footnotemark[2] 
\\
\hline
 $^{12}$C  & $-$4.8E-1 & $-$1.7E-1 & $-$3.7E-2 & $-$1.6E-2 \\
 $^{16}$O  &  1.9E-2 &  6.0E-3 & $-$1.3E-1 & $-$5.2E-2 \\
 $^{20}$Ne &  6.0E-2 &  2.8E-2 & $-$5.8E-2 & $-$2.6E-2 \\
 $^{23}$Na & $-$3.1E-1 & $-$7.9E-2 & $-$6.4E-2 & $-$2.6E-2 \\
 $^{24}$Mg &  3.7E-1 &  9.2E-2 & $-$1.1E-1 & $-$4.6E-2 \\
 $^{25}$Mg & $-$2.8E-1 & $-$4.2E-2 & $-$3.0E-2 & $-$5.0E-3 \\
 $^{26}$Mg & $-$2.8E-1 & $-$5.4E-2 & $-$8.2E-2 & $-$3.7E-2 \\
 $^{26}$Al & $-$4.1E-2 &  4.0E-2 &  1.3E-2 &  6.0E-3 \\
 $^{27}$Al &  2.5E-1 &  7.4E-2 & $-$6.5E-2 & $-$2.7E-2 \\
 $^{28}$Si &  1.1E-2 &  6.0E-3 &  7.2E-2 &  2.9E-2 \\
 $^{29}$Si &  1.3E-1 &  1.6E-2 & $-$2.5E-2 & $-$1.0E-3 \\
 $^{30}$Si &  2.8E-1 &  7.0E-2 & $-$1.6E-1 & $-$7.1E-2 \\
 $^{31}$P  &  3.1E-2 & $-$1.0E-3 & $-$1.3E-2 &  3.0E-3 \\
 $^{32}$S  & $-$1.4E-2 &  0. &  5.7E-2 &  2.2E-2 \\
 $^{33}$S  & $-$1.9E-2 & $-$9.0E-3 &  1.7E-2 &  8.0E-3 \\
 $^{34}$S  & $-$6.7E-2 & $-$5.0E-2 & $-$7.2E-2 & $-$3.0E-2 \\
 $^{35}$Cl & $-$2.5E-1 & $-$6.7E-2 &  1.2E-1 &  6.0E-2 \\
 $^{36}$Ar & $-$1.9E-2 & $-$2.0E-3 &  1.6E-2 &  5.0E-3 \\
 $^{37}$Ar & $-$1.5E-1 & $-$3.0E-2 &  1.8E-1 &  7.8E-2 \\
 $^{38}$Ar & $-$1.9E-1 & $-$4.2E-2 &  1.2E-1 &  4.8E-2 \\
 $^{39}$K  & $-$2.2E-1 & $-$3.6E-2 &  2.3E-1 &  1.0E-1 \\
 $^{40}$Ca & $-$2.3E-2 & $-$6.0E-3 & $-$2.1E-2 & $-$1.1E-2 \\
 $^{44}$Ti & $-$3.1E-2 & $-$9.0E-3 & $-$5.6E-2 & $-$2.8E-2 \\
 $^{48}$V  & $-$2.4E-2 & $-$1.0E-2 & $-$4.1E-2 & $-$1.6E-2 \\
 $^{49}$V  & $-$1.9E-2 & $-$5.0E-3 & $-$2.2E-2 & $-$1.1E-2 \\
 $^{50}$Cr & $-$4.0E-3 &  6.0E-3 &  8.0E-2 &  3.2E-2 \\
 $^{51}$Cr & $-$1.2E-2 & $-$3.0E-3 & $-$9.0E-3 & $-$4.0E-3 \\
 $^{52}$Mn & $-$2.0E-2 & $-$1.2E-2 & $-$4.0E-2 & $-$1.5E-2 \\
 $^{53}$Mn & $-$1.4E-2 & $-$7.0E-3 & $-$2.8E-2 & $-$1.2E-2 \\
 $^{54}$Fe &  0. &  3.0E-3 &  1.3E-2 &  5.0E-3 \\
 $^{55}$Fe &  0. &  0. &  0. &  0. \\
 $^{56}$Fe &  0. &  2.0E-3 &  5.0E-3 &  3.0E-3 \\
 $^{57}$Co & $-$5.0E-3 & $-$4.0E-3 & $-$1.3E-2 & $-$5.0E-3 \\
 $^{56}$Ni & $-$9.0E-3 & $-$9.0E-3 & $-$2.8E-2 & $-$1.2E-2 \\
 $^{58}$Ni &  2.0E-3 &  0. &  0. &  0. \\
 $^{59}$Ni &  1.0E-3 &  0. &  0. &  0. \\
 $^{60}$Ni &  2.0E-3 & $-$1.0E-3 &  1.0E-3 &  0. \\
 $^{61}$Ni &  4.0E-3 &  2.0E-3 &  2.0E-3 & $-$2.0E-3 \\
 $^{62}$Ni &  6.0E-3 &  2.0E-3 &  0. & $-$2.0E-3 \\
\end{tabular}
\end{ruledtabular}
\footnotetext[2]{Enhancement factor function of temperature according to Eq.~\ref{eq2}.}
\end{table*}

\subsection{Rate modified only in the nucleosynthetic code}\label{sensrea}

\begin{figure}[!htb]
 \includegraphics[width=8truecm]{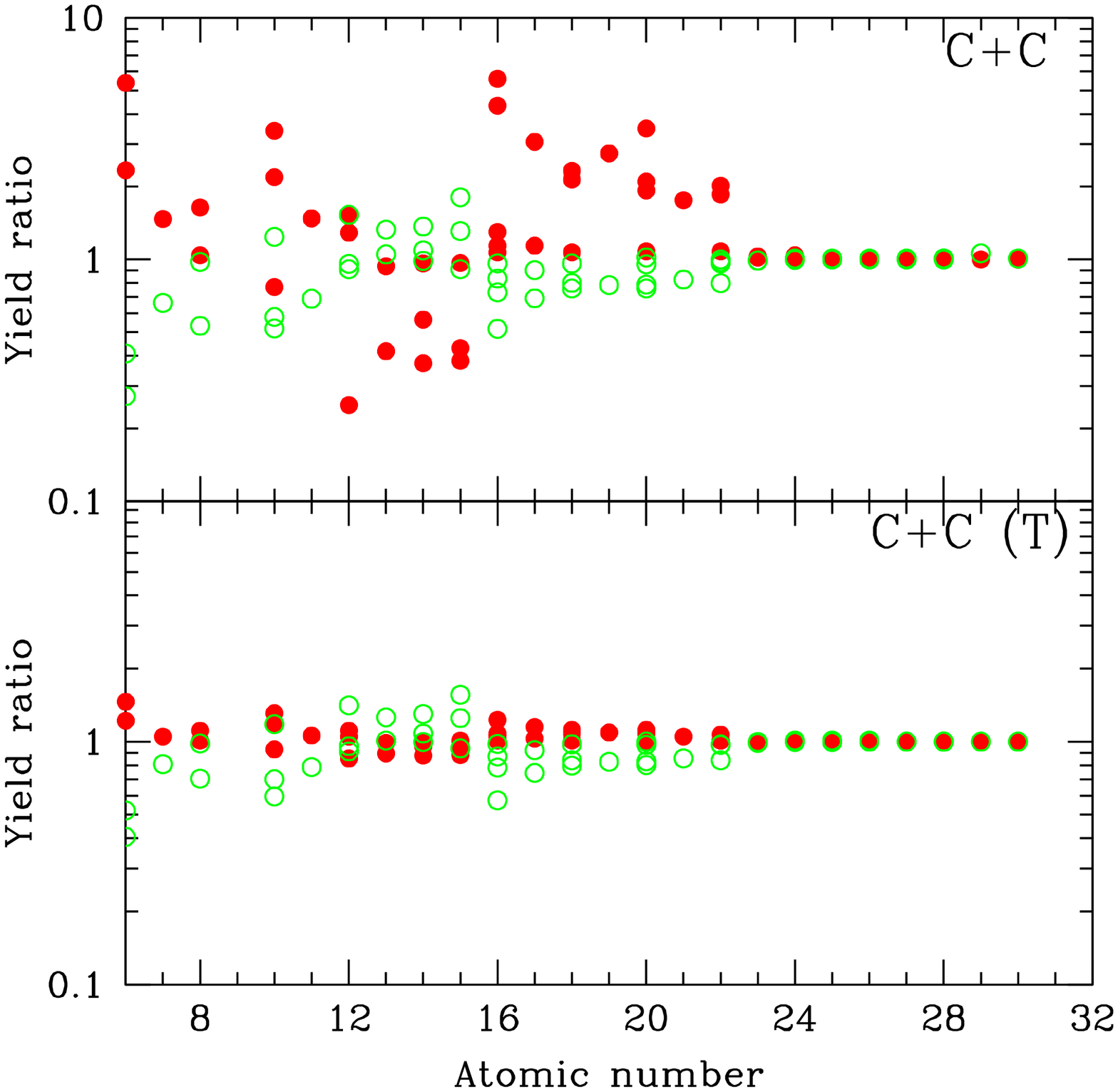}
\caption{(Color online) Ratio of mass ejected for each nuclide with a modified $^{12}$C+$^{12}$C
reaction rate
with respect to the mass ejected in the reference model, as a function of the atomic number of the
product nucleus. We modified the reaction rate only in the nucleosynthetic code.
\textbf{Top}: Rate multiplied by a fixed factor, either $\times10$ (green empty circles)
or $\times0.1$ (red filled circles). \textbf{Bottom}: Rate multiplied by a factor function of
temperature given by Eq.~(\ref{eq2}), with either $f_0=10$ (green empty circles) or $f_0=0.1$
(red
filled circles).
\label{fig9}}
\end{figure}

\begin{figure}[!htb]
 \includegraphics[width=8truecm]{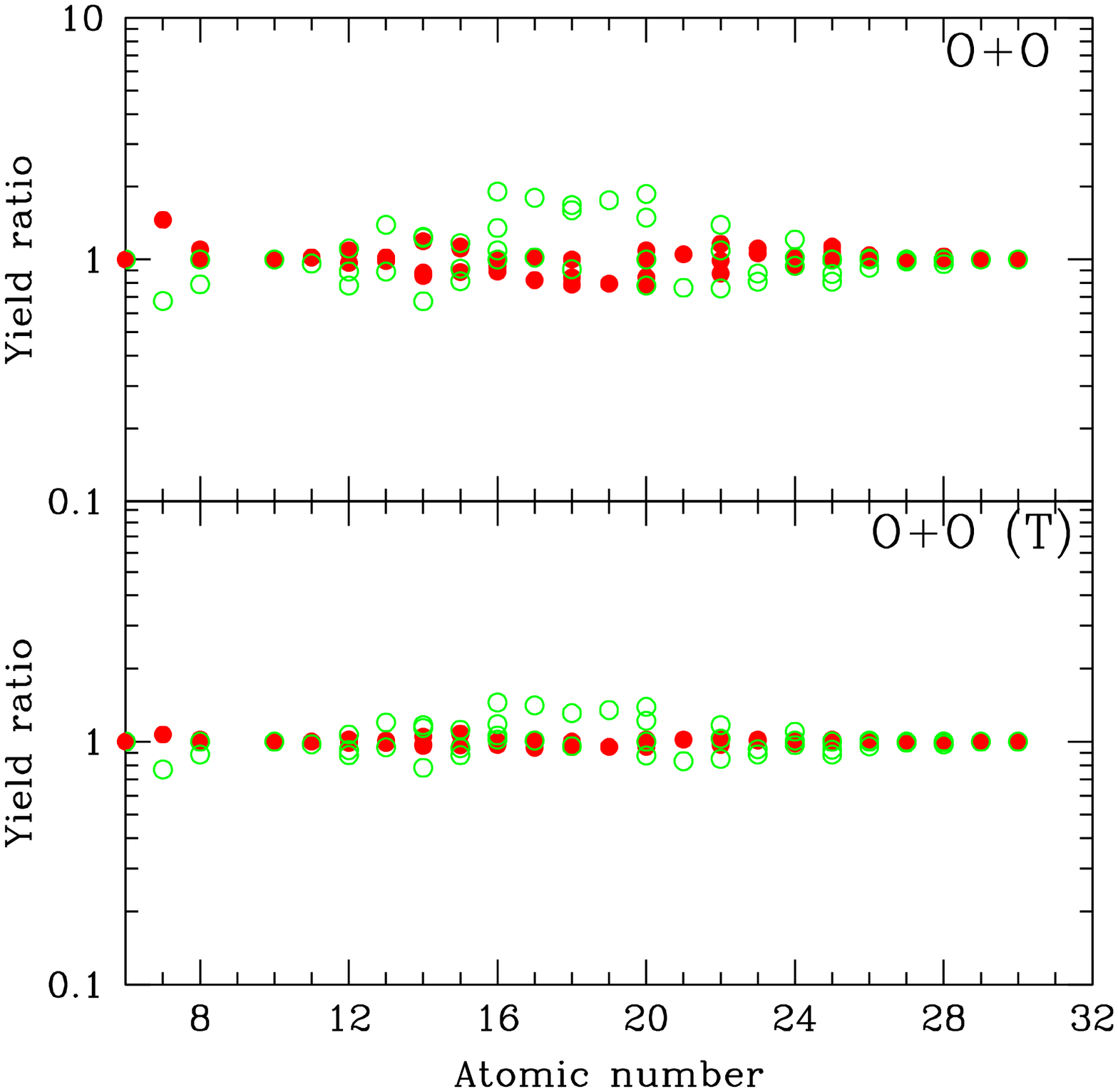}
\caption{(Color online) Ratio of mass ejected for each nuclide with a modified $^{16}$O+$^{16}$O
reaction rate
with respect to the mass ejected in the reference model, as a function of the atomic number of the
product nucleus. We modified the reaction rate only in the nucleosynthetic code.
\textbf{Top}: Rate multiplied by a fixed factor, either $\times10$ (green empty circles)
or $\times0.1$ (red filled circles). \textbf{Bottom}: Rate multiplied by a factor function of
temperature given by Eq.~(\ref{eq2}), with either $f_0=10$ (green empty circles) or $f_0=0.1$
(red
filled circles).
\label{fig10}}
\end{figure}

Figures~\ref{fig9} and \ref{fig10} show the impact of the changes in
the fusion rates of $^{12}$C and of $^{16}$O on the nucleosynthesis of
Type Ia supernovae when the rates are modified only in the
nucleosynthetic code. They can be compared with Figs.~\ref{fig7} and
\ref{fig8}, respectively, to evaluate the relevance of incorporating
the modified rates into the hydrodynamic code. The trends visible in
these figures are qualitatively similar, irrespectively if the
reaction rate has been modified in the hydrodynamic calculations or not. 

\begin{figure}[!htb]
 \includegraphics[width=8truecm]{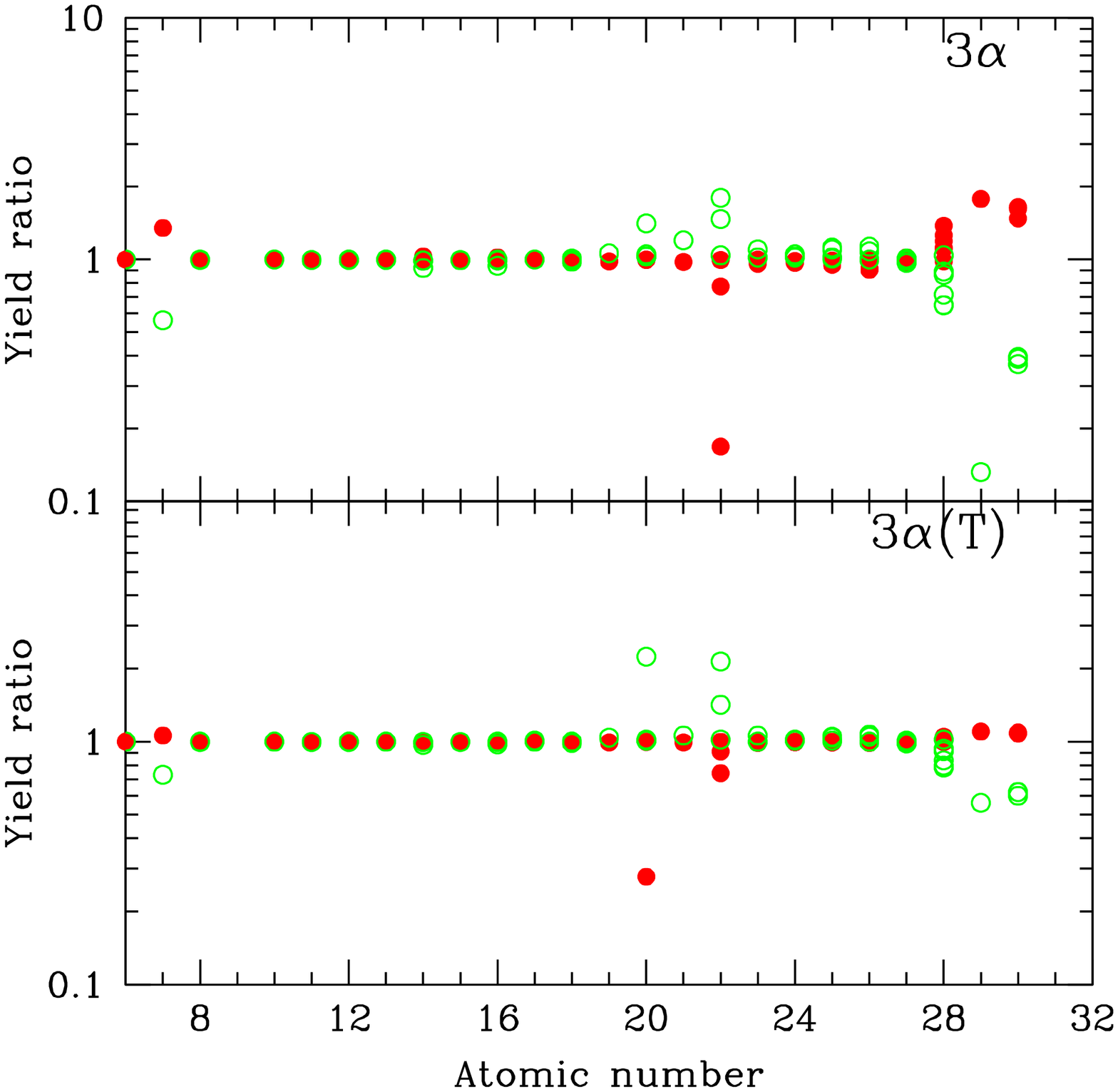}
\caption{(Color online) Ratio of mass ejected for each nuclide when the $3\alpha$ reaction rate is
modified
to the mass ejected in the reference model, as a function of the atomic number of the
product nucleus. We modified the reaction rate only in the nucleosynthetic code.
\textbf{Top}: Rate multiplied by a fixed factor, either $\times10$ (green empty circles)
or $\times0.1$ (red filled circles). \textbf{Bottom}: Rate multiplied by a factor function of
temperature given by Eq.~(\ref{eq2}), with either $f_0=10$ (green empty circles) or $f_0=0.1$
(red
filled circles).
\label{fig11}}
\end{figure}

In Fig.~\ref{fig11} we show the yield ratios belonging to modified rate
of the $3\alpha$ reaction. The influence of the rate of this reaction
focuses on a few elements: Nitrogen, Nickel, Copper, and Zinc 
(specially the isotopes of Nickel and Zinc produced during alpha-rich freeze-out of NSE)
inversely correlate with the factor of enhancement of the $3\alpha$
reaction, while Titanium and, to a lesser extent, Scandium, Manganese,
and Iron
(specially the isotopes produced during explosive Si-burning)
are more abundant when the $3\alpha$ reaction is faster.
These results can be explained by the fact that for a faster rate an alpha-rich freeze-out occurs
at lower temperature and density. As a result, the dotted-line in Fig.~\ref{fig1} (left) shifts
down when the $3\alpha$ reaction is faster, increasing the yield of species made in normal
freeze-out at the expense of alpha-rich freeze-out products.

We give in Table~\ref{tab4b} the sensitivity of the yield of each one
of the species included in Table~\ref{tab2} to the rate of fusion
reactions, when they are modified only in the nucleosynthetic code. 
This table can be compared to Table~\ref{tab4a} to avaluate the importance of running a
hydrodynamic code with the reaction rates modified or take the thermodynamic profiles of a
reference model and modifying the rates only in a post-processing code. One finds that the
sensitivities shown in both tables are qualitatively similar. Although the precise values of $D_i$
for given species are not equal, the rating of the species that are most sensitive to any fusion
reaction rate is the same in both tables. Given the pre-eminence of the fusion reaction rates with
respect to the release of nuclear energy, we 
conclude that \emph{for SNIa} this kind of study can be
safely carried out with a post-processing code, using a set of
thermodynamic trajectories obtained with a supernova hydrodynamics code where
the reaction rates remain unchanged.

\begin{table*}
\caption{Sensitivity of the nucleosynthesis to the rate of fusion reactions: Rate modified only in
the nucleosynthetic code (see also Figs.~\ref{fig9} to \ref{fig11})\footnote{Values of
$D_{i}$ less than 1.0E-3 have been put to 0.}.
\label{tab4b}}
\begin{ruledtabular}
\begin{tabular}{lrrrrrr}
Nucleus & $D_{i}\left(^{12}\text{C}+{}^{12}\text{C}\right)$ &
$D_{i}\left(^{12}\text{C}+{}^{12}\text{C}\right)$\footnotemark[2] &
$D_{i}\left(^{16}\text{O}+{}^{16}\text{O}\right)$ & 
$D_{i}\left(^{16}\text{O}+{}^{16}\text{O}\right)$\footnotemark[2] &
$D_{i}\left(3\alpha\right)$ &
$D_{i}\left(3\alpha\right)$\footnotemark[2]  
\\
\hline
 $^{12}$C  & $-$3.8E-1 & $-$1.9E-1 &  0. &  0. &  0. &  0. \\
 $^{16}$O  & $-$1.3E-2 & $-$6.0E-3 & $-$7.2E-2 & $-$3.1E-2 & $-$1.0E-3 &  0. \\
 $^{20}$Ne &  1.0E-1 &  5.2E-2 &  0. &  0. &  0. &  0. \\
 $^{23}$Na & $-$1.7E-1 & $-$6.5E-2 & $-$1.3E-2 & $-$5.0E-3 & $-$1.0E-3 &  0. \\
 $^{24}$Mg &  3.9E-1 &  1.1E-1 & $-$6.8E-2 & $-$3.2E-2 & $-$1.0E-3 &  0. \\
 $^{25}$Mg & $-$6.4E-2 & $-$2.0E-2 &  3.0E-2 &  1.7E-2 &  0. &  0. \\
 $^{26}$Mg & $-$1.1E-1 & $-$4.3E-2 & $-$4.3E-2 & $-$2.1E-2 & $-$1.0E-3 &  0. \\
 $^{26}$Al &  2.4E-2 &  5.0E-3 &  7.5E-2 &  4.3E-2 &  0. &  0. \\
 $^{27}$Al &  2.5E-1 &  7.4E-2 & $-$3.0E-2 & $-$1.4E-2 & $-$1.0E-3 &  0. \\
 $^{28}$Si &  5.0E-3 &  0. &  7.4E-2 &  3.4E-2 & $-$2.4E-2 & $-$7.0E-3 \\
 $^{29}$Si &  1.4E-1 &  2.7E-2 &  7.9E-2 &  4.3E-2 & $-$2.0E-3 & $-$1.0E-3 \\
 $^{30}$Si &  2.8E-1 &  8.5E-2 & $-$1.2E-1 & $-$6.4E-2 & $-$2.0E-3 & $-$1.0E-3 \\
 $^{31}$P  & $-$1.3E-2 & $-$1.6E-2 &  6.0E-2 &  3.2E-2 & $-$2.0E-3 &  0. \\
 $^{32}$S  & $-$2.3E-2 & $-$7.0E-3 &  3.6E-2 &  1.5E-2 & $-$1.7E-2 & $-$5.0E-3 \\
 $^{33}$S  & $-$9.6E-2 & $-$4.1E-2 &  9.0E-2 &  4.3E-2 & $-$1.0E-3 &  0. \\
 $^{34}$S  & $-$9.6E-2 & $-$7.0E-2 &  1.1E-2 &  7.0E-3 & $-$2.0E-3 & $-$1.0E-3 \\
 $^{35}$Cl & $-$3.2E-1 & $-$9.5E-2 &  1.7E-1 &  8.7E-2 &  1.0E-3 &  3.0E-3 \\
 $^{36}$Ar & $-$2.3E-2 & $-$7.0E-3 & $-$2.0E-2 & $-$9.0E-3 & $-$7.0E-3 & $-$2.0E-3 \\
 $^{37}$Ar & $-$2.1E-1 & $-$5.5E-2 &  1.7E-1 &  7.0E-2 &  2.0E-3 &  0. \\
 $^{38}$Ar & $-$2.4E-1 & $-$7.4E-2 &  1.4E-1 &  6.9E-2 & $-$2.0E-3 &  0. \\
 $^{39}$K  & $-$2.7E-1 & $-$6.0E-2 &  1.7E-1 &  7.6E-2 &  1.7E-2 &  1.0E-2 \\
 $^{40}$Ca & $-$2.6E-2 & $-$9.0E-3 & $-$7.2E-2 & $-$3.3E-2 &  6.0E-3 &  2.0E-3 \\
 $^{44}$Ti & $-$2.4E-2 & $-$9.0E-3 & $-$9.2E-2 & $-$4.2E-2 &  1.4E-1 &  9.6E-2 \\
 $^{48}$V  & $-$7.0E-3 & $-$2.0E-3 & $-$6.8E-2 & $-$3.0E-2 &  2.9E-2 &  1.5E-2 \\
 $^{49}$V  & $-$8.0E-3 & $-$1.0E-3 & $-$4.1E-2 & $-$1.7E-2 &  5.0E-3 &  0. \\
 $^{50}$Cr & $-$6.0E-3 &  3.0E-3 &  5.6E-2 &  2.4E-2 &  1.4E-2 &  5.0E-3 \\
 $^{51}$Cr &  0. &  3.0E-3 & $-$1.8E-2 & $-$7.0E-3 &  1.7E-2 &  5.0E-3 \\
 $^{52}$Mn &  2.0E-3 & $-$1.0E-3 & $-$7.3E-2 & $-$3.1E-2 &  3.5E-2 &  1.3E-2 \\
 $^{53}$Mn &  2.0E-3 &  3.0E-3 & $-$4.4E-2 & $-$1.8E-2 &  3.2E-2 &  1.0E-2 \\
 $^{54}$Fe &  2.0E-3 &  3.0E-3 &  3.0E-3 &  2.0E-3 &  3.5E-2 &  1.3E-2 \\
 $^{55}$Fe &  2.0E-3 &  3.0E-3 & $-$2.6E-2 & $-$1.2E-2 &  4.8E-2 &  1.7E-2 \\
 $^{56}$Fe &  0. &  0. &  0. &  0. &  0. &  0. \\
 $^{57}$Co &  2.0E-3 &  0. & $-$6.0E-3 & $-$2.0E-3 & $-$1.1E-2 & $-$4.0E-3 \\
 $^{56}$Ni &  3.0E-3 &  0. & $-$1.6E-2 & $-$8.0E-3 &  1.2E-2 &  5.0E-3 \\
 $^{58}$Ni &  2.0E-3 &  0. & $-$1.0E-3 &  0. & $-$4.7E-2 & $-$1.7E-2 \\
 $^{59}$Ni &  0. &  0. &  0. &  0. & $-$5.6E-2 & $-$2.2E-2 \\
 $^{60}$Ni &  2.0E-3 &  0. &  0. &  0. & $-$1.1E-1 & $-$4.5E-2 \\
 $^{61}$Ni &  2.0E-3 &  0. &  0. &  0. & $-$1.4E-1 & $-$5.8E-2 \\
 $^{62}$Ni &  2.0E-3 &  0. &  0. &  0. & $-$1.6E-1 & $-$6.4E-2 \\
\end{tabular}
\end{ruledtabular}
\footnotetext[2]{Enhancement factor function of temperature according to Eq.~\ref{eq2}.}
\end{table*}

\section{Sensitivity to the rate of radiative captures and transfer reactions}\label{senspar}

We will discuss in this section the sensitivity of the nucleosynthesis to
changes in the rate of radiative captures and transfer reactions. We
measure the sensitivity in a similar way as with respect to the rate
of the fusion reactions, by defining $D_{i}$
as in Eq.~\ref{eqdi}. The meaning of $D_{i}$ is now the logarithmic
derivative of the mass ejected of species $i$ with respect to the
enhancement factor, $f_0$, of a reaction between particle $k$ and
nucleus $j$, while $m_{i,10}$ is the mass ejected of species $i$ when an
enhancement factor $f_0=10.$ (either fixed or function of temperature)
is applied to the rate of the reactions $j+k\rightarrow l+m$ and
$l+m\rightarrow j+k$, and $m_{i,0.1}$ is the corresponding mass when
$f_0=0.1$.

We start by analyzing the results obtained with a fixed enhancement
factor (our first approach).  In Section~\ref{fdet} we present the
results obtained when we compute the enhancement factor with a decreasing uncertainty (see
Eq.~\ref{eq2}, our second approach). Then, we select the reactions to
which the nucleosynthesis is most sensitive and analyze in
Section~\ref{prescriptions} the results achieved by adopting different
prescriptions for their reaction rates, chosen among the most recent
literature. Finally, in Section~\ref{windows} we analyze the
temperature ranges in which a modification of a given reaction rate
affects most the chemical composition of the supernova ejecta.

\subsection{Fixed rate enhancement factor}\label{fixed}

Tables~\ref{tab5} to \ref{tab20} give, for each reaction pair
$j+k\rightleftarrows l+m$ that has a significant impact on the
nucleosynthesis, the nuclei $i$ for which $|D_{i}|>0.3$ (more than
twofold increase or decrease in the yield when the rate in enhanced or
decreased by a factor of ten), and those for which $0.3>|D_{i}|>0.05$
(relative increase or decrease of the yield between $12\%$ and a
factor of two). Although we only list in the tables the direct reactions,
the inverse reactions contributed as well to the changes in the
nucleosynthetic yield.  There is only one reaction pair for which
$|D_{i}|>1$, it is the
$^{30}\text{Si}+\text{p}\rightleftarrows{}^{31}\text{P}+\gamma$ reaction
and the species whose abundance is mostly affected is $^{35}$S. Each
table shows the reactions belonging to a given type,
e.g. $\left(\text{n},\gamma\right)$, sorted according to the total
mass they processed in our reference model, $\mathfrak{M}_{jk}$ (see
Table~\ref{tab3}).

\begin{table*}
\caption{Sensitivity of the nucleosynthesis to the rate of $\left(\text{n},\gamma\right)$
reactions with the parent nuclide given in the first column.\footnote{The reactions listed are
those that processed more than $10^{-6}$~M$_\odot$ in
the reference model (see Table~\ref{tab3}) \textsl{and} with any
$\text{max}\left(|D_{i}|\right)>0.05$.}
\label{tab5}}
\begin{ruledtabular}
\begin{tabular}{l@{\hspace{0.1truecm}}|d@{\hspace{0.1truecm}}|l@{\hspace{0.1truecm}}|l}
Parent nuclide & \mathfrak{M}_{jk} \left(\text{M}_\odot\right) & Nuclei with $|D_{i}|>0.3$ &
Nuclei with $0.3>|D_{i}|>0.05$ \\
\hline
$^{28}$Si & 0.30 & $^{35}$S  & $^{21}$Ne, $^{25}$Mg, $^{26}$Al, $^{33}$S  \\ 
$^{55}$Fe & 0.20 & & $^{54,55}$Mn, $^{57,58}$Fe, $^{58,59}$Co \\
$^{32}$S  & 0.17 & $^{35}$S  & $^{25}$Mg, $^{26}$Al, $^{29}$Si, $^{32,33}$P, $^{33}$S  \\
$^{36}$Ar & 0.093 & & $^{37}$Cl, $^{37}$Ar \\
$^{44}$Ti & 0.076 & & $^{45}$Sc \\ 
$^{24}$Mg & 0.045 & $^{26}$Al, $^{35}$S  & $^{17}$O, $^{21}$Ne, $^{25}$Mg, $^{32}$P, $^{33}$S\\
$^{25}$Mg & 0.010 & $^{21}$Ne & $^{17}$O, $^{25}$Mg, $^{26}$Al \\
$^{56}$Fe & 0.0093 & & $^{57}$Fe \\
$^{16}$O  & 0.0092 & $^{17}$O  & $^{25}$Mg \\ 
$^{46}$Ti & 0.0068 & & $^{46}$Ti \\ 
$^{29}$Si & 0.0061 & & $^{29}$Si \\ 
$^{20}$Ne & 0.0049 & $^{21}$Ne & \\ 
$^{33}$S  & 0.0033 & & $^{33}$P  \\ 
$^{35}$Cl & 0.0029 & & $^{37}$Cl \\    
$^{12}$C  & 0.0024 & & $^{21,22}$Ne, $^{25}$Mg, $^{45}$Sc \\    
$^{31}$P  & 9.1\times10^{-4} & & $^{32}$P  \\ 
\end{tabular}
\end{ruledtabular}
\end{table*}

The species most sensitive to changes in the rates of
$\left(\text{n},\gamma\right)$ reactions (Table~\ref{tab5}) are
$^{17}$O, $^{26}$Al, $^{21}$Ne, and $^{35}$S. The yields of all these
nuclides are small, of order $10^{-7}$~M$_\odot$. Apart from the
neutron captures on iron isotopes, all the reactions listed in
Table~\ref{tab5} involve IMEs or CNO elements as the parent
nuclides. Among the species with $|D_{i}|>0.05$, the most abundant
are $^{29}$Si and $^{33}$S, both with yields on the order of a few
times $10^{-4}$~M$_\odot$, suggesting that the temperature range where
$\left(\text{n},\gamma\right)$ reaction rates most affect the final
nucleosynthesis is approximately $2\times 10^9 \lesssim T \lesssim
4\times 10^9$~K, in agreement with our analysis in Section~\ref{secsnia}.

\begin{table*}
\caption{Sensitivity of the nucleosynthesis to the rate of $\left(\text{p},\gamma\right)$
reactions with parent nuclide given in the first column. \footnote{The reactions listed are
those that processed more than $10^{-6}$~M$_\odot$ in
the reference model (see Table~\ref{tab3}) \textsl{and} with any
$\text{max}\left(|D_{i}|\right)>0.05$.}
\label{tab6}}
\begin{ruledtabular}
\begin{tabular}{l@{\hspace{0.1truecm}}|d@{\hspace{0.1truecm}}|l@{\hspace{0.1truecm}}|l@{\hspace{
0.1truecm}}|l }
Parent nuclide & \mathfrak{M}_{jk} \left(\text{M}_\odot\right) & Nuclei with $|D_{i}|>1.$ &
Nuclei with $1>|D_{i}|>0.3$ & Nuclei with $0.3>|D_{i}|>0.05$ \\
\hline
$^{29}$Si & 0.67 & & & $^{21}$Ne, $^{25}$Mg, $^{26}$Al, $^{35}$S, $^{43}$Ca, $^{47}$Ti \\
$^{57}$Co & 0.63 & & & $^{54,55}$Mn, $^{58}$Fe \\
$^{58}$Ni & 0.61 & & & $^{63}$Cu \\
$^{34}$S  & 0.61 & & & $^{35}$S  \\
$^{53}$Mn & 0.58 & & & $^{52}$Cr \\
$^{35}$Cl & 0.56 & & & $^{35}$S, $^{37}$Cl, $^{37}$Ar \\
$^{30}$Si & 0.56 & $^{35}$S  & $^{32}$P  &
$^{21}$Ne, $^{23}$Na, $^{24-26}$Mg, $^{26}$Al, $^{29,30}$Si, \\
& & & & $^{31,33}$P, $^{33,34}$S, $^{35,37}$Cl, $^{38}$Ar, $^{42,43}$Ca, \\
& & & & $^{47}$Ti \\
$^{27}$Al & 0.53 & & $^{27}$Al, $^{35}$S, $^{43}$Ca &
$^{23}$Na, $^{24-26}$Mg, $^{26}$Al, $^{32}$P, $^{34}$S, $^{35}$Cl, \\
& & & & $^{47}$Ti \\
$^{55}$Co & 0.47 & & & $^{50}$Cr, $^{55}$Fe, $^{56}$Co \\
$^{39}$K  & 0.39 & & & $^{39}$K, $^{41,43}$Ca \\
$^{59}$Cu & 0.29 & & & $^{59,60}$Ni, $^{63}$Cu, $^{64,65}$Zn \\
$^{56}$Co & 0.28 & & & $^{56}$Co \\
$^{48}$V  & 0.27 & & & $^{46,47}$Ti \\
$^{33}$S  & 0.27 & & & $^{43}$Ca, $^{47}$Ti \\ 
$^{30}$P  & 0.25 & & & $^{43}$Ca \\
$^{51}$Mn & 0.22 & & & $^{50,51}$Cr \\
$^{47}$V  & 0.14 & & & $^{46,47}$Ti \\
$^{47}$Ti & 0.13 & & & $^{46,47}$Ti \\
$^{26}$Al & 0.12 & & $^{26}$Al & \\
$^{54}$Mn & 0.087 & & & $^{54}$Mn \\
$^{44}$Sc & 0.082 & & & $^{47}$Ti \\
$^{59}$Co & 0.079 & & & $^{58}$Fe, $^{59}$Co \\
$^{55}$Mn & 0.065 & & & $^{55}$Mn \\
$^{26}$Mg & 0.061 & & $^{26}$Mg & $^{23}$Na, $^{25}$Mg, $^{26}$Al, $^{29}$Si, $^{35}$S
, $^{43}$Ca, \\
& & & & $^{47}$Ti \\
$^{36}$Cl & 0.026 & & & $^{37}$Cl \\
$^{25}$Mg & 0.023 & & $^{26}$Al & $^{21}$Ne, $^{25}$Mg, $^{35}$S  \\
$^{43}$Sc & 0.021 & & & $^{43}$Ca \\
$^{23}$Na & 0.014 & & & $^{45}$Sc \\
$^{58}$Fe & 0.011 & & & $^{58}$Fe \\
$^{45}$Sc & 0.0081 & & & $^{42}$Ca, $^{45}$Sc, $^{46}$Ti \\
$^{62}$Zn & 0.0068 & & & $^{63}$Cu \\
$^{62}$Cu & 0.0060 & & & $^{63}$Cu, $^{65}$Zn \\
$^{42}$Sc & 0.0042 & & & $^{43}$Ca \\
$^{37}$Cl & 0.0017 & & & $^{37}$Cl \\
$^{21}$Ne & 5.4\times10^{-4} & & & $^{21}$Ne \\ 
$^{17}$F  & 3.9\times10^{-4} & & $^{14}$N  & \\
$^{14}$C  & 1.9\times10^{-4} & & & $^{21}$Ne \\
$^{64}$Ga & 6.3\times10^{-5} & & & $^{63}$Cu, $^{65}$Zn \\
$^{63}$Ga & 5.0\times10^{-5} & & & $^{64}$Zn \\
\end{tabular}
\end{ruledtabular}
\end{table*}

Radiative captures of protons are by far the group of reactions whose
rate most strongly determine the final abundances of the supernova
explosion, as can be deduced from Table~\ref{tab6}. The reaction with
the largest $|D_{i}|$ of the whole network is
$^{30}\text{Si}+\text{p}\rightleftarrows{}^{31}\text{P}+\gamma$, for
which as many as 20 product species have $|D_{i}|>0.05$. The species
most affected by changes in the rates of the proton capture group of
reactions are $^{14}$N, $^{26}$Mg, $^{26}$Al, $^{27}$Al, $^{32}$P,
$^{35}$S, and $^{43}$Ca. Among these, $^{27}$Al is the species with
the largest yield, $m_i = 4.6\times 10^{-4}$~M$_\odot$. Within the species
with $0.3>|D_{i}|>0.05$ there are important products of the
supernova explosion such as $^{24}$Mg, $^{25}$Mg, $^{29}$Si,
$^{30}$Si, $^{31}$P, $^{33}$S, $^{34}$S, $^{35}$Cl, $^{38}$Ar,
$^{50}$Cr, $^{51}$Cr, and $^{55}$Fe. The parent nuclei involved in
these reactions cover a wide range from $^{14}$C to $^{64}$Ga.

\begin{table*}
\caption{Sensitivity of the nucleosynthesis to the rate of $\left(\text{p},\text{n}\right)$
reactions with parent nuclide given in the first column. \footnote{The reactions listed are
those that processed more than $10^{-6}$~M$_\odot$ in
the reference model (see Table~\ref{tab3}) \textsl{and} with any
$\text{max}\left(|D_{i}|\right)>0.05$.}
\label{tab7}}
\begin{ruledtabular}
\begin{tabular}{l@{\hspace{0.1truecm}}|d@{\hspace{0.1truecm}}|l@{\hspace{0.1truecm}}|l}
Parent nuclide & \mathfrak{M}_{jk} \left(\text{M}_\odot\right) & Nuclei with $|D_{i}|>0.3$ &
Nuclei with $0.3>|D_{i}|>0.05$ \\
\hline
$^{53}$Mn & 0.61 & & $^{53}$Mn \\
$^{57}$Co & 0.57 & & $^{57}$Co, $^{61}$Ni, $^{65}$Zn \\
$^{30}$Si & 0.54 & & $^{25}$Mg, $^{26}$Al, $^{29}$Si, $^{32}$P, $^{35}$S  \\
$^{32}$P  & 0.44 & $^{32}$P  & \\
$^{56}$Fe & 0.31 & & $^{55}$Mn, $^{57,58}$Fe \\
$^{58}$Co & 0.27 & & $^{58}$Co \\
$^{49}$V  & 0.26 & & $^{49}$V  \\
$^{47}$Ti & 0.12 & & $^{46}$Ti \\
$^{27}$Al & 0.12 & & $^{26}$Al \\
$^{54}$Mn & 0.11 & & $^{54}$Mn \\
$^{44}$Sc & 0.072 & & $^{47}$Ti \\
$^{55}$Mn & 0.068 & & $^{55}$Mn \\
$^{45}$Sc & 0.064 & & $^{45}$Sc \\
$^{37}$Cl & 0.054 & $^{37}$Cl & \\
$^{62}$Cu & 0.015 & & $^{63}$Cu, $^{65}$Zn \\
$^{26}$Mg & 0.014 & & $^{26}$Al \\
$^{18}$Ne & 7.1\,10^{-4} & & $^{14}$N  \\
$^{35}$S  & 3.4\,10^{-5} & & $^{35}$S  \\
\end{tabular}
\end{ruledtabular}
\end{table*}

The rates of $\left(\text{p},\text{n}\right)$ reactions do not
influence significantly the nucleosynthesis. The most affected species
are $^{32}$P and $^{37}$Cl, both with final yields on the order of
$10^{-7}$~M$_\odot$. Among the species with $|D_{i}|>0.05$, the most
abundant are $^{29}$Si, with $m_i = 4.5\times10^{-4}$~M$_\odot$, $^{25}$Mg
and $^{49}$V, both with yields on the order of $10^{-5}$~M$_\odot$.

\begin{table*}
\caption{Sensitivity of the nucleosynthesis to the rate of $\left(\alpha,\gamma\right)$
reactions with parent nuclide given in the first column. \footnote{The reactions listed are
those that processed more than $10^{-6}$~M$_\odot$ in
the reference model (see Table~\ref{tab3}) \textsl{and} with any
$\text{max}\left(|D_{i}|\right)>0.05$.}
\label{tab8}}
\begin{ruledtabular}
\begin{tabular}{l@{\hspace{0.1truecm}}|d@{\hspace{0.1truecm}}|l@{\hspace{0.1truecm}}|l}
Parent nuclide & \mathfrak{M}_{jk} \left(\text{M}_\odot\right) & Nuclei with $|D_{i}|>0.3$ &
Nuclei with $0.3>|D_{i}|>0.05$ \\
\hline
$^{28}$Si & 0.93 & & $^{30}$Si, $^{31,33}$P, $^{33,34}$S, $^{35}$Cl, $^{38}$Ar\\
$^{32}$S  & 0.39 & & $^{37}$Cl \\
$^{20}$Ne & 0.33 & $^{24,25}$Mg, $^{26,27}$Al, $^{30}$Si, $^{35,37}$Cl,
 & $^{14}$N, $^{20,21}$Ne, $^{23}$Na, $^{26}$Mg, $^{28,29}$Si, $^{32,33}$P, $^{33,34}$S, 
$^{36-38}$Ar,\\
 & & $^{39}$K, $^{41-43}$Ca, $^{46,47}$Ti  &
 $^{40}$Ca, $^{44}$Ti, $^{48,49}$V, $^{52,53}$Mn \\
$^{16}$O  & 0.30 & & $^{14}$N, $^{20,21}$Ne, $^{23}$Na, $^{24}$Mg, $^{27}$Al, $
 ^{29}$Si, $^{32}$P, $^{33,35}$S, $^{45}$Sc \\ 
$^{40}$Ca & 0.20 & & $^{44}$Ti \\
$^{24}$Mg & 0.19 & $^{24}$Mg, $^{35}$S  & $^{23}$Na, $^{25}$Mg, $^{26,27}$Al, $
 ^{30}$Si, $^{31}$P, $^{35}$Cl, $^{38}$Ar, $^{39}$K, $^{41,42}$Ca, \\
 & & & $^{45}$Sc, $^{46,47}$Ti \\
$^{58}$Ni & 0.15 & & $^{62}$Ni, $^{63}$Cu, $^{64-66}$Zn \\
$^{57}$Ni & 0.090 & & $^{61}$Ni \\
$^{12}$C  & 0.074 & $^{45}$Sc & $^{14}$N, $^{28-30}$Si, $^{32}$P, $
 ^{37}$Ar, $^{39}$K, $^{40-42}$Ca, $^{44,46}$Ti, $^{48}$V, \\
 & & & $^{52}$Mn \\
$^{29}$Si & 0.065 & & $^{33}$S  \\
$^{33}$S  & 0.062 & & $^{37}$Cl, $^{37}$Ar \\
$^{30}$Si & 0.047 & $^{35}$S  & $^{30}$Si, $^{31-33}$P, $^{34}$S, $
 ^{35}$Cl, $^{37,38}$Ar, $^{39}$K, $^{42,43}$Ca \\
$^{34}$S  & 0.029 & & $^{38}$Ar, $^{39}$K  \\
$^{41}$Ca & 0.024 & & $^{43}$Ca, $^{47}$Ti \\
$^{42}$Ca & 0.011 & & $^{38}$Ar, $^{39}$K, $^{41,42}$Ca, $^{46,47}$Ti \\
$^{14}$N  & 3.1\,10^{-4} & & $^{21}$Ne \\
$^{62}$Zn & 1.2\,10^{-4} & & $^{66}$Zn \\
$^{17}$O  & 1.0\,10^{-4} & & $^{21}$Ne \\
\end{tabular}
\end{ruledtabular}
\end{table*}

There are several reactions of radiative capture of $\alpha$ particles
that bear a non-negligible influence on the synthesis of large numbers
of species (Table~\ref{tab8}). The most notable is the reaction
$^{20}\text{Ne}+\alpha\rightleftarrows{}^{24}\text{Mg}+\gamma$, for
which there are 13 species with $|D_{i}|>0.3$ and 20 species with
$0.3>|D_{i}|>0.05$, among them species with large abundances such as
$^{20}$Ne, $^{24}$Mg, $^{28}$Si, $^{36}$Ar, $^{40}$Ca, or
$^{52}$Mn. Variations in the rate of $\left(\alpha,\gamma\right)$
reactions on $^{24}$Mg and $^{12}$C also influence the yields of
large numbers of species, although not as much as does the reaction
$^{20}\text{Ne}\left(\alpha,\gamma\right)$, while captures on
$^{28}$Si and $^{32}$S have a much more limited reach. Note that,
because we always modify coherently direct and inverse reaction rates
to maintain detailed balance, the sensitivity to the reactions
$^{20}\text{Ne}\left(\alpha,\gamma\right)^{24}\text{Mg}$ and
$^{24}\text{Mg}\left(\alpha,\gamma\right)^{28}\text{Si}$ also pick up
the effect of variations on
$^{24}\text{Mg}\left(\gamma,\alpha\right)^{20}\text{Ne}$ and
$^{28}\text{Si}\left(\gamma,\alpha\right)^{24}\text{Mg}$,
respectively, which are reactions relevant for silicon burning. Most
of the parent nuclei listed in the table belongs to the IMEs group. An
interesting exception is the reaction
$^{58}\text{Ni}+\alpha\rightleftarrows{}^{62}\text{Zn}+\gamma$, which
plays a relevant role in the alpha-rich freeze-out of incinerated
matter that leaves NSE at densities below $\sim10^8$~g~cm$^{-3}$,
because of the adiabatic expansion of the ejecta.

\begin{table*}
\caption{Sensitivity of the nucleosynthesis to the rate of $\left(\alpha,\text{n}\right)$
reactions with the parent nuclide given in the first column. \footnote{The reactions listed are
those that processed more than $10^{-6}$~M$_\odot$ in
the reference model (see Table~\ref{tab3}) \textsl{and} with any
$\text{max}\left(|D_{i}|\right)>0.05$.}
\label{tab9}}
\begin{ruledtabular}
\begin{tabular}{l@{\hspace{0.1truecm}}|d@{\hspace{0.1truecm}}|l@{\hspace{0.1truecm}}|l}
Parent nuclide & \mathfrak{M}_{jk} \left(\text{M}_\odot\right) & Nuclei with $|D_{i}|>0.3$ &
Nuclei with $0.3>|D_{i}|>0.05$ \\
\hline
 $^{29}$Si & 0.70 & $^{35}$S  & $^{21}$Ne, $^{25}$Mg, $^{26}$Al, $^{29}$Si, $^{34}$S, $
 ^{35}$Cl, $^{43}$Ca, $^{47}$Ti \\
 $^{33}$S  & 0.64 & & $^{37}$Cl, $^{43}$Ca \\
 $^{27}$Al & 0.54 & $^{43}$Ca, $^{47}$Ti & $^{25}$Mg, $^{26,27}$Al, $^{29}$Si, $
 ^{32,33}$P, $^{33,35}$S, $^{37}$Cl \\
 $^{30}$Si & 0.43 & $^{33}$P, $^{43}$Ca & $^{29}$Si, $^{32}$P, $^{33,35}$S, $
 ^{37}$Cl, $^{47}$Ti \\
 $^{25}$Mg & 0.28 & $^{26}$Al & $^{17}$O, $^{21}$Ne, $^{25}$Mg, $^{32}$P, $^{35}$S  \\
 $^{34}$S  & 0.10 & $^{37}$Cl & $^{37}$Ar \\
 $^{26}$Mg & 0.075 & $^{26}$Mg, $^{35}$S  & $^{21}$Ne, $^{25}$Mg, $^{26,27}$Al,
$ ^{29}$Si, $^{32}$P, $^{43}$Ca, $^{47}$Ti \\
 $^{46}$Ti & 0.058 & & $^{46}$Ti \\
 $^{41}$Ca & 0.055 & & $^{43}$Ca, $^{47}$Ti \\
 $^{38}$Ar & 0.042 & & $^{41}$Ca \\
 $^{21}$Ne & 0.014 & $^{21}$Ne & \\
 $^{22}$Ne & 0.013 & & $^{17}$O, $^{22}$Ne \\
 $^{17}$O  & 0.011 & $^{17}$O  & $^{21}$Ne \\
 $^{14}$C  & 7.5\times 10^{-4} & & $^{17}$O  \\
\end{tabular}
\end{ruledtabular}
\end{table*}

The species most affected by variations on the rate of
$\left(\alpha,\text{n}\right)$ reactions, as well as the parent species
listed in Table~\ref{tab9}, belong to the IMEs group. Among the species
with $|D_{i}|>0.3$, the most abundant is $^{26}$Mg, whose yield is
$2.7\times 10^{-5}$~M$_\odot$.

\begin{table*}
\caption{Sensitivity of the nucleosynthesis to the rate of $\left(\alpha,\text{p}\right)$
reactions with parent nuclide given in the first column. \footnote{The reactions listed are
those that processed more than $10^{-6}$~M$_\odot$ in
the reference model (see Table~\ref{tab3}) \textsl{and} with any
$\text{max}\left(|D_{i}|\right)>0.05$.}
\label{tab20}}
\begin{ruledtabular}
\begin{tabular}{l@{\hspace{0.1truecm}}|d@{\hspace{0.1truecm}}|l@{\hspace{0.1truecm}}|l}
Parent nuclide & \mathfrak{M}_{jk} \left(\text{M}_\odot\right) & Nuclei with $|D_{i}|>0.3$ &
Nuclei with $0.3>|D_{i}|>0.05$ \\
\hline
 $^{28}$Si & 0.77 & & $^{26}$Mg, $^{31-33}$P, $^{35}$S, $^{37}$Cl \\
 $^{30}$P  & 0.68 & & $^{43}$Ca \\
 $^{32}$S  & 0.65 & & $^{32,33}$P, $^{35}$S, $^{35,37}$Cl \\
 $^{27}$Al & 0.63 & $^{27}$Al, $^{35}$S  & $^{24-26}$Mg, $^{26}$Al, 
  $^{29,30}$Si, $^{32,33}$P, $^{33}$S, $^{37}$Cl, $^{43}$Ca, $^{47}$Ti \\
 $^{31}$P  & 0.61 & & $^{30}$Si, $^{31-33}$P, $^{34,35}$S, $^{35}$Cl \\
 $^{24}$Mg & 0.58 & $^{35}$S, $^{43}$Ca & $^{24-26}$Mg, $^{26,27}$Al, $^{30}$Si, $^{32}$P,
  $^{34}$S, $^{35,37}$Cl, $^{47}$Ti \\
 $^{56}$Ni & 0.58 & & $^{43}$Ca, $^{47}$Ti, $^{59,60}$Ni, $^{63}$Cu, $^{64,65}$Zn\\
 $^{33}$S  & 0.55 & & $^{43}$Ca \\
 $^{39}$K  & 0.48 & & $^{43}$Ca, $^{47}$Ti \\
 $^{40}$Ca & 0.44 & & $^{43}$Ca \\
 $^{29}$Si & 0.44 & & $^{32}$P  \\
 $^{13}$N  & 0.39 & & $^{14}$N, $^{28}$Si, $^{37,38}$Ar, $^{40,43}$Ca, 
  $^{45}$Sc, $^{44}$Ti, $^{48,49}$V, $^{50}$Cr, $^{52,53}$Mn \\
 $^{35}$Cl & 0.38 & & $^{38}$Ar, $^{39}$K, $^{41,42}$Ca, $^{47}$Ti \\
 $^{20}$Ne & 0.37 & $^{23}$Na & $^{17}$O, $^{21}$Ne, $^{26}$Mg, $^{35}$S, $^{43}$Ca \\
 $^{25}$Mg & 0.24 & & $^{26}$Al \\
 $^{58}$Ni & 0.23 & & $^{62}$Ni, $^{63}$Cu, $^{64,66}$Zn \\
 $^{44}$Ti & 0.18 & & $^{14}$N, $^{45}$Sc, $^{44,47}$Ti, $^{48,49}$V  \\
 $^{57}$Ni & 0.17 & & $^{61}$Ni, $^{65}$Zn \\
 $^{48}$Cr & 0.13 & & $^{48}$V, $^{49}$V  \\
 $^{45}$Ti & 0.12 & & $^{45}$Sc \\
 $^{23}$Na & 0.12 & $^{26}$Mg, $^{43}$Ca & $^{14}$N, $^{21}$Ne, $^{23}$Na, $^{29}$Si, $
 ^{32}$P, $^{33}$S, $^{37}$Cl, $^{40}$Ca, $^{45}$Sc, $^{44,47}$Ti \\
 $^{41}$Ca & 0.068 & & $^{43}$Ca, $^{47}$Ti \\
 $^{46}$Ti & 0.068 & & $^{46}$Ti \\
 $^{35}$Ar & 0.061 & & $^{43}$Ca \\
 $^{34}$S  & 0.054 & $^{37}$Cl & \\
 $^{39}$Ca & 0.049 & & $^{47}$Ti \\
 $^{30}$Si & 0.032 & & $^{33}$P  \\
 $^{42}$Ca & 0.020 & & $^{46}$Ti \\
 $^{60}$Zn & 0.0077 & & $^{63}$Cu, $^{64}$Zn \\
 $^{62}$Zn & 0.0054 & & $^{66}$Zn \\
 $^{42}$Sc & 0.0024 & & $^{43}$Ca \\
 $^{61}$Zn & 0.0020 & & $^{65}$Zn \\
 $^{17}$Ne & 7.6\,10^{-4} & $^{14}$N  & \\
 $^{43}$Ti & 4.1\,10^{-4} & & $^{47}$Ti \\
 $^{18}$Ne & 3.4\,10^{-4} & & $^{14}$N  \\
 $^{21}$Ne & 1.6\,10^{-4} & & $^{21}$Ne \\
 $^{47}$Cr & 8.9\,10^{-5} & $^{47}$Ti & \\
\end{tabular}
\end{ruledtabular}
\end{table*}

In Table~\ref{tab20}, we give the sensitivities to the rates of
$\left(\alpha,p\right)$ reactions. Abundant species most
affected by variations of this type of reactions are $^{23}$Na,
$^{26}$Mg, and $^{27}$Al, the last with a yield of
$4.6\times 10^{-4}$~M$_\odot$. As with $\left(p,\gamma\right)$
reactions, the parent species cover a wide range of baryon numbers,
from $^{13}$N to $^{62}$Zn. The most influential reactions are
$^{13}\text{N}+\alpha\rightleftarrows{}^{16}\text{O}+\text{p}$,
$^{20}\text{Ne}+\alpha\rightleftarrows{}^{23}\text{Na}+\text{p}$,
$^{23}\text{Na}+\alpha\rightleftarrows{}^{26}\text{Mg}+\text{p}$,
$^{24}\text{Mg}+\alpha\rightleftarrows{}^{27}\text{Al}+\text{p}$, and
$^{27}\text{Al}+\alpha\rightleftarrows{}^{30}\text{Si}+\text{p}$. There
are present as well several reactions relevant for the alpha-rich
freeze-out of NSE, such as
$^{56}\text{Ni}+\alpha\rightleftarrows{}^{59}\text{Cu}+\text{p}$ and
$^{58}\text{Ni}+\alpha\rightleftarrows{}^{61}\text{Cu}+\text{p}$.

Several reactions appear in the tables that are responsible for bridging
the gap between the quasi-statistical equilibrium groups (the QSE groups of silicon and iron) in
silicon burning \cite{hix96}, such as
$^{45}\text{Sc}+\text{p}\rightleftarrows{}^{46}\text{Ti}+\gamma$,
$^{42}\text{Ca}+\alpha\rightleftarrows{}^{46}\text{Ti}+\gamma$, and
$^{44}\text{Ti}+\alpha\rightleftarrows{}^{47}\text{V}+\text{p}$. It is
remarkable that the set of abundances significantly affected by the
modification of the rates of these reactions is limited to species that
fall into the QSE groups gap, but there appear very few species
belonging to the QSE groups. When one of
these reactions is modified, the increase or decrease it produces in
the flux between QSE groups is offset against a slight adjustment in
the abundances of other species located within the gap, therefore
leaving the abundances of the species in QSE nearly unchanged.
For instance, modifying the
rate of $^{42}\text{Ca}+\alpha\rightleftarrows{}^{46}\text{Ti}+\gamma$ by a factor of $\times10$,
the global flux carried by all the reactions that bridge the QSE gap changes by less than 5\%, in
spite of an increase in the flux carried by the
$^{42}\text{Ca}+\alpha\rightleftarrows{}^{46}\text{Ti}+\gamma$ reaction by a factor of $\times7$
(together with a decrease of the final yield of $^{42}$Ca by a factor of $\times0.7$ and an
increase of the final yield of $^{46}$Ti by a factor of $\times1.8$). The larger flux carried by
$^{42}\text{Ca}+\alpha\rightleftarrows{}^{46}\text{Ti}+\gamma$ is offset by a decrease in the
fluxes due to $^{44}\text{Ti}+\alpha\rightleftarrows{}^{47}\text{V}+\text{p}$, and other reactions
within the gap.

\begin{figure*}[!htb]
 \includegraphics[width=8
truecm]{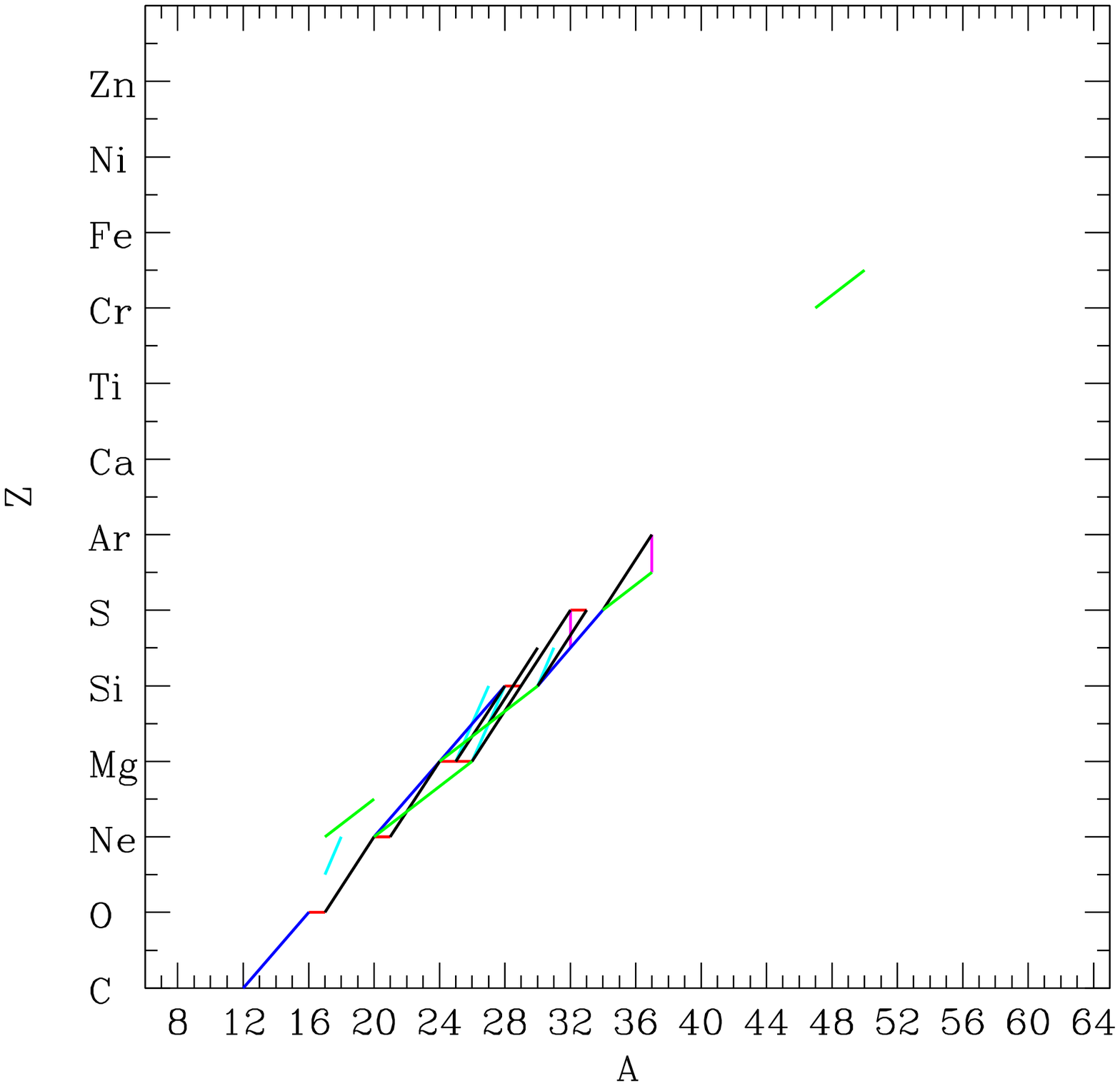}
 \includegraphics[width=8truecm]{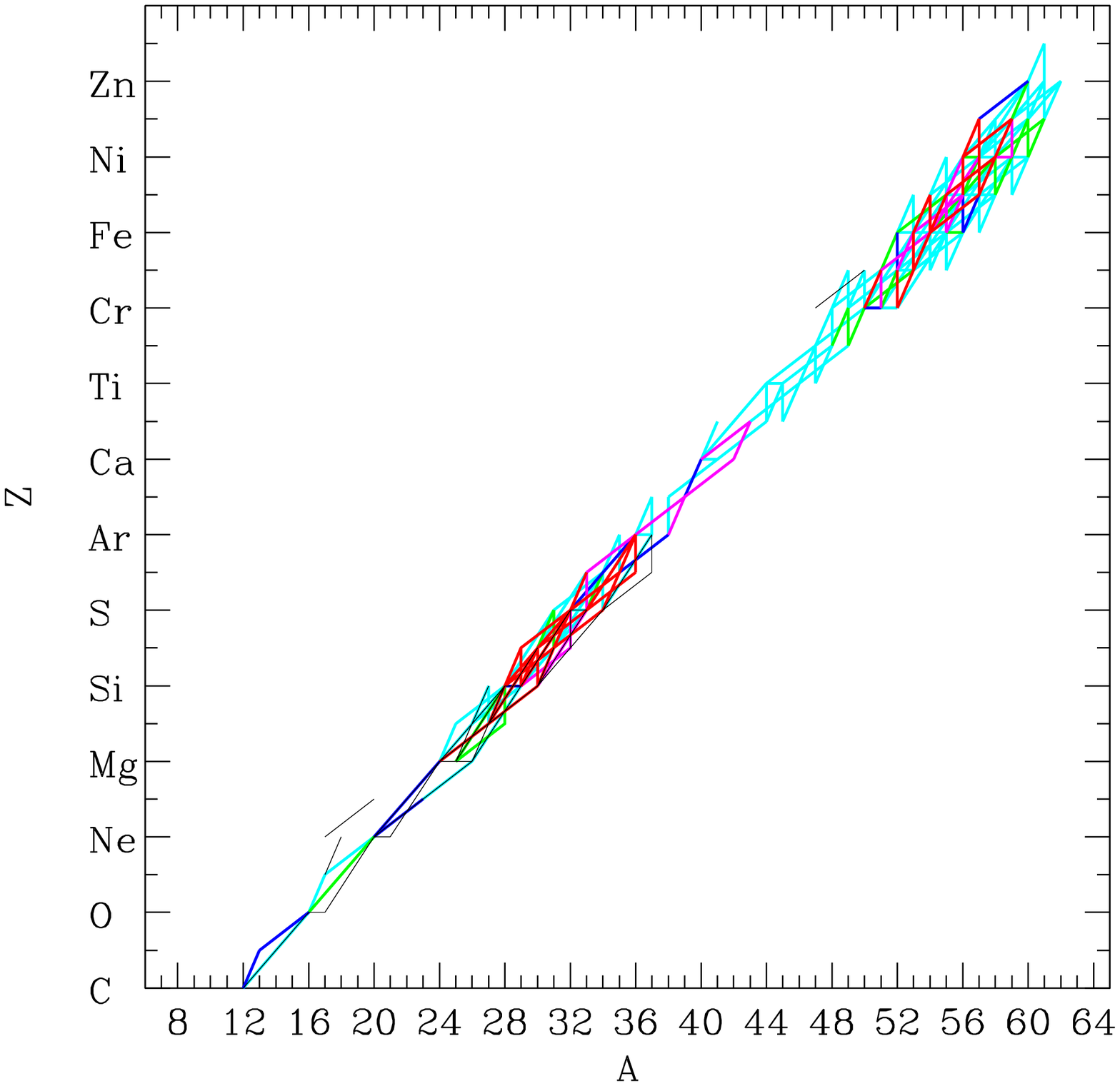}
\caption{
(Color online) Chart of the most influential reactions compared to the maximum mass flow path.
\textbf{Left:} Reactions with $\text{max}\left(|D_{i}|\right)>0.3$ in
Tables~\ref{tab5} to \ref{tab20}. The type of each reaction can be deduced from the differences of
atomic and baryonic number of the nuclei it connects or from the color in which they are drawn:
$\left(\text{n},\gamma\right)$ reactions in red, $\left(\text{p},\gamma\right)$ reactions in cyan,
$\left(\text{p},\text{n}\right)$ reactions in magenta, $\left(\alpha,\gamma\right)$ reactions in
blue, $\left(\alpha,\text{n}\right)$ reactions in black, and $\left(\alpha,\text{p}\right)$
reactions in green. We recall that direct and inverse reaction rates were modified simultaneously.
\textbf{Right:} Reactions that carry a large mass flow. In this plot, the mass flow has been
color coded according to the value of $\mathfrak{M}_{jk}$: red for
$\mathfrak{M}_{jk}>0.5~\mathrm{M}_\odot$, magenta for
$0.5\ge\mathfrak{M}_{jk}>0.4~\mathrm{M}_\odot$, blue for
$0.4\ge\mathfrak{M}_{jk}>0.3~\mathrm{M}_\odot$, green for
$0.3\ge\mathfrak{M}_{jk}>0.2~\mathrm{M}_\odot$, and cyan for
$0.2\ge\mathfrak{M}_{jk}>0.01~\mathrm{M}_\odot$. The chart of most influential reactions has been
superimposed as black thin lines.
\label{fig12}}
\end{figure*}

We have plotted in Fig.~\ref{fig12} the reactions with the largest
$\text{max}\left(|D_{i}|\right)$ in Tables~\ref{tab5} to
\ref{tab20}. It is remarkable that no reaction appears involving
the main products of SNIa, i.e. elements from the Fe-group nuclei,
except
$^{47}\text{Cr}+\alpha\rightleftarrows{}^{50}\text{Mn}+\text{p}$. This
reaction might play a role in the freeze-out from incomplete silicon
burning \cite{hix99}, as one of the last links between the two main
QSE groups. The rest of the reactions in the plot sketch a connected
network from $^{12}$C up to $^{37}$Ar. The pattern displayed in the
figure suggests that we can talk not only of the reactions whose
rates are most influential in the supernova yields, but also of a path
in the nuclear chart that is most influential. The main
stream involves $\left(\alpha,\gamma\right)$ reactions from $^{12}$C
to $^{28}$Si, although the step from $^{16}$O to $^{20}$Ne is not
covered by the
$^{16}\text{O}+\alpha\rightleftarrows{}^{20}\text{Ne}+\gamma$ reaction
pair but by the combination of
$^{16}\text{O}+\text{n}\rightleftarrows{}^{17}\text{O}+\gamma$ and
$^{17}\text{O}+\alpha\rightleftarrows{}^{20}\text{Ne}+\text{n}$. Above
Mg, there appear many branches due to a number of
$\left(\alpha,\text{p}\right)$ and $\left(\alpha,\text{n}\right)$
reactions and their inverses, which shift the stream to the side of
moderately neutron-rich nuclei. 
The path ends in a loop involving
the reactions
$^{34}\text{S}+\alpha\rightleftarrows{}^{37}\text{Ar}+\text{n}$,
$^{37}\text{Ar}+\text{n}\rightleftarrows{}^{37}\text{Cl}+\text{p}$, and
$^{37}\text{Cl}+\text{p}\rightleftarrows{}^{34}\text{S}+\alpha$.

The right panel of Fig.~\ref{fig12} shows the path defined by the reactions that carry the largest
mass flow during the SNIa explosion, computed through Eq.~\ref{eqmassflow}, together with the path
of most influential reactions. A couple of points have to be retained to make a
meaningful comparison between both paths. First, as explained in Section~\ref{secselec}, in shells
achieving NSE, the mass flow prior to equilibrium has not been accounted for. Hence,
Fig.~\ref{fig12} does not reflect the nuclear flow from carbon and oxygen up to the Fe-group in
incinerated shells, giving the impression that the Si-group and the Fe-group are disconnected. 
Second, our consideration of a nuclear reaction as highly influential is based on the relative
variation of the yield of any species that has $\text{max}\left(|D_{i}|\right)>0.3$. Thus, it is
possible for a nuclear reaction off the maximum mass flow path to affect significatively the
abundance of trace species.
The maximum mass flow path follows the $Z=N$ line from carbon to the Si-group, but deviates to the
neutron rich side within the Fe-group, that accumulates the neutron excess due to the initial
metallicity and the electron captures close to the center of the star (see Fig.~\ref{fig1}).
Consequently, and due to the small abundance of neutrons, the connection between both groups is
provided mainly by $\left(\alpha,\text{p}\right)$ reactions.
It is evident from Fig.~\ref{fig12} that the most influential path occupies a region slightly more
neutron rich as compared to the maximum mass flow path, especially below $A\sim28$. The species
most affected by reactions that lay off the maximum mass flow path are trace species, e.g.
$^{17}$O,
$^{21}$Ne, $^{25}$Mg, $^{37}$Cl, and $^{47}$Ti. Reactions within the Si-group form a dense network
in which the rate of a particular reaction loses relevance, and the same applies to reactions
within the Fe-group.

Tables~\ref{tab21} and \ref{tab22} give, for the most important
product nuclei, $i$, the reactions $j+k\rightleftarrows l+m$ that have
the largest impact on its yield, $|D_{i}|$. We have included in
these tables only the species with mass fraction greater than
$10^{-5}$, or that are interesting radioactive isotopes, and with
$\text{max}\left(|D_{i}|\right)>0.01$. We show as well the
production factor of these species (in our reference model), 
to help in evaluating the relevance of the modifications to the yield of
each nuclide. For radioactive isotopes, we have calculated the
production factor taking as a reference the solar abundance of the end
product of the disintegration chain. For each nucleus, we show a
maximum of ten reactions, sorted by $|D_{i}|$.

\begin{turnpage}
\begin{table*}
\caption{Rates that influence the yields of each product species, from Carbon to
Chlorine.\footnote{Only the
species with
mass fraction greater than $10^{-5}$ or that are interesting radioactive isotopes \textsl{and}
with $\text{max}\left(|D_{i}|\right)>0.01$ are shown here. For each nucleus, we show a
maximum of ten reactions.}
\label{tab21}}
\begin{ruledtabular}
\begin{tabular}{l@{\hspace{0.1truecm}}|l@{\hspace{0.1truecm}}|d@{\hspace{0.1truecm}}|l}
Nucleus & End product\footnote{For radioactive nuclei, here it is shown the end product of the
disintegration chain as well as the longest half-life in the decay chain.} &
\text{Production} & Reaction and $D_{i}$ \\
 & & \text{factor}\footnote{Mass fraction of the species in the supernova ejecta
normalized to its solar mass
fraction. For radioactive nuclides the normalization is with respect to the solar mass fraction
of the end product of the disintegration chain.} & \\
\hline
 $^{16}$O & &  9.49 & $^{20}$Ne($\alpha$, $\gamma$):-0.03; $^{12}$C (n,$\gamma$):+0.02; $^{12}$C
($\alpha$,$\gamma$):-0.02 \\

 $^{20}$Ne & &  1.17 & $^{20}$Ne($\alpha$,$\gamma$):-0.29; $^{16}$O ($\alpha$,$\gamma$):-0.11;
$^{20}$Ne($\alpha$,p):+0.03; $^{23}$Na($\alpha$,p):-0.03;\\
& & & $^{24}$Mg($\alpha$,$\gamma$):+0.02;
$^{16}$O (n,$\gamma$):+0.02; $^{23}$Na(p,$\gamma$):-0.01 \\

 $^{23}$Na & & 0.29 & $^{20}$Ne($\alpha$,p):-0.46; $^{23}$Na($\alpha$,p):-0.23;
$^{30}$Si(p,$\gamma$):+0.15; $^{20}$Ne($\alpha$,$\gamma$):+0.12; $^{27}$Al(p,$\gamma$):+0.08;\\
& & & $^{26}$Mg(p,$\gamma$):+0.07; $^{16}$O ($\alpha$,$\gamma$):+0.07;
$^{24}$Mg($\alpha$,$\gamma$):+0.06; $^{27}$Al($\alpha$,p):-0.05; $^{12}$C (n,$\gamma$):+0.04 \\

 $^{24}$Mg & & 22.5 & $^{20}$Ne($\alpha$,$\gamma$):+0.70; $^{24}$Mg($\alpha$,$\gamma$):-0.42;
$^{24}$Mg($\alpha$,p):-0.17; $^{27}$Al(p,$\gamma$):-0.11; $^{30}$Si(p,$\gamma$):-0.09;\\
& & & $^{27}$Al($\alpha$,p):-0.08; $^{16}$O ($\alpha$,$\gamma$):-0.07; $^{24}$Mg(n,$\gamma$):-0.05;
$^{12}$C (n,$\gamma$):-0.05; $^{26}$Mg($\alpha$,n):-0.04 \\

 $^{25}$Mg & & 0.14 & $^{20}$Ne($\alpha$,$\gamma$):+0.37; $^{25}$Mg(p,$\gamma$):-0.29;
$^{25}$Mg($\alpha$,n):-0.22; $^{24}$Mg(n,$\gamma$):+0.22; $^{30}$Si(p,$\gamma$):+0.20;\\
& & & $^{24}$Mg($\alpha$,p):-0.17; $^{25}$Mg(n,$\gamma$):-0.17; $^{27}$Al($\alpha$,p):-0.16;
$^{29}$Si($\alpha$,n):+0.15; $^{26}$Mg($\alpha$,n):+0.10 \\

 $^{26}$Mg & & 0.22 & $^{26}$Mg(p,$\gamma$):-0.34; $^{26}$Mg($\alpha$,n):-0.34;
$^{23}$Na($\alpha$,p):+0.31; $^{30}$Si(p,$\gamma$):+0.23; $^{20}$Ne($\alpha$,p):-0.17;\\
& & & $^{27}$Al($\alpha$,p):-0.13; $^{20}$Ne($\alpha$,$\gamma$):+0.12; $^{27}$Al(p,$\gamma$):+0.10;
$^{24}$Mg($\alpha$,p):-0.07; $^{28}$Si($\alpha$,p):+0.06 \\

 $^{26}$Al & $^{26}$Mg ($7.2\times 10^5$~yr)& 9.4\times 10^{-4} & $^{26}$Al(p,$\gamma$):-0.67;
$^{24}$Mg(n,$\gamma$):+0.46;
$^{25}$Mg($\alpha$,n):-0.39; $^{25}$Mg(p,$\gamma$):+0.38; $^{20}$Ne($\alpha$,$\gamma$):+0.37;\\
& & & $^{30}$Si(p,$\gamma$):+0.26; $^{29}$Si($\alpha$,n):+0.21; $^{24}$Mg($\alpha$,p):-0.19;
$^{27}$Al($\alpha$,p):-0.17; $^{26}$Mg($\alpha$,n):+0.15 \\ 

 $^{27}$Al & & 4.98 & $^{20}$Ne($\alpha$,$\gamma$):+0.48; $^{27}$Al($\alpha$,p):-0.44;
$^{27}$Al(p,$\gamma$):-0.35; $^{24}$Mg($\alpha$,$\gamma$):-0.28; $^{24}$Mg($\alpha$,p):+0.26;\\
& & & $^{30}$Si(p,$\gamma$):+0.09; $^{27}$Al($\alpha$,n):-0.06; $^{16}$O ($\alpha$,$\gamma$):-0.06;
$^{26}$Mg($\alpha$,n):-0.05; $^{28}$Si($\alpha$,p):+0.04 \\

 $^{28}$Si & & 220. & $^{13}$N ($\alpha$,p):-0.09; $^{20}$Ne($\alpha$,$\gamma$):+0.07;
$^{12}$C
($\alpha$,$\gamma$):-0.06; $^{23}$Na($\alpha$,p):+0.03;\\
& & & $^{16}$O ($\alpha$,$\gamma$):-0.02;
$^{12}$C (n,$\gamma$):-0.01; $^{28}$Si($\alpha$,$\gamma$):-0.01 \\

 $^{29}$Si & & 8.26 & $^{20}$Ne($\alpha$,$\gamma$):+0.21; $^{27}$Al($\alpha$,n):+0.18;
$^{29}$Si($\alpha$,n):-0.16; $^{30}$Si(p,$\gamma$):+0.13; $^{27}$Al($\alpha$,p):-0.11;\\
& & & $^{23}$Na($\alpha$,p):+0.11; $^{30}$Si(p,n):-0.09; $^{26}$Mg($\alpha$,n):+0.07; $^{32}$S
(n,$\gamma$):-0.06; $^{30}$Si($\alpha$,n):-0.06 \\

 $^{30}$Si & & 22.2 & $^{20}$Ne($\alpha$,$\gamma$):+0.37; $^{30}$Si(p,$\gamma$):-0.27;
$^{30}$Si($\alpha$,$\gamma$):-0.16; $^{27}$Al($\alpha$,p):+0.14;
$^{28}$Si($\alpha$,$\gamma$):-0.11;\\
& & &  $^{31}$P ($\alpha$,p):-0.09;
$^{24}$Mg($\alpha$,$\gamma$):-0.08; $^{24}$Mg($\alpha$,p):+0.08; $^{12}$C
($\alpha$,$\gamma$):-0.06; $^{26}$Mg($\alpha$,n):-0.05 \\

 $^{31}$P  & & 30.0 & $^{28}$Si($\alpha$,p):-0.20; $^{30}$Si(p,$\gamma$):+0.19;
$^{28}$Si($\alpha$,$\gamma$):-0.10; $^{30}$Si($\alpha$,$\gamma$):-0.09; $^{31}$P
($\alpha$,p):-0.06;\\
& & &  $^{24}$Mg($\alpha$,$\gamma$):+0.06; $^{27}$Al(p,$\gamma$):+0.04;
$^{27}$Al($\alpha$,p):+0.04; $^{31}$P (p,$\gamma$):-0.03; $^{28}$Si(n,$\gamma$):-0.03 \\

 $^{32}$S  & & 240. & $^{13}$N ($\alpha$,p):-0.04; $^{20}$Ne($\alpha$,$\gamma$):-0.03;
$^{28}$Si($\alpha$,$\gamma$):+0.01; $^{16}$O ($\alpha$,$\gamma$):-0.01 \\

 $^{33}$S  & & 35.4 & $^{32}$S (n,$\gamma$):+0.18; $^{30}$Si($\alpha$,n):-0.17;
$^{20}$Ne($\alpha$,$\gamma$):-0.17; $^{30}$Si(p,$\gamma$):-0.16; $^{27}$Al($\alpha$,p):+0.12;\\
& & & $^{28}$Si($\alpha$,$\gamma$):+0.09; $^{24}$Mg(n,$\gamma$):+0.07; $^{23}$Na($\alpha$,p):-0.06;
$^{28}$Si(n,$\gamma$):+0.05; $^{27}$Al($\alpha$,n):-0.05 \\

 $^{34}$S  & & 36.5 & $^{30}$Si(p,$\gamma$):+0.14; $^{20}$Ne($\alpha$,$\gamma$):-0.10;
$^{30}$Si($\alpha$,$\gamma$):+0.09; $^{24}$Mg($\alpha$,p):-0.09;
$^{28}$Si($\alpha$,$\gamma$):+0.07;\\
& & &  $^{31}$P ($\alpha$,p):+0.07; $^{27}$Al(p,$\gamma$):+0.06;
$^{29}$Si($\alpha$,n):+0.05; $^{24}$Mg($\alpha$,$\gamma$):+0.05; $^{32}$S ($\alpha$,p):+0.05 \\

 $^{35}$Cl & & 11.0 & $^{20}$Ne($\alpha$,$\gamma$):-0.35; $^{30}$Si($\alpha$,$\gamma$):+0.20;
$^{31}$P ($\alpha$,p):+0.12; $^{30}$Si(p,$\gamma$):+0.11; $^{24}$Mg($\alpha$,p):-0.07;\\
& & & $^{24}$Mg($\alpha$,$\gamma$):+0.07; $^{32}$S ($\alpha$,p):-0.07; $^{29}$Si($\alpha$,n):+0.07;
$^{27}$Al(p,$\gamma$):+0.07; $^{28}$Si($\alpha$,$\gamma$):+0.05 \\
\end{tabular}
\end{ruledtabular}
\end{table*}
\end{turnpage}

\begin{turnpage}
\begin{table*}
\caption{Rates that influence the yields of each product species, from Argon to Nickel
(continuation
of
Table~\ref{tab21}).\footnote{Only the species with
mass fraction greater than $10^{-5}$ or that are interesting radioactive isotopes \textsl{and}
with $\text{max}\left(|D_{i}|\right)>0.01$ are shown here. For each nucleus, we show a
maximum of ten reactions.} 
\label{tab22}}
\begin{ruledtabular}
\begin{tabular}{l@{\hspace{0.1truecm}}|l@{\hspace{0.1truecm}}|d@{\hspace{0.1truecm}}|l}
Nucleus & End product\footnote{For radioactive nuclei here it is shown the end product of the
disintegration chain as well as the longest half-life in the decay chain.} &
\text{Production} & Reaction and $D_{i}$ \\
 & & \text{factor}\footnote{Mass fraction of the species in the supernova ejecta
normalized to its solar mass
fraction. For radioactive nuclides the normalization is with respect to the solar mass fraction
of the end product of the disintegration chain.} & \\
\hline
 $^{36}$Ar & & 270. & $^{20}$Ne($\alpha$,$\gamma$):-0.09; $^{12}$C ($\alpha$,$\gamma$):+0.04;
$^{13}$N ($\alpha$,p):+0.03;\\
& & &  $^{23}$Na($\alpha$,p):-0.03; $^{44}$Ti($\alpha$,p):-0.02;
$^{23}$Na(p,$\gamma$):-0.02 \\

 $^{37}$Ar & $^{37}$Cl (35 d) & 7.37 & $^{20}$Ne($\alpha$,$\gamma$):-0.25; $^{34}$S
($\alpha$,n):-0.15;
$^{36}$Ar(n,$\gamma$):+0.11; $^{12}$C ($\alpha$,$\gamma$):+0.07;
$^{30}$Si($\alpha$,$\gamma$):+0.06;\\
& & &  $^{35}$Cl(p,$\gamma$):+0.06; $^{33}$S
($\alpha$,$\gamma$):+0.06; $^{13}$N ($\alpha$,p):-0.05; $^{24}$Mg($\alpha$,$\gamma$):+0.05;
$^{42}$Ca($\alpha$,$\gamma$):-0.05 \\

 $^{38}$Ar & & 13.6 & $^{20}$Ne($\alpha$,$\gamma$):-0.22; $^{35}$Cl($\alpha$,p):+0.11; $^{34}$S
($\alpha$,$\gamma$):+0.09; $^{30}$Si(p,$\gamma$):+0.09; $^{30}$Si($\alpha$,$\gamma$):+0.08;\\
& & & $^{24}$Mg($\alpha$,$\gamma$):+0.08; $^{28}$Si($\alpha$,$\gamma$):+0.06; $^{13}$N
($\alpha$,p):-0.05; $^{42}$Ca($\alpha$,$\gamma$):-0.05; $^{24}$Mg($\alpha$,p):-0.05 \\

 $^{39}$K  & & 7.81 & $^{20}$Ne($\alpha$,$\gamma$):-0.35; $^{12}$C ($\alpha$,$\gamma$):+0.16;
$^{35}$Cl($\alpha$,p):+0.09; $^{42}$Ca($\alpha$,$\gamma$):-0.07;
$^{24}$Mg($\alpha$,$\gamma$):+0.07;\\
& & &  $^{30}$Si($\alpha$,$\gamma$):+0.07; $^{39}$K
(p,$\gamma$):-0.07; $^{34}$S ($\alpha$,$\gamma$):+0.05; $^{45}$Sc(p,$\gamma$):-0.05;
$^{30}$Si(p,$\gamma$):+0.04 \\

 $^{40}$Ca & & 370. & $^{20}$Ne($\alpha$,$\gamma$):-0.15; $^{13}$N ($\alpha$,p):+0.11; $^{12}$C
($\alpha$,$\gamma$):+0.10; $^{23}$Na($\alpha$,p):-0.05; $^{23}$Na(p,$\gamma$):-0.03;\\
& & &  $^{16}$O
($\alpha$,$\gamma$):+0.03; $^{44}$Ti($\alpha$,p):-0.03; $^{21}$Na($\alpha$,p):-0.01; $^{12}$C
(n,$\gamma$):+0.01 \\ 

 $^{44}$Ti & $^{44}$Ca (60 yr) & 14.3 & $^{20}$Ne($\alpha$,$\gamma$):-0.17;
$^{44}$Ti($\alpha$,p):-0.17; $^{13}$N
($\alpha$,p):+0.16; $^{12}$C ($\alpha$,$\gamma$):+0.14; $^{40}$Ca($\alpha$,$\gamma$):-0.08;\\
& & & $^{23}$Na($\alpha$,p):-0.06; $^{16}$O ($\alpha$,$\gamma$):+0.04; $^{23}$Na(p,$\gamma$):-0.03;
$^{21}$Na($\alpha$,p):-0.02; $^{12}$C (n,$\gamma$):+0.01 \\

 $^{48}$V  & $^{48}$Ti (16 d) & 96.8 & $^{13}$N ($\alpha$,p):+0.11; $^{48}$Cr($\alpha$,p):-0.10;
$^{20}$Ne($\alpha$,$\gamma$):-0.09; $^{44}$Ti($\alpha$,p):+0.06; $^{12}$C
($\alpha$,$\gamma$):+0.05;\\
& & &  $^{23}$Na($\alpha$,p):-0.04; $^{16}$O ($\alpha$,$\gamma$):+0.03;
$^{49}$V (p,n):-0.02; $^{52}$Fe($\alpha$,p):-0.01; $^{23}$Na(p,$\gamma$):-0.01 \\

 $^{49}$V  & $^{49}$Ti (330 d) & 59.8 & $^{20}$Ne($\alpha$,$\gamma$):-0.08; $^{49}$V (p,n):-0.08;
$^{48}$Cr($\alpha$,p):-0.07; $^{13}$N ($\alpha$,p):+0.07; $^{44}$Ti($\alpha$,p):+0.06;\\
& & &  $^{12}$C
($\alpha$,$\gamma$):+0.05; $^{53}$Mn(p,n):-0.04; $^{23}$Na($\alpha$,p):-0.02; $^{49}$V
(p,$\gamma$):-0.02; $^{16}$O ($\alpha$,$\gamma$):+0.02 \\

 $^{50}$Cr & & 100. & $^{51}$Mn(p,$\gamma$):-0.11; $^{55}$Co(p,$\gamma$):+0.10; $^{13}$N
($\alpha$,p):-0.06; $^{20}$Ne($\alpha$,$\gamma$):+0.04; $^{53}$Mn(p,n):-0.03;\\
& & & $^{23}$Na($\alpha$,p):+0.02; $^{52}$Fe($\alpha$,p):-0.02; $^{50}$Cr(n,$\gamma$):-0.02;
$^{52}$Mn(p,$\gamma$):-0.02; $^{20}$Ne($\alpha$,p):-0.01 \\

 $^{51}$Cr & $^{51}$V (28 d) & 73.8 & $^{51}$Mn(p,$\gamma$):-0.17;
$^{20}$Ne($\alpha$,$\gamma$):-0.05; $^{13}$N
($\alpha$,p):+0.03; $^{50}$Cr(p,$\gamma$):+0.03; $^{53}$Mn(p,n):-0.03;\\
& & &  $^{44}$Ti($\alpha$,p):+0.03;
$^{52}$Fe($\alpha$,p):-0.02; $^{12}$C ($\alpha$,$\gamma$):+0.02; $^{16}$O
($\alpha$,$\gamma$):+0.01 \\

 $^{52}$Mn & $^{52}$Cr (5.6 d) & 280. & $^{13}$N ($\alpha$,p):+0.12;
$^{20}$Ne($\alpha$,$\gamma$):-0.07; $^{12}$C
($\alpha$,$\gamma$):+0.05; $^{16}$O ($\alpha$,$\gamma$):+0.04; $^{23}$Na($\alpha$,p):-0.04;\\
& & & $^{44}$Ti($\alpha$,p):+0.03; $^{52}$Fe($\alpha$,p):-0.03; $^{53}$Mn(p,n):-0.03;
$^{42}$Ca($\alpha$,$\gamma$):+0.01; $^{45}$Sc(p,$\gamma$):+0.01 \\

 $^{53}$Mn & $^{53}$Cr ($3.7\times 10^6$~yr) & 240. & $^{53}$Mn(p,n):-0.12; $^{13}$N
($\alpha$,p):+0.07;
$^{20}$Ne($\alpha$,$\gamma$):-0.05; $^{12}$C ($\alpha$,$\gamma$):+0.03; $^{52}$Mn(p,n):+0.03;\\
& & & $^{16}$O ($\alpha$,$\gamma$):+0.02; $^{23}$Na($\alpha$,p):-0.02; $^{44}$Ti($\alpha$,p):+0.02;
$^{52}$Fe($\alpha$,p):-0.02; $^{53}$Mn(p,$\gamma$):-0.01 \\

 $^{55}$Fe & $^{55}$Mn (2.7 yr) & 260. & $^{55}$Co(p,$\gamma$):-0.11; $^{13}$N ($\alpha$,p):+0.04;
$^{20}$Ne($\alpha$,$\gamma$):-0.03;\\
& & &  $^{12}$C ($\alpha$,$\gamma$):+0.02; $^{16}$O
($\alpha$,$\gamma$):+0.02; $^{23}$Na($\alpha$,p):-0.01 \\

 $^{57}$Co & $^{57}$Fe (270 d) & 300. & $^{57}$Co(p,n):-0.07; $^{13}$N ($\alpha$,p):+0.01 \\

 $^{56}$Ni & $^{56}$Fe (77 d) & 360. & $^{13}$N ($\alpha$,p):+0.03;
$^{20}$Ne($\alpha$,$\gamma$):-0.01; $^{12}$C($\alpha$,$\gamma$):+0.01 \\

 $^{58}$Ni & & 410. & $^{57}$Co(p,n):+0.01 \\

 $^{59}$Ni & $^{59}$Co ($7.6\times 10^4$~yr) & 83.4 & $^{59}$Cu(p,$\gamma$):-0.18;
$^{56}$Ni($\alpha$,p):+0.13; $^{58}$Ni(n,$\gamma$):-0.01; $^{57}$Co(p,n):+0.01 \\

 $^{60}$Ni & & 210. & $^{56}$Ni($\alpha$,p):+0.13; $^{59}$Cu(p,$\gamma$):+0.07;
$^{56}$Ni($\alpha$,$\gamma$):+0.01 \\

 $^{61}$Ni & & 110. & $^{57}$Ni($\alpha$,p):+0.16; $^{57}$Co(p,n):-0.08;
$^{57}$Ni($\alpha$,$\gamma$):+0.07; $^{60}$Cu(p,$\gamma$):+0.02; $^{21}$Na($\alpha$,p):-0.01 \\

 $^{62}$Ni & & 160. & $^{58}$Ni($\alpha$,$\gamma$):+0.12; $^{58}$Ni($\alpha$,p):+0.10;
$^{60}$Zn($\alpha$,p):+0.05; $^{57}$Co(p,n):+0.04;\\
& & &  $^{21}$Na($\alpha$,p):-0.02;
$^{59}$Cu($\alpha$,p):+0.02; $^{61}$Cu(p,$\gamma$):+0.01 \\
\end{tabular}
\end{ruledtabular}
\end{table*}
\end{turnpage}

Among the species with the largest production factors, 
the yields of $^{28}$Si and $^{32}$S are hardly affected by any
rate (Table~\ref{tab21}), the maximum $|D_{i}|$ being 0.09 and 0.04
respectively (both due to the reaction
$^{13}\text{N}+\alpha\rightleftarrows{}^{16}\text{O}+\text{p}$),
implying relative variations on their yields of 23\% and 10\% when the
rates change by a factor of ten. The same applies to $^{36}$Ar,
whose maximum $|D_{i}|$ is 0.09 (Table~\ref{tab22}), while the
yield of $^{40}$Ca is slightly more dependent on the rates of the
radiative $\alpha$ captures on $^{20}$Ne and $^{12}$C and on the
$\left(\alpha,\text{p}\right)$ reaction on $^{13}\text{N}$, with
$|D_{i}|$ up to 0.15 (variation of up to 40\% of the yield for a
rate change by a factor of ten). Isotopes belonging to the
Fe-group with production factor larger than 100 have a similar level
of sensitivity to the variation of the reaction rates with maximum
$|D_{i}|$ slightly above $\sim0.1$, with the exception of
$^{56}$Ni, $^{58}$Ni, $^{54}$Fe, and $^{56}$Fe (the last two do not
appear in the table), whose yields are quite robust.  

The list of
reactions that are most influential on $^{44}$Ti synthesis has some
points in common with that found by \cite{the98} in the context of
core-collapse supernovae as, for instance,
$^{44}\text{Ti}+\alpha\rightleftarrows{}^{47}\text{V}+\text{p}$,
$^{40}\text{Ca}+\alpha\rightleftarrows{}^{44}\text{Ti}+\gamma$,
$^{12}\text{C}+\alpha\rightleftarrows{}^{16}\text{O}+\gamma$, and the
$3\alpha$ reaction. However, in SNIa there are no reactions involving
nuclei heavier than Ti that affect significantly the yield of
$^{44}$Ti, at variance with what was found by \cite{the98}
for core-collapse supernovae. The main reason is that, in our models, $^{44}$Ti is made in
moderately neutronized matter ($\eta\gtrsim0.001$). In QSE, while the composition of the Si-group
is nearly independent of the neutron excess, that of the Fe-group is strongly affected
\cite{hix96}. Thus, an increase in the neutron excess favours the equilibrium linking of $^{44}$Ti
(a $\eta=0$ nucleus) to the Si-group, leading to a low sensitivity of its abundance to the rate of
the reactions within the Fe-group. Such a progressive decrease of the importance of the Fe-group
reactions for the production of $^{44}$Ti as $\eta$ increases, can also be deduced from comparison
of Tables 4, 7, and 8 in \cite{the98}.

\subsection{Enhancement factor function of temperature}\label{fdet} 

Tables~\ref{tab23} to \ref{tab28} are the same as Tables~\ref{tab5} to \ref{tab20}, except that
the enhancement factor $f$ is a function of temperature, given by Eq.~\ref{eq2}.

\begin{table*}
\caption{Sensitivity of the nucleosynthesis to the rate of
  $\left(\text{n},\gamma\right)$ reactions with the parent nuclide given in the first column, with
enhancement factor given by
Eq.~\ref{eq2}.\footnote{The reactions listed are those
that processed more than $10^{-6}$~M$_\odot$ in the reference model (see Table~\ref{tab3})
\textsl{and} with any $\text{max}\left(|D_{i}|\right)>0.05$.}
\label{tab23}}
\begin{ruledtabular}
\begin{tabular}{l@{\hspace{0.1truecm}}|l@{\hspace{0.1truecm}}|l}
Parent nuclide & Nuclei with $|D_{i}|>0.3$ & Nuclei with $0.3>|D_{i}|>0.05$ \\
\hline
 $^{28}$Si & & $^{21}$Ne,$^{25}$Mg,$^{26}$Al,$^{35}$S  \\
 $^{55}$Fe & & $^{55}$Mn \\
 $^{32}$S  & & $^{32,33}$P,$^{33,35}$S  \\
 $^{36}$Ar & & $^{37}$Cl,$^{37}$Ar \\
 $^{24}$Mg & & $^{21}$Ne,$^{25}$Mg,$^{26}$Al,$^{35}$S  \\
 $^{25}$Mg & & $^{17}$O,$^{21}$Ne,$^{25}$Mg,$^{26}$Al \\
 $^{56}$Fe & & $^{57}$Fe \\
 $^{16}$O  & & $^{17}$O  \\
 $^{46}$Ti & & $^{46}$Ti \\
 $^{20}$Ne & $^{21}$Ne & \\
 $^{12}$C  & & $^{21}$Ne \\
 $^{31}$P  & & $^{32}$P  \\
\end{tabular}
\end{ruledtabular}
\end{table*}

\begin{table*}
\caption{Sensitivity of the nucleosynthesis to the rate of $\left(\text{p},\gamma\right)$
reactions with the parent nuclide given in the first column, with enhancement factor given by
Eq.~\ref{eq2}.\footnote{The reactions listed are those
that processed more than $10^{-6}$~M$_\odot$ in the reference model (see Table~\ref{tab3})
\textsl{and} with any $\text{max}\left(|D_{i}|\right)>0.05$.}
\label{tab24}}
\begin{ruledtabular}
\begin{tabular}{l@{\hspace{0.1truecm}}|l@{\hspace{0.1truecm}}|l}
Parent nuclide & Nuclei with $|D_{i}|>0.3$ & Nuclei with $0.3>|D_{i}|>0.05$ \\
\hline
 $^{29}$Si & & $^{26}$Al,$^{35}$S,$^{43}$Ca \\
 $^{35}$Cl & & $^{37}$Cl \\
 $^{30}$Si & $^{35}$S  &
$^{23}$Na,$^{25,26}$Mg,$^{26}$Al,$^{29,30}$Si,$^{31,32}$P,$^{33,34}$S,$^{35}$Cl,$^{43}
$Ca,$^{47}$Ti \\
 $^{27}$Al & & $^{23}$Na,$^{24-26}$Mg,$^{26,27}$Al,$
 ^{32}$P,$^{35}$S,$^{43}$Ca,$^{47}$Ti \\
 $^{59}$Cu & & $^{59}$Ni \\
 $^{56}$Co & & $^{56}$Co \\
 $^{33}$S  & & $^{43}$Ca \\
 $^{51}$Mn & & $^{50,51}$Cr \\
 $^{47}$V  & & $^{47}$Ti \\
 $^{26}$Al & $^{26}$Al & \\
 $^{59}$Co & & $^{59}$Co \\
 $^{26}$Mg & & $^{26}$Mg,$^{26}$Al,$^{35}$S,$^{43}$Ca,$^{47}$Ti \\
 $^{25}$Mg & & $^{21}$Ne,$^{25}$Mg,$^{26}$Al,$^{35}$S  \\
 $^{58}$Fe & & $^{58}$Fe \\
 $^{45}$Sc & & $^{45}$Sc \\
 $^{62}$Cu & & $^{63}$Cu,$^{65}$Zn \\
 $^{37}$Cl & & $^{37}$Cl \\
 $^{17}$F  & & $^{14}$N  \\
 $^{64}$Ga & & $^{63}$Cu,$^{65}$Zn \\
\end{tabular}
\end{ruledtabular}
\end{table*}

\begin{table*}
\caption{Sensitivity of the nucleosynthesis to the rate of $\left(\text{p},\text{n}\right)$
reactions with the parent nuclide given in the first column, with enhancement factor given by
Eq.~\ref{eq2}.\footnote{The reactions listed are those
that processed more than $10^{-6}$~M$_\odot$ in the reference model (see Table~\ref{tab3})
\textsl{and} with any $\text{max}\left(|D_{i}|\right)>0.05$.}
\label{tab25}}
\begin{ruledtabular}
\begin{tabular}{l@{\hspace{0.1truecm}}|l@{\hspace{0.1truecm}}||l@{\hspace{0.1truecm}}|l}
Parent nuclide & Nuclei with $|D_{i}|>0.05$ & Parent nuclide & Nuclei with $|D_{i}|>0.05$ \\
\hline
 $^{53}$Mn & $^{53}$Mn &  $^{32}$P  & $^{32}$P \\
 $^{30}$Si & $^{35}$S  & $^{45}$Sc & $^{45}$Sc \\
 $^{58}$Co & $^{58}$Co &  $^{37}$Cl & $^{37}$Cl \\
 $^{54}$Mn & $^{54}$Mn & $^{62}$Cu & $^{63}$Cu,$^{65}$Zn \\
 $^{55}$Mn & $^{55}$Mn &  $^{47}$Ti & $^{46}$Ti \\
\end{tabular}
\end{ruledtabular}
\end{table*}

\begin{table*}
\caption{Sensitivity of the nucleosynthesis to the rate of $\left(\alpha,\gamma\right)$
reactions with the parent nuclide given in the first column, with enhancement factor given by
Eq.~\ref{eq2}.\footnote{The reactions listed are those
that processed more than $10^{-6}$~M$_\odot$ in the reference model (see Table~\ref{tab3})
\textsl{and} with any $\text{max}\left(|D_{i}|\right)>0.05$.}
\label{tab26}}
\begin{ruledtabular}
\begin{tabular}{l@{\hspace{0.1truecm}}|l@{\hspace{0.1truecm}}|l}
Parent nuclide & Nuclei with $|D_{i}|>0.3$ & Nuclei with $0.3>|D_{i}|>0.05$ \\
\hline
 $^{28}$Si & & $^{30}$Si \\
 $^{32}$S  & & $^{37}$Cl \\
 $^{20}$Ne & & $^{20,21}$Ne,$^{23}$Na,$^{24-26}$Mg,$
 ^{26,27}$Al,$^{29,30}$Si,$^{32,33}$P,$^{33,34}$S,$^{35,37}$Cl,$^{37,38}$Ar,$^{39}$K,$^{40-43}$Ca,
$^{45}$Sc,\\
 & & $^{44-47}$Ti \\
 $^{16}$O  & & $^{21}$Ne,$^{23}$Na \\
 $^{24}$Mg & $^{43}$Ca & $^{24}$Mg,$^{27}$Al,$^{35}$S,$^{47}$Ti \\
 $^{58}$Ni & & $^{62}$Ni,$^{63}$Cu,$^{64}$Zn \\
 $^{12}$C  & & $^{39}$K,$^{41,42}$Ca,$^{45}$Sc,$^{44,46}$Ti \\
 $^{33}$S  & & $^{37}$Cl \\
 $^{30}$Si & & $^{30}$Si,$^{32,33}$P,$^{34,35}$S,$^{35}$Cl \\
 $^{41}$Ca & & $^{43}$Ca,$^{47}$Ti \\
 $^{42}$Ca & & $^{46}$Ti \\
 $^{62}$Zn & & $^{66}$Zn \\
\end{tabular}
\end{ruledtabular}
\end{table*}

\begin{table*}
\caption{Sensitivity of the nucleosynthesis to the rate of $\left(\alpha,\text{n}\right)$
reactions with the parent nuclide given in the first column, with enhancement factor given by
Eq.~\ref{eq2}.\footnote{The reactions listed are those
that processed more than $10^{-6}$~M$_\odot$ in the reference model (see Table~\ref{tab3})
\textsl{and} with any $\text{max}\left(|D_{i}|\right)>0.05$.}
\label{tab27}}
\begin{ruledtabular}
\begin{tabular}{l@{\hspace{0.1truecm}}|l@{\hspace{0.1truecm}}|l}
Parent nuclide & Nuclei with $|D_{i}|>0.3$ & Nuclei with $0.3>|D_{i}|>0.05$ \\
\hline
 $^{29}$Si & & $^{25}$Mg,$^{26}$Al,$^{29}$Si,$^{35}$S,$^{43}$Ca \\
 $^{33}$S  & & $^{37}$Cl,$^{43}$Ca \\
 $^{27}$Al & $^{43}$Ca & $^{25}$Mg,$^{29}$Si,$^{32,33}$P,$^{35}$S,$
 ^{47}$Ti \\
 $^{30}$Si & & $^{33}$P,$^{33,35}$S,$^{37}$Cl,$^{43}$Ca,$^{47}$Ti \\
 $^{25}$Mg & & $^{17}$O,$^{21}$Ne,$^{25}$Mg,$^{26}$Al,$^{35}$S  \\
 $^{34}$S  & & $^{37}$Cl,$^{37}$Ar \\
 $^{26}$Mg & & $^{21}$Ne,$^{25,26}$Mg,$^{26}$Al,$^{32}$P,$^{35}$S,$
 ^{43}$Ca,$^{47}$Ti \\
 $^{41}$Ca & & $^{43}$Ca,$^{47}$Ti \\
 $^{21}$Ne & $^{21}$Ne & \\
 $^{17}$O  & & $^{17}$O,$^{21}$Ne \\
\end{tabular}
\end{ruledtabular}
\end{table*}

\begin{table*}
\caption{Sensitivity of the nucleosynthesis to the rate of $\left(\alpha,\text{p}\right)$
reactions with the parent nuclide given in the first column, with enhancement factor given by
Eq.~\ref{eq2}.\footnote{The reactions listed are those
that processed more than $10^{-6}$~M$_\odot$ in the reference model (see Table~\ref{tab3})
\textsl{and} with any $\text{max}\left(|D_{i}|\right)>0.05$.}
\label{tab28}}
\begin{ruledtabular}
\begin{tabular}{l@{\hspace{0.1truecm}}|l@{\hspace{0.1truecm}}||l@{\hspace{0.1truecm}}|l}
Parent nuclide & Nuclei with $|D_{i}|>0.05$ & Parent nuclide & Nuclei with $|D_{i}|>0.05$ \\
\hline
 $^{28}$Si & $^{31,33}$P & $^{44}$Ti & $^{44}$Ti  \\
 $^{30}$P  & $^{43}$Ca &  $^{57}$Ni & $^{61}$Ni,$^{65}$Zn\\
 $^{32}$S  & $^{32,33}$P,$^{35}$S,$^{37}$Cl &  $^{23}$Na &
$^{21}$Ne,$^{23}$Na,$^{26}$Mg,$^{29}$Si,$^{43}$Ca,$^{47}$Ti\\
 $^{27}$Al&$^{25,26}$Mg,$^{26,27}$Al,$^{29,30}$Si,$^{32}$P,$^{33,35}$S,
$^{43}$Ca & $^{41}$Ca & $^{43}$Ca \\
 $^{31}$P  & $^{33}$P,$^{35}$S,$^{35}$Cl & $^{34}$S  & $^{37}$Cl\\
 $^{24}$Mg & $^{24,25}$Mg,$^{26,27}$Al,$^{35}$S,$^{43}$Ca,$^{47}$Ti & $^{39}$Ca & $^{43}$Ca \\
 $^{56}$Ni & $^{43}$Ca,$^{59,60}$Ni,$^{63}$Cu,$^{64}$Zn &  $^{30}$Si & $^{33}$P \\
 $^{39}$K  & $^{43}$Ca &  $^{42}$Ca & $^{46}$Ti \\
 $^{40}$Ca & $^{43}$Ca &  $^{62}$Zn & $^{66}$Zn \\
 $^{29}$Si & $^{32}$P &  $^{61}$Zn & $^{65}$Zn \\
 $^{13}$N  & $^{45}$Sc,$^{44}$Ti &  $^{17}$Ne & $^{14}$N  \\
 $^{35}$Cl & $^{38}$Ar &  $^{43}$Ti & $^{43}$Ca,$^{47}$Ti \\
 $^{20}$Ne & $^{17}$O,$^{21}$Ne,$^{23}$Na,$^{26}$Mg & $^{47}$Cr & $^{47}$Ti \\
 $^{58}$Ni & $^{62}$Ni,$^{63}$Cu,$^{64}$Zn & & \\
\end{tabular}
\end{ruledtabular}
\end{table*}

A general result that applies to all the rates shown in
Tables~\ref{tab23} to \ref{tab28} is that the sensitivities drop (in
absolute value) strongly as compared to the case of fixed enhancement
factor. Very few reactions have $|D_{i}|>0.3$ when we compute the
enhancement factor using Eq.~\ref{eq2}. The list of species sensitive
to the rates of $\left(\text{n},\gamma\right)$ reactions is much
shorter, and the only product species with $|D_{i}|>0.3$ in this
list is $^{21}$Ne. Radiative captures of protons suffer as well from a
reduction of their influence on the supernova yields:
$^{30}\text{Si}+\text{p}\rightleftarrows{}^{31}\text{P}+\gamma$
continues being the most influential reaction, but the number and
importance of the species affected by its rate is much lower than with
a fixed rate enhancement factor. The only species with $|D_{i}|>0.3$
are $^{26}$Al and $^{35}$S. The influence of
$\left(\text{p},\text{n}\right)$ reactions on the supernova yields is
marginal when Eq.~\ref{eq2} is used to determine the enhancement
factor of the rates, and the same applies to
$\left(\alpha,\text{p}\right)$ reactions.  Among the
$\left(\alpha,\gamma\right)$ reactions, the capture on $^{20}$Ne
continues being the reaction with the largest list of product species
with $|D_{i}|>0.05$. The only species with $|D_{i}|>0.3$ due to
variations on the rate of this type of reactions is the trace species
$^{43}$Ca. Finally, the only species with $|D_{i}|>0.3$ with respect
to variations of $\left(\alpha,\text{n}\right)$ reaction rates in
Table~\ref{tab27} are $^{21}$Ne and $^{43}$Ca.

The rate enhancement factor computed with Eq.~\ref{eq2} differs most from the
fixed enhancement factor at temperatures $T\gtrsim3\text{--}5\times
10^9$~K. Thus, it has a stronger effect on 
reactions whose main role is played at high temperatures, as highlighted by
comparing, once more, the list of reactions (parent nuclei) in
Tables~\ref{tab23} to \ref{tab28} with that in Tables~\ref{tab5} to \ref{tab20}. For instance, as
most $\left(\text{n},\gamma\right)$
reactions influence the yields at temperatures of order $2\times
10^9\lesssim T\lesssim4\times 10^9$~K, the
list of reactions in Tables~\ref{tab5} and \ref{tab23} is quite similar. On the other side, the
list of $\left(\text{p},\gamma\right)$ reactions in Table~\ref{tab24} is much shorter than in
Table~\ref{tab6} because the reduction of the enhancement factor at high temperatures affects most
reactions with Fe-group nuclei, while reactions with IMEs are less affected. 

\begin{turnpage}
\begin{table*}
\caption{Rates that influence the yields of each product species, from Carbon to Chlorine, when the
enhancement
factor is given by Eq.~\ref{eq2}.\footnote{Only the species with
mass fraction greater than $10^{-5}$ or that are interesting radioactive isotopes \textsl{and}
with $\text{max}\left(|D_{i}|\right)>0.01$ are shown here. For each nucleus, we show a
maximum of ten reactions.}
\label{tab29}}
\begin{ruledtabular}
\begin{tabular}{l@{\hspace{0.1truecm}}|l}
Nucleus & Reaction and $D_{i}$ \\
\hline
 $^{16}$O  &  $^{20}$Ne($\alpha$,$\gamma$):-0.01; $^{12}$C (n,$\gamma$):+0.01 \\        
                                                                                                   
 $^{20}$Ne & $^{20}$Ne($\alpha$,$\gamma$):-0.15; $^{23}$Na($\alpha$,p):-0.01; $^{16}$O
($\alpha$,$\gamma$):-0.01; $^{24}$Mg($\alpha$,$\gamma$):+0.01; $^{16}$O (n,$\gamma$):+0.01 \\     
                          
 $^{23}$Na & $^{20}$Ne($\alpha$,p):-0.24; $^{23}$Na($\alpha$,p):-0.13;
$^{20}$Ne($\alpha$,$\gamma$):+0.08; $^{30}$Si(p,$\gamma$):+0.07; $^{16}$O
($\alpha$,$\gamma$):+0.06; $^{27}$Al(p,$\gamma$):+0.05; $^{24}$Mg($\alpha$,$\gamma$):+0.04;
$^{26}$Mg(p,$\gamma$):+0.04; \\
 & $^{27}$Al($\alpha$,p):-0.03; $^{12}$C (n,$\gamma$):+0.02 \\        
                                                                             
 $^{24}$Mg &  $^{20}$Ne($\alpha$,$\gamma$):+0.25; $^{24}$Mg($\alpha$,$\gamma$):-0.18;
$^{24}$Mg($\alpha$,p):-0.09; $^{27}$Al(p,$\gamma$):-0.06; $^{30}$Si(p,$\gamma$):-0.05;
$^{27}$Al($\alpha$,p):-0.04; $^{16}$O ($\alpha$,$\gamma$):-0.03; $^{12}$C (n,$\gamma$):-0.03;\\
 & $^{24}$Mg(n,$\gamma$):-0.02; $^{29}$Si($\alpha$,n):-0.02 \\                                     
                                                
 $^{25}$Mg & $^{20}$Ne($\alpha$,$\gamma$):+0.19; $^{25}$Mg(p,$\gamma$):-0.17;
$^{25}$Mg($\alpha$,n):-0.13; $^{30}$Si(p,$\gamma$):+0.12; $^{24}$Mg(n,$\gamma$):+0.10;
$^{29}$Si($\alpha$,n):+0.09; $^{27}$Al($\alpha$,p):-0.09; $^{25}$Mg(n,$\gamma$):-0.09;\\
 & $^{24}$Mg($\alpha$,p):-0.07;$^{27}$Al(p,$\gamma$):+0.05 \\                                     
                                                              
 $^{26}$Mg & $^{26}$Mg(p,$\gamma$):-0.20; $^{26}$Mg($\alpha$,n):-0.20;
$^{23}$Na($\alpha$,p):+0.17; $^{30}$Si(p,$\gamma$):+0.10; $^{20}$Ne($\alpha$,$\gamma$):+0.10;
$^{27}$Al($\alpha$,p):-0.07; $^{27}$Al(p,$\gamma$):+0.07; $^{20}$Ne($\alpha$,p):-0.06;\\ &
$^{24}$Mg($\alpha$,p):-0.03; $^{24}$Mg($\alpha$,$\gamma$):+0.03 \\                            
                                                               
 $^{26}$Al & $^{26}$Al(p,$\gamma$):-0.33; $^{25}$Mg($\alpha$,n):-0.22;
$^{24}$Mg(n,$\gamma$):+0.18; $^{20}$Ne($\alpha$,$\gamma$):+0.18; $^{25}$Mg(p,$\gamma$):+0.17;
$^{30}$Si(p,$\gamma$):+0.14; $^{29}$Si($\alpha$,n):+0.12; $^{27}$Al($\alpha$,p):-0.10;\\ & 
$^{24}$Mg($\alpha$,p):-0.08; $^{26}$Mg(p,$\gamma$):-0.08 \\                                   
                                                                
 $^{27}$Al & $^{27}$Al($\alpha$,p):-0.23; $^{27}$Al(p,$\gamma$):-0.18;
$^{20}$Ne($\alpha$,$\gamma$):+0.17; $^{24}$Mg($\alpha$,$\gamma$):-0.13;
$^{24}$Mg($\alpha$,p):+0.06; $^{26}$Mg($\alpha$,n):-0.03; $^{27}$Al($\alpha$,n):-0.03; $^{16}$O
($\alpha$,$\gamma$):-0.02;\\ & $^{26}$Mg(p,$\gamma$):+0.02; $^{30}$Si($\alpha$,$\gamma$):-0.02 \\ 
                                                                             
 $^{28}$Si & $^{13}$N ($\alpha$,p):-0.03; $^{20}$Ne($\alpha$,$\gamma$):+0.02; $^{12}$C
($\alpha$,$\gamma$):-0.02; $^{23}$Na($\alpha$,p):+0.01 \\                                         
                           
 $^{29}$Si &  $^{27}$Al($\alpha$,n):+0.09; $^{29}$Si($\alpha$,n):-0.08;
$^{20}$Ne($\alpha$,$\gamma$):+0.07; $^{30}$Si(p,$\gamma$):+0.06; $^{27}$Al($\alpha$,p):-0.06;
$^{23}$Na($\alpha$,p):+0.06; $^{30}$Si(p,n):-0.04; $^{26}$Mg($\alpha$,n):+0.04;\\ &
$^{26}$Mg(p,$\gamma$):-0.03; $^{32}$S (n,$\gamma$):-0.03 \\                                   
                                                                       
 $^{30}$Si & $^{30}$Si(p,$\gamma$):-0.14; $^{20}$Ne($\alpha$,$\gamma$):+0.13;
$^{30}$Si($\alpha$,$\gamma$):-0.08; $^{27}$Al($\alpha$,p):+0.06;
$^{28}$Si($\alpha$,$\gamma$):-0.05; $^{31}$P ($\alpha$,p):-0.04; $^{32}$S ($\alpha$,p):-0.03;
$^{12}$C ($\alpha$,$\gamma$):-0.02; \\ & $^{26}$Mg($\alpha$,n):-0.02;
$^{24}$Mg($\alpha$,$\gamma$):-0.02 \\                                                             
           
 $^{31}$P  & $^{30}$Si(p,$\gamma$):+0.07; $^{28}$Si($\alpha$,p):-0.06;
$^{28}$Si($\alpha$,$\gamma$):-0.05; $^{20}$Ne($\alpha$,$\gamma$):-0.04;
$^{30}$Si($\alpha$,$\gamma$):-0.04; $^{31}$P ($\alpha$,p):-0.03;
$^{24}$Mg($\alpha$,$\gamma$):+0.03; $^{27}$Al(p,$\gamma$):+0.02; \\ & $^{27}$Al($\alpha$,p):+0.02;
$^{32}$S ($\alpha$,p):-0.02 \\                                                                    
           
 $^{32}$S  &  $^{13}$N ($\alpha$,p):-0.02; $^{20}$Ne($\alpha$,$\gamma$):-0.01 \\        
                        
 $^{33}$S  &  $^{20}$Ne($\alpha$,$\gamma$):-0.11; $^{32}$S (n,$\gamma$):+0.09;
$^{30}$Si($\alpha$,n):-0.08; $^{27}$Al($\alpha$,p):+0.06; $^{30}$Si(p,$\gamma$):-0.06;
$^{28}$Si($\alpha$,$\gamma$):+0.05; $^{24}$Mg(n,$\gamma$):+0.04; $^{23}$Na($\alpha$,p):-0.03;\\ &
$^{27}$Al($\alpha$,n):-0.03; $^{28}$Si(n,$\gamma$):+0.03 \\                                   
                                                         
 $^{34}$S  & $^{20}$Ne($\alpha$,$\gamma$):-0.12; $^{30}$Si(p,$\gamma$):+0.05;
$^{30}$Si($\alpha$,$\gamma$):+0.05; $^{28}$Si($\alpha$,$\gamma$):+0.04; $^{31}$P
($\alpha$,p):+0.04; $^{32}$S ($\alpha$,p):+0.03; $^{27}$Al(p,$\gamma$):+0.03;
$^{24}$Mg($\alpha$,p):-0.03;\\ & $^{29}$Si($\alpha$,n):+0.03; $^{24}$Mg($\alpha$,$\gamma$):+0.02\\ 
                                                                            
 $^{35}$Cl & $^{20}$Ne($\alpha$,$\gamma$):-0.15; $^{30}$Si($\alpha$,$\gamma$):+0.11;
$^{31}$P ($\alpha$,p):+0.06; $^{30}$Si(p,$\gamma$):+0.05; $^{27}$Al(p,$\gamma$):+0.04;
$^{29}$Si($\alpha$,n):+0.04; $^{24}$Mg($\alpha$,$\gamma$):+0.02;
$^{28}$Si($\alpha$,$\gamma$):+0.02; $^{23}$Na($\alpha$,p):-0.02;\\ & $^{24}$Mg($\alpha$,p):-0.02 \\
\end{tabular}
\end{ruledtabular}
\end{table*}
\end{turnpage}
                                                                              
\begin{turnpage}
\begin{table*}
\caption{Rates that influence the yields of each product species, from Argon to Nickel, when the
enhancement factor is
given by Eq.~\ref{eq2} (continuation of Table~\ref{tab29}).\footnote{Only the species with
mass fraction greater than $10^{-5}$ or that are interesting radioactive isotopes \textsl{and}
with $\text{max}\left(|D_{i}|\right)>0.01$ are shown here. For each nucleus, we show a
maximum of ten reactions.}
\label{tab30}}
\begin{ruledtabular}
\begin{tabular}{l@{\hspace{0.1truecm}}|l}
Nucleus & Reaction and $D_{i}$ \\
\hline
 $^{36}$Ar & $^{20}$Ne($\alpha$,$\gamma$):-0.04; $^{12}$C ($\alpha$,$\gamma$):+0.02;
$^{13}$N ($\alpha$,p):+0.01 \\                                                                    
                            
 $^{37}$Ar & $^{20}$Ne($\alpha$,$\gamma$):-0.09; $^{34}$S ($\alpha$,n):-0.08;
$^{36}$Ar(n,$\gamma$):+0.06; $^{12}$C ($\alpha$,$\gamma$):+0.04; $^{13}$N ($\alpha$,p):-0.03;
$^{35}$Cl(p,$\gamma$):+0.03; $^{30}$Si($\alpha$,$\gamma$):+0.03; $^{33}$S
($\alpha$,$\gamma$):+0.03;\\ & $^{27}$Al($\alpha$,p):+0.02; $^{37}$Cl(p,n):-0.02 \\                
                                                                     
 $^{38}$Ar & $^{20}$Ne($\alpha$,$\gamma$):-0.10; $^{35}$Cl($\alpha$,p):+0.05; $^{34}$S
($\alpha$,$\gamma$):+0.05; $^{30}$Si(p,$\gamma$):+0.04; $^{30}$Si($\alpha$,$\gamma$):+0.04;
$^{24}$Mg($\alpha$,$\gamma$):+0.03; $^{28}$Si($\alpha$,$\gamma$):+0.03;
$^{27}$Al(p,$\gamma$):+0.02;\\ & $^{13}$N ($\alpha$,p):-0.02; $^{29}$Si($\alpha$,n):+0.02 \\       
                                                                
 $^{39}$K  & $^{20}$Ne($\alpha$,$\gamma$):-0.11; $^{12}$C ($\alpha$,$\gamma$):+0.08;
$^{35}$Cl($\alpha$,p):+0.04; $^{30}$Si($\alpha$,$\gamma$):+0.03; $^{13}$N ($\alpha$,p):-0.03;
$^{39}$K (p,$\gamma$):-0.03; $^{24}$Mg($\alpha$,$\gamma$):+0.03;
$^{42}$Ca($\alpha$,$\gamma$):-0.03;\\ & $^{34}$S ($\alpha$,$\gamma$):+0.02;
$^{30}$Si(p,$\gamma$):+0.02 \\                                                                    
                                                                                                 
 $^{40}$Ca & $^{20}$Ne($\alpha$,$\gamma$):-0.07; $^{13}$N ($\alpha$,p):+0.04; $^{12}$C
($\alpha$,$\gamma$):+0.04; $^{23}$Na($\alpha$,p):-0.02; $^{44}$Ti($\alpha$,p):-0.01 \\            
                          
 $^{44}$Ti & $^{20}$Ne($\alpha$,$\gamma$):-0.07; $^{44}$Ti($\alpha$,p):-0.07; $^{13}$N
($\alpha$,p):+0.07; $^{12}$C ($\alpha$,$\gamma$):+0.05; $^{23}$Na($\alpha$,p):-0.02;
$^{40}$Ca($\alpha$,$\gamma$):-0.02 \\                                                             
                                         
 $^{48}$V  & $^{48}$Cr($\alpha$,p):-0.04; $^{13}$N ($\alpha$,p):+0.04;
$^{20}$Ne($\alpha$,$\gamma$):-0.04; $^{44}$Ti($\alpha$,p):+0.03; $^{12}$C
($\alpha$,$\gamma$):+0.02; $^{23}$Na($\alpha$,p):-0.02 \\                                         
                                                                    
 $^{49}$V  & $^{49}$V (p,n):-0.04; $^{20}$Ne($\alpha$,$\gamma$):-0.03;
$^{48}$Cr($\alpha$,p):-0.03; $^{13}$N ($\alpha$,p):+0.02; $^{44}$Ti($\alpha$,p):+0.02;
$^{53}$Mn(p,n):-0.02; $^{12}$C ($\alpha$,$\gamma$):+0.02 \\                                       
                                                       
 $^{50}$Cr &  $^{51}$Mn(p,$\gamma$):-0.05; $^{55}$Co(p,$\gamma$):+0.03;
$^{20}$Ne($\alpha$,$\gamma$):+0.02; $^{13}$N ($\alpha$,p):-0.02; $^{23}$Na($\alpha$,p):+0.01 \\   
                                          
 $^{51}$Cr &  $^{51}$Mn(p,$\gamma$):-0.09; $^{20}$Ne($\alpha$,$\gamma$):-0.02;
$^{55}$Co(p,$\gamma$):-0.01; $^{50}$Cr(p,$\gamma$):+0.01; $^{52}$Fe($\alpha$,p):-0.01;
$^{44}$Ti($\alpha$,p):+0.01 \\                                                                    
                                                
 $^{52}$Mn & $^{13}$N ($\alpha$,p):+0.04; $^{20}$Ne($\alpha$,$\gamma$):-0.03; $^{12}$C
($\alpha$,$\gamma$):+0.02; $^{23}$Na($\alpha$,p):-0.02; $^{52}$Fe($\alpha$,p):-0.01;
$^{44}$Ti($\alpha$,p):+0.01 \\                                                                    
                                         
 $^{53}$Mn & $^{53}$Mn(p,n):-0.05; $^{20}$Ne($\alpha$,$\gamma$):-0.02; $^{13}$N
($\alpha$,p):+0.02; $^{12}$C ($\alpha$,$\gamma$):+0.01 \\                                         
                                 
 $^{55}$Fe &  $^{55}$Co(p,$\gamma$):-0.05; $^{13}$N ($\alpha$,p):+0.01;
$^{20}$Ne($\alpha$,$\gamma$):-0.01 \\                                                             
                                          
 $^{57}$Co &  $^{57}$Co(p,n):-0.03 \\                                                   
                        
 $^{56}$Ni & $^{13}$N ($\alpha$,p):+0.01 \\                                            
                        
 $^{59}$Ni &  $^{59}$Cu(p,$\gamma$):-0.09; $^{56}$Ni($\alpha$,p):+0.08 \\               
                         
 $^{60}$Ni & $^{56}$Ni($\alpha$,p):+0.07; $^{59}$Cu(p,$\gamma$):+0.02 \\               
                        
 $^{61}$Ni & $^{57}$Ni($\alpha$,p):+0.09; $^{57}$Ni($\alpha$,$\gamma$):+0.04;
$^{57}$Co(p,n):-0.03 \\                                                                           
                                   
 $^{62}$Ni &  $^{58}$Ni($\alpha$,$\gamma$):+0.07; $^{58}$Ni($\alpha$,p):+0.06;
$^{60}$Zn($\alpha$,p):+0.02; $^{57}$Co(p,n):+0.01 \\                                              
\end{tabular}
\end{ruledtabular}
\end{table*}
\end{turnpage}

Tables~\ref{tab29} and \ref{tab30} are similar to Tables~\ref{tab21} and \ref{tab22}, except that
the enhancement factor $f$ is now a function of temperature, given by Eq.~\ref{eq2}. The
maximum $|D_{i}|$ achieved with Eq.~\ref{eq2} for a given species is in general a factor of two
smaller than when using a fixed rate enhancement factor, while all the species with a production
factor larger than 100 have maximum $|D_{i}|<0.1$.

\subsection{Sensitivity to different rate prescriptions}\label{prescriptions}

In this Section, we analyze the changes in the yields obtained using different prescriptions
for the rates of a few selected reactions. To this end, we have accessed the JINA REACLIB
Database to compare the most recent rates
for
each one of the selected reactions. 
We discuss in the following the prescriptions for the reactions that appear in Tables~\ref{tab5} to
\ref{tab20} with maximum $|D_{i}|>0.3$. The results are presented in Table~\ref{tab31} in the
form of percent variations of the yield of product species when two different prescriptions are
used for each reaction rate. We give in the table as well the sources of the rates of each
reaction.
The reference rate (i.e. that used in the denominator of the calculation of the relative
variation of the yield) is always that cited in second place in the table. All the references that
appear in
this Section are taken from the JINA webpage.

\subsubsection{$\left(\text{n},\gamma\right)$ reactions}

The rates of the reactions we consider are fits to either theoretical or experimental results
published in \cite{cyb10, ka02, rath, baka}.

The discrepancy between the different rates of the reactions
$^{32}\text{S}+\text{n}\rightleftarrows{}^{33}\text{S}+\gamma$,
$^{28}\text{Si}+\text{n}\rightleftarrows{}^{29}\text{Si}+\gamma$,
$^{24}\text{Mg}+\text{n}\rightleftarrows{}^{25}\text{Mg}+\gamma$,
$^{25}\text{Mg}+\text{n}\rightleftarrows{}^{26}\text{Mg}+\gamma$, and
$^{20}\text{Ne}+\text{n}\rightleftarrows{}^{21}\text{Ne}+\gamma$ computed from the above references
is
less than a factor of ten for $T\gtrsim10^9$~K. As this uncertainty is within the range
explored in Section~\ref{fixed}, we do not deem it necessary to further analyze these reaction
rates.

On the other hand, the rate of the reaction
\mbox{$^{16}\text{O}+n\rightleftarrows{}^{17}\text{O}+\gamma$}
computed from the two references in JINA, \cite{ka02} and \cite{baka},
shows a discrepancy of more than two orders of
magnitude between these two cases. We have computed the nucleosynthesis of our SNIa model with
both rates and compared the results in the first row of
Table~\ref{tab31}. Aside from the trace product $^{17}$O, whose yield
decreases by two orders of magnitude when using the rate from
\cite{baka}, the effect on each abundance is smaller than 27\%. We
conclude that $(n,\gamma)$ reaction rates, in
general, are not critical for obtaining accurate yields from SNIa
models.

\begin{table*}
\caption{Sensitivity of the nucleosynthesis to different rate prescriptions for the 
reactions in the first column.
\label{tab31}}
\begin{ruledtabular}
\begin{tabular}{l@{\hspace{0.1truecm}}|l}
Reaction & Product nuclei and percent variation\footnote{We only show species for which the percent
variation is largest than 10\% in absolute value and whose yield is $\gtrsim10^{-8}$~M$_\odot$.} \\
\hline
$^{16}\text{O}+\text{n}\rightleftarrows{}^{17}\text{O}+\gamma$ \footnote{Rate from \cite{baka} vs.
 rate from \cite{ka02}} & $^{17}$O:-99\%; $^{14}$N:-27\%; $^{25}$Mg:-25\%; $^{22}$Ne:+22\%; 
 $^{43}$Ca:+18\%; $^{21}$Ne:+15\%; $^{35}$S:+14\%; $^{26}$Al:-11\% \\
$^{17}\text{F}+\text{p}\rightleftarrows{}^{18}\text{Ne}+\gamma$ \footnote{The reference rate from
 \cite{ba00} is compared to a rate that incorporates several contributions from M. Wiescher, as
 given in the JINA Database} & $^{14}$N:+79\% \\
$^{30}\text{Si}+\text{p}\rightleftarrows{}^{31}\text{P}+\gamma$ \footnote{Rate from \cite{cyb10}
vs.
 rate from \cite{il01}} & $^{35}$S:+100\%; $^{32}$P:+36\%; $^{43}$Ca:+33\%; $^{30}$Si:-30\%;
 $^{26}$Al:+29\%; $^{25}$Mg:+28\%; $^{29}$Si:+23\%; $^{26}$Mg:+22\%; \\
 & $^{31}$P:+18\%; $^{33}$S:-17\%; $^{34}$S:+16\%; $^{35}$Cl:+13\%; $^{23}$Na:+12\%;
 $^{38}$Ar:+10\% \\
$^{30}\text{Si}+\text{p}\rightleftarrows{}^{31}\text{P}+\gamma$ \footnote{Rate from \cite{rath} vs.
 rate from \cite{il01}} & $^{35}$S:+68\%; $^{32}$P:+26\%; $^{43}$Ca:+24\%; $^{30}$Si:-22\%;
 $^{26}$Al:+21\%; $^{25}$Mg:+20\%; $^{29}$Si:+16\%; $^{26}$Mg:+16\%; \\
 & $^{33}$S:-14\%; $^{31}$P:+13\%; $^{34}$S:+12\% \\
$^{20}\text{Ne}+\alpha\rightleftarrows{}^{24}\text{Mg}+\gamma$ \footnote{Rate from \cite{cyb10} vs.
 rate from \cite{il10}} & $^{24}$Mg:+156\%; $^{27}$Al:+90\%; $^{40}$K:-80\%; $^{43}$Ca:-75\%;
$^{48}$Ti:-74\%; $^{44}$Ca:-73\%; $^{26}$Al:+72\%; $^{30}$Si:+71\%; \\
 & $^{35}$S:-57\%; $^{36}$S:+53\%; $^{35}$Cl:-53\%; $^{25}$Mg:+52\%; $^{33}$P:+49\%;
$^{42}$Ca:-49\%; $^{38}$Ar:-43\%; $^{47}$Ti:-41\%; \\
 & $^{45}$Sc:-39\%; $^{20}$Ne:-38\%; $^{39}$K:-38\%; $^{37}$Ar:-37\%; $^{41}$Ca:-37\%;
$^{46}$Ti:-36\%; $^{32}$P:+34\%; $^{34}$S:-34\%; \\
 & $^{33}$S:-30\%; $^{29}$Si:+29\%; $^{37}$Cl:-29\%; $^{44}$Ti:-27\%; $^{23}$Na:-21\%;
$^{14}$N:-16\%; $^{21}$Ne:-16\%; $^{40}$Ca:-16\%; \\
 & $^{26}$Mg:+11\%; $^{28}$Si:+10\%; $^{31}$P:-10\% \\
$^{20}\text{Ne}+\alpha\rightleftarrows{}^{24}\text{Mg}+\gamma$ \footnote{Rate from \cite{nacr} vs.
 rate from \cite{il10}} & $^{24}$Mg:+84\%; $^{40}$K:-58\%; $^{43}$Ca:-56\%; $^{27}$Al:+52\%;
$^{48}$Ti:-51\%; $^{30}$Si:+48\%; $^{44}$Ca:-48\%; $^{35}$S:-37\%; \\
 & $^{42}$Ca:-37\%; $^{45}$Sc:-37\%; $^{33}$P:+36\%; $^{36}$S:+35\%; $^{35}$Cl:-35\%;
$^{47}$Ti:-32\%; $^{44}$Ti:-30\%; $^{26}$Al:+29\%; \\
 & $^{32}$P:+29\%; $^{38}$Ar:-29\%; $^{41}$Ca:-28\%; $^{39}$K:-26\%; $^{46}$Ti:-26\%;
$^{37}$Ar:-24\%; $^{25}$Mg:+21\%; $^{29}$Si:+21\%; \\
 & $^{34}$S:-20\%; $^{40}$Ca:-19\%; $^{20}$Ne:-16\%; $^{33}$S:-16\%; $^{14}$N:-15\%;
$^{37}$Cl:-14\%; $^{28}$Si:+12\%; $^{48}$V:-11\%; \\
$^{24}\text{Mg}+\alpha\rightleftarrows{}^{27}\text{Al}+\text{p}$ \footnote{Rate from \cite{laur}
vs.
 rate from \cite{il01}} & $^{35}$S:+52\%; $^{24}$Mg:+44\%; $^{26}$Al:+35\%; $^{25}$Mg:+26\%;
 $^{27}$Al:-19\%; $^{37}$Cl:+16\%; $^{43}$Ca:+15\%; $^{34}$S:+13\%; \\
 & $^{33}$P:+12\%; $^{26}$Mg:+12\%; $^{35}$Cl:+12\%; $^{32}$P:+10\% \\
\end{tabular}
\end{ruledtabular}
\end{table*}

\subsubsection{$\left(\text{p},\gamma\right)$ reactions}\label{laspg}

The rates of the reactions we consider are fits to either theoretical
or experimental results published in \cite{cyb10, rath, il01, laur,
  ba00}.

The discrepancy between the different rates of the reactions
$^{25}\text{Mg}+\text{p}\rightleftarrows{}^{26}\text{Al}+\gamma$,
$^{26}\text{Mg}+\text{p}\rightleftarrows{}^{27}\text{Al}+\gamma$,
$^{26}\text{Al}+\text{p}\rightleftarrows{}^{27}\text{Si}+\gamma$,
$^{27}\text{Al}+\text{p}\rightleftarrows{}^{28}\text{Si}+\gamma$, and
$^{30}\text{Si}+\text{p}\rightleftarrows{}^{31}\text{P}+\gamma$
computed from the above references is 
less than a factor of ten for $T\gtrsim 10^9$~K, well within the range
explored in Section~\ref{fixed}. 
The reaction
$^{45}\text{Sc}+\text{p}\rightleftarrows{}^{46}\text{Ti}+\gamma$, which
contributes to the linking of QSE groups in silicon burning, is only
evaluated in JINA through three somewhat different theoretical
models. In the temperature range of interest, the rates given by these
models match perfectly.  

The rate of the reaction
$^{17}\text{F}+\text{p}\rightleftarrows{}^{18}\text{Ne}+\gamma$ is
given in JINA for two different fits to experimental rates, which
differ by more than one order of magnitude for temperatures in the
range $10^9\lesssim T\lesssim10^{10}$~K. We have recomputed the
nucleosynthesis with both evaluations of the rate of this reaction and
show the results in the second row of Table~\ref{tab31}. Only the
yield of $^{14}$N, a marginal product of the supernova
nucleosynthesis, changes by more than 10\%.

The reaction
$^{30}\text{Si}+\text{p}\rightleftarrows{}^{31}\text{P}+\gamma$ has
the largest $D_{i}$, consequently we have recomputed the
nucleosynthesis using the three most recent evaluations of its rate
from JINA. The results are shown in the third and fourth rows of
Table~\ref{tab31}, in which we have taken as a reference the
recommended rate from \cite{il01}, which is compared to two other
evaluations due to \cite{cyb10} and \cite{rath}. In the temperature
range of interest, the rates computed from these sources differ by
less than a factor of three, and the same applies to the rate computed
from the older reference \cite{cf88}. The species whose yields are most
sensitive to the different prescriptions for this rate are more or
less the same as already noted in Tables~\ref{tab6} and
\ref{tab24}. However, the changes in the yields are more consistent
with those shown in Table~\ref{tab24}, indicating that the use of the
enhancement factor function of temperature, as in Eq.~\ref{eq2}, might
describe better the actual uncertainties in the nucleosynthesis than
using a fixed enhancement factor, 
at least at the current level of knowledge of this reaction rate (but see Sections~\ref{windows}
and \ref{conc}).

\subsubsection{$\left(\text{p},\text{n}\right)$ reactions}

Only two $\left(\text{p},\text{n}\right)$ reactions in
Table~\ref{tab7} have any $|D_{i}|>0.3$, these are
\mbox{$^{32}\text{P}+\text{p}\rightleftarrows{}^{32}\text{S}+\text{n}$}
and
\mbox{$^{37}\text{Cl}+\text{p}\rightleftarrows{}^{37}\text{Ar}+\text{n}$}. In
the JINA library there are only theoretical rates of these reactions,
all of them obtained from the NON-SMOKER code, using different nuclear
inputs. In the range of temperatures of interest, the different rates
for these reactions match each other perfectly.

\subsubsection{$\left(\alpha,\gamma\right)$ reactions}\label{lasag}

The rates of the reactions
$^{24}\text{Mg}+\alpha\rightleftarrows{}^{28}\text{Si}+\gamma$ and
$^{30}\text{Si}+\alpha\rightleftarrows{}^{34}\text{S}+\gamma$ are
derived from different evaluations obtained with the NON-SMOKER code
with different nuclear inputs, and from \cite{cf88}. All these rates,
for a given reaction, agree within a factor smaller than the
enhancement factor we have explored earlier in this paper, hence we do not
continue with the analysis of these two reactions.

The different prescriptions for the rate of the reaction
$^{20}\text{Ne}+\alpha\rightleftarrows{}^{24}\text{Mg}+\gamma$ are
from \cite{il10, cyb10, rath, nacr, cf88}. Within the temperature range of
interest, these rates show discrepancies of nearly an order of
magnitude. Thus, as this reaction is one of the most influential for
SNIa nucleosynthesis, we have recomputed the yields for the three most
recent prescriptions of its rate. The results are shown in the fifth
and sixth rows in Table~\ref{tab31}. The list of species whose yields are most sensitive to
the prescription used for this reaction is quite similar to that found
in Table~\ref{tab8}.
The first point to note is the long list of species whose yield varies more than 10\% when using
either of the rates from \cite{il10}, \cite{cyb10}, or \cite{nacr}. The species most sensitive to
the changes in the $^{20}\text{Ne}+\alpha\rightleftarrows{}^{24}\text{Mg}+\gamma$ rates is
$^{24}$Mg, whose yield changes by 156\% when using the rate from \cite{cyb10} instead that from
\cite{il10}, and by 84\% when using the rate from \cite{nacr}. Several other species, like
$^{27}$Al and $^{30}$Si, experience changes in the range 70-90\%. The yields obtained using the
theoretical rate in \cite{rath} (not shown in Table~\ref{tab31}) agree quite well
with those obtained using the experimental rate in \cite{nacr}. On the contrary,
using the theoretical rates in \cite{cyb10}, obtained with the same
code as in \cite{rath} but with different nuclear inputs, gives yields
that differ from those belonging to the rates from \cite{nacr} by as
much as 42\%. 

The three most recent evaluations of the rate of the reaction
$^{12}\text{C}+\alpha\rightleftarrows{}^{16}\text{O}+\gamma$ in JINA are
from \cite{nacr}, \cite{bu96}, and \cite{kun02}. We have recomputed
the supernova nucleosynthesis for these three prescriptions of the
rate. Not a single species experiences a change of abundance larger
than 10\%.

The reaction
$^{58}\text{Ni}+\alpha\rightleftarrows{}^{62}\text{Zn}+\gamma$ is
important for the alpha-rich freeze-out of NSE, which affects a
large portion of SNIa ejecta. Hence, we have recomputed the
nucleosynthesis with the three most recent rates given in JINA, from
\cite{cyb10} and from \cite{rath} using different nuclear inputs. Not
a single species experiences a change of abundance larger than 10\%.

\subsubsection{$\left(\alpha,\text{n}\right)$ reactions}

The different prescriptions for the reactions that have any
$|D_{i}|>0.3$ in Table~\ref{tab9} give rates that agree with each
other within the factor of ten explored in this paper. These reactions
are: $^{17}\text{O}+\alpha\rightleftarrows{}^{20}\text{Ne}+\text{n}$,
$^{21}\text{Ne}+\alpha\rightleftarrows{}^{24}\text{Mg}+\text{n}$,
$^{25}\text{Mg}+\alpha\rightleftarrows{}^{28}\text{Si}+\text{n}$,
$^{26}\text{Mg}+\alpha\rightleftarrows{}^{29}\text{Si}+\text{n}$,
$^{27}\text{Al}+\alpha\rightleftarrows{}^{30}\text{P}+\text{n}$,
$^{29}\text{Si}+\alpha\rightleftarrows{}^{32}\text{S}+\text{n}$,
$^{30}\text{Si}+\alpha\rightleftarrows{}^{33}\text{S}+\text{n}$, and
$^{34}\text{S}+\alpha\rightleftarrows{}^{37}\text{Ar}+\text{n}$. The
rates are from a variety of sources, most of them from calculations
with the NON-SMOKER code using different nuclear inputs, but there
are as well rates based on experimental measurements by \cite{cf88,
  sm93, nacr}.

\subsubsection{$\left(\alpha,\text{p}\right)$ reactions}\label{lasap}

The different prescriptions for the rates of the reactions
$^{17}\text{Ne}+\alpha\rightleftarrows{}^{20}\text{Na}+\text{p}$,
$^{20}\text{Ne}+\alpha\rightleftarrows{}^{23}\text{Na}+\text{p}$,
$^{23}\text{Na}+\alpha\rightleftarrows{}^{26}\text{Mg}+\text{p}$,
$^{27}\text{Al}+\alpha\rightleftarrows{}^{30}\text{Si}+\text{p}$,
$^{34}\text{S}+\alpha\rightleftarrows{}^{37}\text{Cl}+\text{p}$, and
$^{47}\text{Cr}+\alpha\rightleftarrows{}^{50}\text{Mn}+\text{p}$ all
agree within a factor of ten for the temperature range of interest to
us. These reaction rates come from several calculations with the
NON-SMOKER code using different nuclear inputs, as well as
experimental measurements in \cite{sm93, ha04, nacr}.
The different rates of the reactions
$^{58}\text{Ni}+\alpha\rightleftarrows{}^{61}\text{Cu}+\text{p}$ and
$^{56}\text{Ni}+\alpha\rightleftarrows{}^{59}\text{Cu}+\text{p}$, both
of importance for the alpha-rich freeze-out from NSE, agree quite well
within the temperature range of interest.  Thus, we do not continue
with the analysis of these reaction rates.

The three most recent evaluations of the rate of the reaction
$^{24}\text{Mg}+\alpha\rightleftarrows{}^{27}\text{Al}+\text{p}$ in the JINA database
are from \cite{cyb10, il01, laur}. Within the temperature range of
interest, the rate from the last reference differs from the other
rates by as much as three orders of magnitude. Hence, we have
recomputed the nucleosynthesis for the three different prescriptions
of this rate. There is no significant difference in the chemical
composition obtained with the rates of \cite{cyb10} and \cite{il01},
i.e. not a single species experiences a change of abundance larger
than 10\%, which is a consequence of the match between the
rates from these two references. However, when comparing the yields
obtained with the rate from \cite{laur} with those obtained with the
rate from \cite{il01}, many important variations in the
yield of species with a significant abundance show up.

The three most recent evaluations of the rate of the reaction
$^{28}\text{Si}+\alpha\rightleftarrows{}^{31}\text{P}+\text{p}$ in JINA
are from the same references as the reaction on $^{24}$Mg. Although
the discrepancies of the rates are not as large as for Mg, they
reach one order of magnitude. Hence, we have recomputed as well the
supernova nucleosynthesis for the three prescriptions of the rate of
the reaction on $^{28}$Si. However, not a single species experiences a
change of abundance larger than 10\%.

The reaction
$^{44}\text{Ti}+\alpha\rightleftarrows{}^{47}\text{V}+\text{p}$ is
important for bridging the gap between QSE groups in silicon
burning. Hence, we have recomputed the nucleosynthesis with the three
most recent rates given in JINA, from \cite{cyb10} and from
\cite{rath} using different nuclear inputs. Even though these rates
differ up to near an order of magnitude in the range of temperatures
of interest, not a single species experiences a change of abundance
larger than 10\%.

\subsubsection{Summary of the sensitivity of the yields to different rate
prescriptions}\label{subsummary}

The most notable reactions in Table~\ref{tab31} are
$^{30}\text{Si}+\text{p}\rightleftarrows{}^{31}\text{P}+\gamma$,
$^{20}\text{Ne}+\alpha\rightleftarrows{}^{24}\text{Mg}+\gamma$, and
$^{24}\text{Mg}+\alpha\rightleftarrows{}^{27}\text{Al}+\text{p}$. Within
the set of most influential reactions and most influenced species,
there are two important details. First, very few
reactions appear whose parent nuclei belong to the Fe-group or that
are important for the bridging of the QSE groups in silicon
burning. Second, in Table~\ref{tab31} there are no product species
belonging to the Fe-group whose yield depends significantly on the
explored reaction rates.  This is most remarkable because the elements
of the Fe-group constitute the main nucleosynthetic products of SNIa.

\subsection{Sensitivity to different temperature ranges}\label{windows}

\begin{figure}[!htb]
 \includegraphics[width=8truecm]{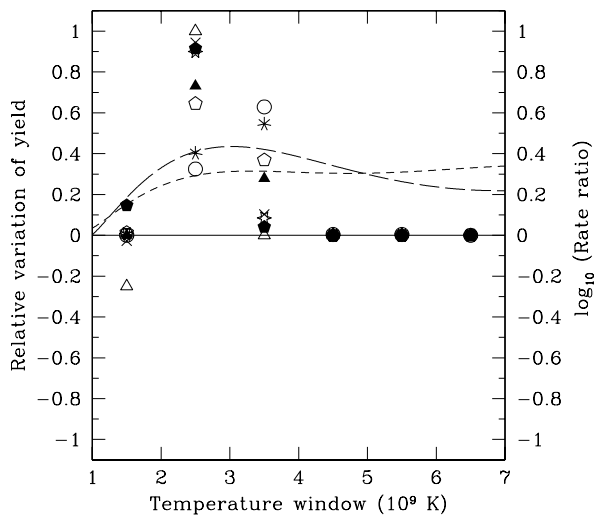}
\caption{Sensitivity of the yield of selected species to variations by a factor of $\times10$ in
the
rate of the reactions $^{30}\text{Si}+\text{p}\rightleftarrows{}^{31}\text{P}+\gamma$ in different
temperature ranges of size $10^9$~K. The points give the difference between the yield
of a species for an enhanced rate in a temperature window and its yield in our reference
model, normalized by the difference between the yield for an enhanced rate at all temperatures and
the yield of our reference model, see Eq.~\ref{la7}. The points are centered on each temperature
window, and each symbol represents a product nucleus as follows: open triangles stand for
$^{20}$Ne, solid triangles for
$^{24}$Mg, crosses for $^{26}$Al, open pentagons for $^{30}$Si, solid pentagons for $^{32}$P, stars
for $^{35}$S, asterisks with seven vertices for $^{38}$Ar, and open circles for $^{47}$Ti. The
dashed lines (scaled according to the right axis) give the logarithm of the ratio of the different
rates of the reaction $^{30}\text{Si}+\text{p}\rightarrow^{31}\text{P}+\gamma$ in JINA. The
short-dashed line belongs to the ratio of the rate from \cite{cyb10} to that from \cite{il01},
while the long-dashed line belongs to the ratio of the rate from \cite{rath} to that from
\cite{il01}. The horizontal solid line marks the zero of axes, i.e. no variation of the yield and
rate ratio equal to one.
\label{fig13}}
\end{figure}

\begin{figure}[!htb]
 \includegraphics[width=8truecm]{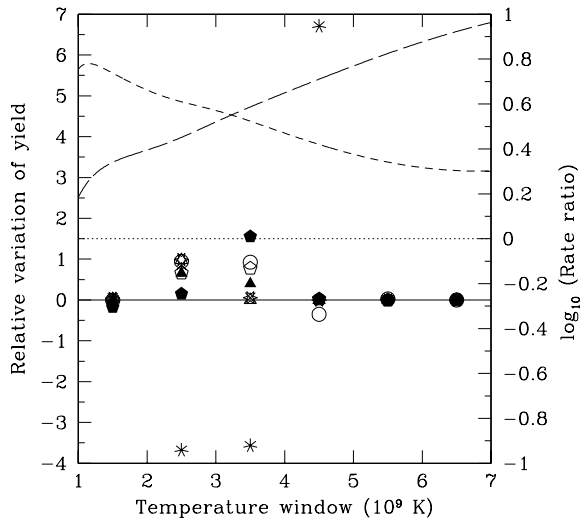}
\caption{Sensitivity of the yield of selected species to variations by a factor of $\times10$ in
the
rate of the reactions $^{20}\text{Ne}+\alpha\rightleftarrows{}^{24}\text{Mg}+\gamma$ in different
temperature windows of size $10^9$~K. The meaning of the points is the same as in Fig.~\ref{fig13}.
The dashed lines (scaled according to the right axis) give the logarithm of the ratio of the
different rates of the reaction $^{20}\text{Ne}+\alpha\rightarrow^{24}\text{Mg}+\gamma$ in JINA.
The short-dashed line belongs to the ratio of the rate from \cite{cyb10} to that from \cite{il10},
while the long-dashed line belongs to the ratio of the rate from \cite{nacr} to that from
\cite{il10}. The two horizontal lines mark the zero of the left axis, i.e. no variation of the
yield (solid line), and the zero of the right axis, i.e. rate ratio equal to one (dotted line).
\label{fig14}}
\end{figure}

\begin{figure}[!htb]
 \includegraphics[width=8truecm]{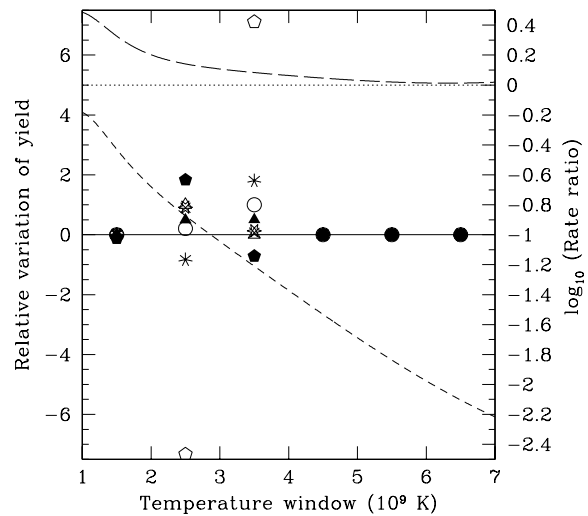}
\caption{Sensitivity of the yield of selected species to variations by a factor of $\times10$ in
the
rate of the reactions $^{24}\text{Mg}+\alpha\rightleftarrows{}^{27}\text{Al}+\text{p}$ in different
temperature windows of size $10^9$~K. The meaning of the points is the same as in Fig.~\ref{fig13}.
The dashed lines (scaled according to the right axis) give the logarithm of the ratio of the
different rates of the reaction $^{24}\text{Mg}+\alpha\rightarrow^{27}\text{Al}+\text{p}$ in JINA.
The long-dashed line belongs to the ratio of the rate from \cite{cyb10} to that from \cite{il01},
while the short-dashed line belongs to the ratio of the rate from \cite{laur} to that from
\cite{il01}. The two horizontal lines mark the zero of the left axis, i.e. no variation of the
yield (solid line), and the zero of the right axis, i.e. rate ratio equal to one (dotted line).
\label{fig15}}
\end{figure}

As a final step in our present study, we now perform an
analysis of the temperature dependence of the sensitivity of the
yields to the reaction rates. We analyze here the three most notable
rates found in Section~\ref{subsummary}. To this end, we consider
again a fixed enhancement factor of the rates, a factor of ten, but this
time we limit it to a temperature window $10^9$~K wide. We explore
windows centered on temperatures 1.5, 2.5, 3.5, 4.5, 5.5, and
$6.5\times10^9$~K. We show the results in Figs.~\ref{fig13} to
\ref{fig15}, where we plot for selected species the relative variation
of their yield, defined as
\begin{equation}
 r_{ij}=\frac{m_{ij}-m_i}{m'_i-m_i}\,,
\label{la7}
\end{equation}
where $m_i$ is the mass of nucleus $i$ ejected in our reference model
(Table~\ref{tab2}), $m'_i$ is the yield of the same species when the
rate of the reaction being analyzed is multiplied by a factor of ten
independent of temperature, and $m_{ij}$ is the yield when
the rate of the reaction is multiplied by the same factor only in the
window $j\times10^9\leq T\leq \left(j+1\right)\times10^9$~K. We
selected the species to plot in the figures from among those with
non-negligible abundances that present a large difference between
$m_i$ and $m'_i$. We also required that the species covered a wide
range of $Z$. Finally, we choose the same species to explore the
sensitivities of all three reactions being considered: $^{20}$Ne,
$^{24}$Mg, $^{26}$Al, $^{30}$Si, $^{32}$P, $^{35}$S, $^{38}$Ar, and
$^{47}$Ti.

We show in Fig.~\ref{fig13} the results for the reaction
\mbox{$^{30}\text{Si}+\text{p}\rightleftarrows{}^{31}\text{P}+\gamma$}. We
note that all the nuclei follow the same behavior with respect to
the temperature window in which the reaction rate is modified. There
is a modest variation of order 20\% of the yields for the window
centered on $1.5\times 10^9$~K, and a large increase for the next window,
centered on $2.5\times 10^9$~K. In the window centered on $3.5\times 10^9$~K
there are some species which experience large variations of their
yields while others are scarcely affected at all. The yield of
$^{20}$Ne (open triangles) shows a peculiar behavior, with $r_{ij}<0$
in the first thermal window, meaning that increasing the rate of the
reaction only at low temperatures ($T\leq2\times 10^9$~K) results in a
variation of the yield of $^{20}$Ne of opposite sign as that obtained
if the reaction rate is increased for any temperature. Note that the
sign of $r_{ij}$ of $^{20}$Ne in the next window is positive, and it
has the largest $r_{ij}$ among the species shown in the figure:
increasing the reaction rate only in the interval $2\times 10^9\leq
T\leq3\times 10^9$~K produces a change of the yield of this species that is
as much as that obtained by increasing the reaction rate for all
temperatures.  Modifying the rate on thermal windows above $4\times 10^9$~K
has no effect on any of the final abundances of the species.

We show as well in Fig.~\ref{fig13} the ratio of the rates belonging to
the three prescriptions adopted for the rate of the reaction
$^{30}\text{Si}+\text{p}\rightleftarrows{}^{31}\text{P}+\gamma$, which
were discussed in Section~\ref{laspg} and in Table~\ref{tab31}. The
uncertainty in the rates derived from these different prescriptions is
more or less uniform for temperatures above $\sim2\times 10^9$~K. The rate
from \cite{rath} differs most from that based on \cite{il01} at
temperatures where the yields are most sensitive to this reaction
rate. However, the discrepancy between these rates is much lower than
the factor of ten used in our simulations, thus we believe that
the supernova yields should not be affected by any reasonable future
change of this reaction rate.

Figure~\ref{fig14} summarizes the results for the reaction
$^{20}\text{Ne}+\alpha\rightleftarrows{}^{24}\text{Mg}+\gamma$. The
most noticeable difference with respect to Fig.~\ref{fig13} is the
behavior and range of the variations of the yield of $^{38}$Ar
(asterisks). The maximum sensitivity of this species occurs in the
temperature window $4\times 10^9\leq T\leq5\times 10^9$~K, where the
change of its yield reaches a value seven times larger than the change
with a rate modified at all temperatures. This is compensated by the
fact that modifying the rate at temperatures in the interval
$2\times 10^9\leq T\leq4\times 10^9$~K produces a change of the yield of
$^{38}$Ar of opposite sign. The rest of nuclei plotted show a behavior
similar to the one in Fig.~\ref{fig13}, with maximum $|r_{ij}|\sim1.5$
($^{32}$P, solid pentagons).

As revealed by Fig.~\ref{fig14}, the different prescriptions for the
rate of the reaction
$^{20}\text{Ne}+\alpha\rightarrow{}^{24}\text{Mg}+\gamma$ show a maximum
discrepancy by a factor of $\sim10$ in the temperature range $10^9$ - $10^{10}$~K. However, both
the rate from \cite{cyb10} and that from \cite{nacr} differ from the rate given by \cite{il10} by
a similar factor in the interval $2\times10^9\leq T\leq4\times10^9$~K.
 
Finally, we show in Fig.~\ref{fig15} the results for the reaction
$^{24}\text{Mg}+\alpha\rightleftarrows{}^{27}\text{Al}+\text{p}$. It
highlights the behavior of $^{30}$Si (open pentagons) whose yield
experiences variations up to seven times larger in the temperature
window $3\times 10^9\leq T\leq4\times 10^9$~K than when the rate is
modified for all temperatures. Furthermore, the variation of its yield
changes sign if the thermal window is $2\times 10^9\leq T\leq3\times
10^9$~K, still reaching $|r_{ij}|\sim7$. The rest of the nuclei show a
behavior similar to that in Fig.~\ref{fig13}, although their maximum
$|r_{ij}|$ is now a bit larger,
$\text{max}\left(|r_{ij}|\right)\sim2$.

The ratio of the rates of the reaction
\mbox{$^{24}\text{Mg}+\alpha\rightleftarrows{}^{27}\text{Al}+\text{p}$} from
\cite{cyb10} and \cite{il01} agree quite well for $T\gtrsim 2\times 10^9$~K
(see Fig.~\ref{fig15}). On the other hand, the rate from \cite{laur}
differs from the other two by more than two orders of magnitude at
high temperatures.  However, in the range of temperatures where the
abundances plotted are most sensitive to this reaction, $2\times 10^9\leq
T\leq4\times 10^9$~K, their discrepancy is less than a factor of $\sim30$. It
is interesting to note that, using these rates, the change of the
sensitivity of the yield of $^{30}$Si in the two temperature windows
with $|r_{ij}|\sim7$ almost compensates each other, with the result
that the final yield of this species is negligibly affected by using
the rate from \cite{laur} instead of that from \cite{il01} (thus, it
does not appear in the row reserved for this reaction in
Table~\ref{tab31}).

\section{Conclusions}\label{conc}

We have computed the chemical composition of a reference SNIa model
with a nucleosynthetic post-processing code that takes as inputs the
nuclear data and the thermodynamic trajectories of each mass
shell. Our reference SNIa model is the one-dimensional
delayed-detonation model DDTc in \cite{bad05c}. 
Our nucleosynthetic calculations include 3138 reactions during the
integration of the nuclear evolutionary equations 
but only 1096 of them can contribute significantly to the
nucleosynthesis of the supernova model. In this paper, we have
explored the sensitivity of the SNIa explosive nucleosynthesis to
simple variations on nuclear reaction rates (either a fixed
enhancement factor or one that decreases monotonously with
temperature) and comparisons between different theoretical and
experimental prescriptions of the rates.

The nucleosynthesis resulting from our Type Ia supernova model is
quite robust with respect to variations of nuclear reaction rates,
with the exception of the fusion of two $^{12}$C
nuclei. The energy of the explosion changes by less than $\sim4\%$
when the rates of the reactions $^{12}\text{C}+{}^{12}\text{C}$ or
$^{16}\text{O}+{}^{16}\text{O}$ are multiplied by a factor of 10 or
0.1. The changes in the nucleosynthesis due to the modification of the
rates of these fusion reactions are also quite modest, for instance
no species with a mass fraction larger than 0.02 experiences a
variation of its yield larger than a factor of two. 
The robustness of the production of $^{56}$Ni and many other Fe-group
isotopes stands out. If the enhancement factor of the rates is a decreasing
function of temperature, the effect on the yields of all the
species is even less than with a fixed enhancement factor. For
instance, no species experiences a
variation of its yield larger than 30\% with respect to
the rate of $^{16}\text{O}+{}^{16}\text{O}$ or with respect to the
$3\alpha$ rate. We have checked that the modifications in the
nucleosynthesis produced when the fusion rates are modified only in
the nucleosynthetic code are quite similar to those obtained when the
rates are modified in the full supernova simulation.

We provide the sensitivities of the yield of each relevant nuclear species
ejected in the supernova with respect to those nuclear reactions that affect it.  In general, the
yields of
Fe-group nuclei are less sensitive than the yields of intermediate-mass
elements. However, the yields of $^{28}$Si, and $^{32}$S, as well as
$^{54}$Fe, $^{56}$Ni, and $^{58}$Ni do not change appreciably within the range
of enhancement factors of the nuclear reaction rates explored
here. 
The only reactions for which the relative change of the abundance of any species is larger (in
absolute value) than the relative change in the rate $(|D_{i}|>1)$ is
$^{30}\text{Si}+\text{p}\rightleftarrows{}^{31}\text{P}+\gamma$.  In
general, radiative captures of protons are the group of reactions with
the largest influence on the supernova yields. Other important groups
of reactions are radiative captures of $\alpha$ particles, most
notably the reactions
$^{20}\text{Ne}+\alpha\rightleftarrows{}^{24}\text{Mg}+\gamma$, for
which there are 33 species whose yields change by more than 12\%, and the
$\left(\alpha,\text{p}\right)$ reactions
$^{13}\text{N}+\alpha\rightleftarrows{}^{16}\text{O}+\text{p}$,
$^{20}\text{Ne}+\alpha\rightleftarrows{}^{23}\text{Na}+\text{p}$,
$^{23}\text{Na}+\alpha\rightleftarrows{}^{26}\text{Mg}+\text{p}$,
$^{24}\text{Mg}+\alpha\rightleftarrows{}^{27}\text{Al}+\text{p}$, and
$^{27}\text{Al}+\alpha\rightleftarrows{}^{30}\text{Si}+\text{p}$.

We have discussed the sensitivity of the nucleosynthesis to the rates
of reactions that take part in the most relevant nucleosynthetic processes
in SNIa. Modifying the rates of the reactions that bridge the gap
between QSE groups in explosive silicon burning has a very limited
effect on the yields. They affect mainly nuclei that belong to this
gap. Changing the rate of reactions relevant for
the alpha-rich freeze-out does not produce important changes on the
abundances of nuclear species either. 
This can be explained both by the small amount of matter that goes through alpha-rich freeze-out
from NSE and by the small excess of alpha particles at freeze-out (see Fig.~\ref{fig1}, where the
maximum value of $Y_\alpha$ does not attain $0.01$~mol~g$^{-1}$).
It is as well remarkable that the
reactions involving nuclei with $Z>22$ have a tiny
influence on the supernova nucleosynthesis.

We have relied on the JINA REACLIB Database to estimate realistic
uncertainties of the most relevant reaction rates, by comparing the
most recent prescriptions for the rates of these reactions. We have
paid special attention to the reactions
$^{30}\text{Si}+\text{p}\rightleftarrows{}^{31}\text{P}+\gamma$,
$^{20}\text{Ne}+\alpha\rightleftarrows{}^{24}\text{Mg}+\gamma$, and
$^{24}\text{Mg}+\alpha\rightleftarrows{}^{27}\text{Al}+\text{p}$,
especially the last one for which there is a discrepancy of up to
three orders of magnitude between the rates due to \cite{il01} and
\cite{laur}. In spite of this large difference of rates, the maximum
change in the yields is only 52\%, belonging to $^{35}$S.

There are two main reasons for the small relative impact of the
uncertainties of individual nuclear reaction rates on the supernova
yields. First, the nuclear flows that determine the final abundances
during the supernova explosion are driven collectively by many
reactions, which are much faster than the hydrodynamic explosion
timescale because of the high temperatures involved. The relevance of
any individual rate is much diluted within this large pool of
reactions. 
A similar conclusion was reached by \cite{hof99} in the context of Type II supernovae. They cite
three major causes, which we can adapt to nucleosynthesis in SNIa: 1) the dominant nuclear flows
are governed by the fusion reactions of the fuel, carbon and oxygen, while the rest are only
perturbations on the main stream, 2) the nuclear flow follows the path of least resistance, i.e.
if one reaction rate drops by a large factor there is always another reaction capable of playing
its role, and 3) if the freeze-out from high temperatures is fast enough, the rates of individual
reactions are much less important than the properties of nuclei (binding energy, partition
function).

Second, there are narrow temperature ranges where the yields
are more sensitive to the rates. For instance, the temperatures at
which a modification of the rate of the above-mentioned three
reactions has a larger impact are in the range $2\times 10^9\lesssim
T\lesssim4\times 10^9$~K (see Figs.~\ref{fig13} to \ref{fig15}).
One kind of rate uncertainty we have not explored is that due to the
erroneous location of a resonance. Such a kind of error might
originate an increase of the rate (with respect to the presently
recommended one) in a temperature range and a decrease in a contiguous
one. If this were the case, the changes of the yields of some species
might be exacerbated. Thus, this kind of error in the
nuclear reaction rates might be the most relevant with respect to the
supernova yields.

We conclude that the explosion model chiefly determines the element
production of Type Ia supernovae, and derived quantities like their
luminosity, while the individual nuclear reaction rates used in the
simulations have a small influence on the kinetic energy and final
chemical composition of the ejecta.  Often, it is argued that
discrepancies of up to a factor of two between isotopic ratios in SNIa
ejecta and those in the solar system, especially within the Fe-group,
can be attributed to uncertainties in nuclear reaction rates.  Our
results show that the uncertainty in individual thermonuclear reaction
rates cannot account for this factor. It remains to be seen if the
yields are more sensitive to uncertainties in nuclear masses, weak
interaction rates, or to the simultaneous modification of the bulk of
thermonuclear reaction rates. The sensitivity of the supernova
nucleosynthesis to simultaneous random modifications in the bulk of
thermonuclear reaction rates will be the subject of future work. In
this respect, it is interesting our finding that the most influential
reactions depict a clear path in a plot $Z$ vs. $A$ (Fig.~\ref{fig12}),
going from $^{12}$C up to $^{37}$Ar through many branches involving
mainly reactions with $\alpha$ particles plus the fusion reaction
$^{12}\text{C}+{}^{12}\text{C}$. Modifications of these rates 'in phase'
may have interesting consequences for the chemical composition of
supernova ejecta.

Finally, it is worth noting that reaction rate variations may also have an impact on 
the hydrostatic evolution of the progenitor of the exploding white dwarf. Given the robustness of
the explosive yields, it may well be that changes 
in progenitor evolution are the largest source of reaction rate 
sensitivity in thermonuclear supernovae.

\begin{acknowledgments}
  This work has been partially supported by a MEC grant, by the
  European Union ERDF funds, and by the Generalitat de Catalunya. GMP
  is partly supported by the Deutsche Forschungsgemeinschaft through
  contract SFB 634, by the Helmholtz International Center for FAIR
  within the framework of the LOEWE program launched by the state of
  Hesse, and by the Helmholtz Association through the Nuclear Astrophysics
Virtual Institute (VH-VI-417).
\end{acknowledgments}


%

\end{document}